\documentclass[fleqn,10pt]{wlscirep}
\pdfoutput=1
\usepackage[T1]{fontenc}
\usepackage{mathtools}
\usepackage{amsthm}
\usepackage{tikz}
\usetikzlibrary{positioning, calc}
\usepackage{pgfplots}
\pgfplotsset{compat=1.18}
\usepackage{subcaption}
\usetikzlibrary{decorations.pathreplacing, decorations.markings, calc, arrows, shapes.geometric}
\usetikzlibrary{backgrounds}
\usepackage{url}
\usepackage{cleveref}
\usepackage[switch]{lineno}

\newcommand{\U}{\mathcal{U}}
\newcommand{\V}{\mathcal{V}}
\newcommand{\G}{\mathcal{G}}
\newcommand{\Enc}{\mathcal{E}}
\newcommand{\Dec}{\mathcal{D}}
\newcommand{\Proc}{\mathcal{P}}
\renewcommand{\d}{\,\textup{d}}
\newcommand{\R}{\mathbb{R}}

\renewcommand{\thefootnote}{\fnsymbol{footnote}}

\csname title\endcsname{Missing Physics Discovery through Fully Differentiable Finite Element-Based Machine Learning}
\author[1,2,3]{Ado Farsi\thanks{Email: \href{mailto:ado.farsi@imperial.ac.uk}{\nolinkurl{ado.farsi@imperial.ac.uk}}}}
\author[3,4,5]{Nacime Bouziani\thanks{Email: \href{mailto:n.bouziani18@imperial.ac.uk}{\nolinkurl{n.bouziani18@imperial.ac.uk}}. This work was conducted independently of Amazon.}}
\author[4]{David A.~Ham\thanks{Email: \href{mailto:david.ham@imperial.ac.uk}{\nolinkurl{david.ham@imperial.ac.uk}}}}
\affil[1]{Imperial College London, Earth Science and Engineering Department, London, United Kingdom}
\affil[2]{University College London, Earth Sciences Department, London, United Kingdom}
\affil[3]{Tanuki Technologies, London, United Kingdom}
\affil[4]{Imperial College London, Department of Mathematics, London, United Kingdom}
\affil[5]{Amazon Science UK, London, United Kingdom}

\begin{abstract}
Modelling physical systems with partial differential equations (PDEs) is central to science and engineering, yet in most real applications the PDE model is incomplete: relationships such as constitutive or thermal laws are unknown. Existing surrogate approaches close this gap by learning the PDE solution from data, sometimes with added physical constraints, but they remain tied to a specific configuration (geometry, boundary conditions, discretisation) and recover the solution rather than the missing physics itself.
We introduce FEML, an end-to-end differentiable framework that couples the known PDE (the system's known physics) with a machine-learned operator for the missing physics. Embedding the PDE solver into training lets this operator be learned directly from the PDE solution, even when its own output cannot be measured---for example, stress when learning constitutive laws. Because the operator, unlike the PDE model, is independent of the system configuration, a law learned in one setting transfers zero-shot to new geometries, boundary conditions, and discretisations, and can be inspected by domain specialists. FEML represents the operator with structure-preserving operator networks (SPONs), which retain key continuous properties at the discrete level and enable learning over complex geometries and meshes.
We demonstrate FEML across solid mechanics and thermal transport. From synthetic experiments we progressively discover an elastoplastic law---the nonlinear elastic response, then the plastic hardening law---and compose the two operators into a foundation constitutive model that transfers zero-shot to a three-dimensional torsion problem. Moving to real data, we learn coupled plastic-hardening and ductile-damage laws directly from a benchmark shear-coupon test, reproducing the measured response, including post-peak softening, to within the experimental scatter. Finally, we recover a temperature-dependent conductivity from transient heat-flow data and apply symbolic regression to the learned operator to extract a closed-form law matching the ground truth.
\end{abstract}

\begin{document}

\flushbottom
\maketitle
\setcounter{footnote}{0}
\renewcommand{\thefootnote}{\arabic{footnote}}
\csname thispagestyle\endcsname{empty}
\section*{Introduction}

Many science and engineering problems rest on well-understood physics yet contain unresolved or missing relationships that are unknown or cannot be readily expressed in mathematical form.

Modelling complex physical systems is traditionally achieved through physics-based models, such as partial differential equations (PDEs). However, such models require exact physical understanding and a fully specified mathematical formulation to accurately describe the physics of interest-conditions that are often difficult to satisfy in practice. Data-driven approaches, particularly machine learning (ML) models, offer an alternative by inferring input–output relationships directly from observable data, enabling prediction of system behaviour even when the governing physics are unknown or incomplete. However, purely data-driven models have their own drawbacks, including the need for large training datasets, predictive power limited to regimes represented in the training data, and reduced interpretability.

Various strategies have been proposed to incorporate physical and mathematical knowledge into machine-learning algorithms \cite{karniadakis2021physics, bouziani_diffprogram_2024, Quarteroni2025, belbute-peres_combining_2020, bouziani_physics-driven_2023, crilly2024learning}. In general, these approaches seek to address the shortcomings of purely data-driven models by training surrogate ML models with both data and physical constraints. Although this reduces the volume of data required, the resulting models typically learn the \emph{solution of a particular PDE configuration} rather than the \emph{unknown operator itself}.

A distinct but related line of research focuses on the direct discovery of physical laws from data. Equation-discovery methods, such as the Sparse Identification of Nonlinear Dynamical Systems (SINDy) framework \cite{brunton_discovering_2016}, use sparse regression over a library of candidate nonlinear functions to recover governing equations in closed symbolic form from time-series measurements. Recent extensions have incorporated physics-informed priors, encoding known conservation laws or symmetries into neural-network architectures, to discover nonlinear PDE dynamics from scarce or noisy data \cite{rao_discovering_2023}. In parallel, physics-informed neural networks (PINNs) have been used to calibrate constitutive material models by embedding conservation laws into the loss function \cite{haghighat_constitutive_2023}. While these approaches share with the present work the objective of uncovering hidden physical relationships, they differ in scope and methodology. Equation-discovery methods require a pre-specified dictionary of candidate terms and are most effective when the governing variables are directly observable; extending them to PDE-governed systems where the operator of interest (e.g., a stress--strain relation) is not directly measurable, and can only be inferred through the PDE solution, is considerably more challenging. PINN-based constitutive calibration, as formulated by Haghighat et~al.~\cite{haghighat_constitutive_2023}, operates at the material-point level and requires direct stress--strain data from homogeneous tests as training input; however, stress is an internal quantity that cannot be directly measured in most experimental settings. By contrast, the framework introduced in this work embeds the unknown operator inside a full boundary-value problem solved by the finite element method, and can therefore learn the constitutive law from indirect, experimentally accessible measurements such as displacement fields or boundary forces, without ever requiring direct observation of stress. These lines of work are nonetheless complementary: for instance, having identified a material law via the present framework, one can subsequently apply sparse-identification or symbolic-regression techniques to learn a closed-form expression from the learned neural operator, as we demonstrate in Section~\ref{sec:symbolic_regression_thermal}.

In many applications, the missing physics does not correspond to an explicit input–output map, but rather manifests as an \textit{internal operator}---a hidden functional relationship (such as a constitutive law or thermal material property) that governs interactions within the PDE system itself. This operator acts locally within the governing equations, coupling quantities (e.g., stresses or heat fluxes) through unknown dependencies that cannot be directly observed or isolated experimentally. Existing methods such as Fourier Neural Operators (FNO) or Physics-Informed Neural Operators (PINO) \cite{li_fourier_2021, li_physics-informed_2023} address a different learning problem: they learn the full \emph{solution map} from PDE inputs to PDE solutions, producing surrogate models for fast evaluation within a given training distribution. In addition, they rely on pointwise discretisations of input–output fields that can discard the underlying continuous function-space structure, leading to inconsistencies and degraded operator approximation \cite{bartolucci_representation_2023}. While effective as surrogate models, these approaches are designed to approximate the entire configuration-specific mapping rather than to identify the hidden operator itself; consequently, they do not produce a reusable physical law that can be transferred across different problem configurations (e.g., different geometries, boundary conditions, or discretisations) or inspected by domain specialists for scientific interpretation. Similarly, Raissi~et~al.~\cite{raissi_physics-informed_2019} showed that physics-informed neural networks (PINNs) can be used for inverse coefficient identification, but their formulation is primarily a pointwise residual minimisation method in strong form, not a variational (weak-form) method. Moreover, it learns the solution and unknown coefficients jointly for a single inverse instance, rather than amortising a reusable operator over a family of configurations; in that sense it does not leverage the function-space and operator-learning structure that later enables discretisation transfer and rapid reuse across configurations. The objective of the present work is instead to recover only the unknown relationship, such that (i) the learned operator can be transferred to other problem configurations, and (ii) the resulting model can be inspected and constrained by domain experts for downstream analysis. Learning such operators from data requires coupling traditional PDE solvers with machine learning models. Since the operator of interest is embedded within the PDE, measurable physical quantities, from which a training loss can be defined, can be obtained by solving the governing equations. This coupling necessitates \textit{end-to-end differentiability} through both the PDE solver and the embedded ML operator, enabling gradients to propagate across the entire system during training \cite{bouziani_diffprogram_2024}.

In this work, we introduce Finite Element-Based Machine Learning (FEML), a general end-to-end differentiable framework for learning missing physics in systems with partially known governing laws. FEML cleanly separates known physics, expressed as PDEs in weak form and discretised with the finite element method (FEM), from unknown relationships, which are represented by ML operators embedded within the FEM solver. By embedding the PDE solver in the training loop and providing end-to-end differentiability, FEML enables learning from indirect measurements when operator inputs/outputs are not directly observable. Our framework employs \emph{structure-preserving operator networks} (SPONs) to model the missing-physics operator. SPONs preserve key continuous properties at the discrete level, enable learning over complex geometries and meshes, and permit zero-shot generalisation across different discretisations (mesh resolutions and/or finite element choices). In addition, SPONs provide theoretical guarantees on operator approximation for a given training discretisation \cite{bouziani2024structure}.

The FEML framework is also particularly suited when some of the quantities involved in the unknown relationship are not directly measurable, allowing one to learn from data on related, measurable quantities within the same system. An example is the discovery of material constitutive laws. Here, the unknown relationship is that between strains (i.e.\ material deformation) and stresses (the internal forces reacting to deformation); the latter, although it can be estimated under simplified loading conditions, is not directly measurable. The idea is to use known physical relationships to learn the constitutive law from measurable quantities such as displacements and applied forces (this will be covered in Section~\ref{sec:learning_constitutive_models}). One may then wish to examine the ML operator on its own to uncover the previously hidden relationships, for example by applying symbolic regression or sparse-identification techniques (see Section~\ref{sec:symbolic_regression_thermal}). In the context of constitutive modelling, this could enable the extrapolation of generalised stress–strain laws. One might also leverage the learned operator as a foundation model, trained on data from simplified laboratory tests, to predict material behaviour under different geometries and loading conditions without retraining (see Section~\ref{sec:torque_holed_plate}). This has potential practical applications in many fields, such as subsurface engineering, where integrating sparse prior knowledge with limited field measurements is critical.

Our framework relies on Firedrake \cite{FiredrakeUserManual, bouziani_escaping_2021} and allows for end-to-end differentiable coupling with ML architectures implemented in PyTorch \cite{PyTorch} and JAX \cite{jax2018github}. To the best of our knowledge, the FEML framework is the first to support the bidirectional coupling of state-of-the-art general finite element solvers and arbitrary machine learning architectures in an end-to-end differentiable manner. In contrast, most existing efforts have focused on deploying the algorithmic differentiation pipelines of machine learning frameworks to yield differentiable physics constraints, often specialised to particular applications (such as \textit{XLB} \cite{ataei_xlb_2024}, \textit{PhiFlow} \cite{holl_learning_2020}, \textit{Adept} \cite{joglekar_machine_2023}). A similar approach to ours has been presented in \cite{rackauckas_universal_2021}, although it lacks the adjoint capabilities required for differentiation through constraints based on PDEs, and in \cite{belbute-peres_combining_2020}, although it is limited to a restricted set of fluid simulations.

This work is organised as follows. Section~\ref{sec:framework} presents our FEML framework for learning missing physics by embedding trainable ML operators within PDE systems. Section~\ref{sec:learning_constitutive_models} illustrates its application to solid mechanics through a sequence of experiments that demonstrate how FEML enables the progressive discovery of new physics: a learned physical law can be composed into a new model to discover further, previously inaccessible physical laws. Concretely, the elastic constitutive law learned in a first experiment is frozen and embedded as known physics in a second model, whose sole trainable component is the plastic hardening law that governs material behaviour beyond the elastic limit. The two pretrained operators are then composed into a foundation constitutive model and deployed zero-shot on an unseen three-dimensional problem, illustrating how individually discovered laws can be assembled into increasingly complete physical descriptions. As a demonstration of real-world applicability, the framework is applied to \emph{real} experimental data from a benchmark shear-coupon test, where both a plastic-hardening law and a ductile-damage law are learned directly from the measured response, extending the discovered description to post-peak softening within a finite-strain formulation. Section~\ref{sec:learning_thermal_properties} extends the examples to transient thermodynamics by learning nonlinear thermal conductivities from noisy temperature fields, and further demonstrates how a closed-form expression can be learned from the neural operator via symbolic regression (Section~\ref{sec:symbolic_regression_thermal}). Overall, the examples increase in complexity to progressively build intuition for the proposed framework, from an initial setting with simple geometry and boundary/loading conditions to a final case involving a complex multi-component domain and a time-dependent governing equation with fluctuating boundary conditions. Finally, Section~\ref{sec:conclusions} summarises our findings and outlines directions for future work. All examples presented herein can be implemented and executed using the Firedrake finite element framework.

\section{Results}

\subsection{Framework}
\label{sec:framework}






We introduce the Finite Element-Based Machine Learning (FEML) framework, a general and end-to-end differentiable approach for learning missing physics, that is, unknown relationships between quantities in systems where partial physical knowledge is available. Our main motivation is to learn operators that model such unknown relations. We consider an operator $\G:\U\to \V$, where $\U$ and $\V$ are (typically infinite-dimensional) Hilbert spaces of functions defined on a bounded domain $\Omega\subset \R^d$ in spatial dimension $d\in\{1,2,3\}$. For sake of simplicity, we consider $\U$ and $\V$ to be defined on the same domain $\Omega$, but different bounded domains and spatial dimensions can be considered. We introduce a framework for approximating $\G$ by a learnable operator $\G_{\theta}:\U_{h} \to \V_{h}$ of parameters $\theta$, where $\U_{h}$ and $\V_{h}$ are finite element spaces arising from the discretisation of the spaces $\U$ and $\V$, respectively.

Such operators can traditionally be learned in a supervised manner from direct observations, i.e. a set of input-output pairs $\{(x_{i}, \mathcal{G}(x_{i}))\}_{i}$, with $x_{i} \in \U$. However, in practice, many input–output signals from unknown operators cannot be accessed directly via experiments, since they form only part of a larger physical model governed by a PDE and consequently, we cannot learn them directly. For example, constitutive laws describing the stress-strain relationship of a material are generally unknown, and the resulting stress tensor, which encodes internal forces under deformation, cannot be measured directly. In contrast, the displacement field is measurable and satisfies a PDE that incorporates this constitutive relation. Therefore, to learn the constitutive law, we embed our ML model within the PDE solver: during training, the ML model predicts the stress tensor required by the solver to compute the corresponding displacement, and we compute a loss against the measured displacement to update the model.

Our aim is to learn $\G_{\theta}$ under some loss $\mathcal{L}(u_{\theta}, u^{obs})$, where $u_{\theta}$ is a PDE solution and $u^{obs}$ are observable data, i.e. we have:

\begin{equation}
    \label{eq:var_form}
    \begin{aligned}
        F(u_{\theta}, \G_{\theta}(u_{\theta}); v) = 0 & \quad \forall v \in U_{h}
    \end{aligned}
\end{equation}

where \eqref{eq:var_form} is the variational form of the PDE with $F$ the residual. This PDE can be linear or nonlinear, steady or time-dependent. In practice, one can also equip it with boundary conditions but we have simplified the setup description for sake of simplicity.

We propose a general framework and the loss function \( \mathcal{L} \) can be defined in different ways depending on the specific problem. Figure~\ref{fig:framework_schematic} shows a schematic of the proposed framework. For example, in the case of learning a constitutive law in Section~\ref{sec:load_controlled}, $F$ is the residual of the PDE that describes the equilibrium equation with some boundary conditions in the simulated domain. Moreover, \( \mathcal{L} \) quantifies the discrepancy between experimental displacement measurements and the corresponding solution of the FEM solver, and \( \mathcal{G}_\theta \) is an unknown constitutive law.

Our framework combines PDE modelling of the physical problem of interest with ML modelling of the operator to approximate. Minimising the loss $\mathcal{L}(u_{\theta}, u^{obs})$ for training requires computing the gradient of the loss with respect to the parameters, i.e. $\frac{\d \mathcal{L}}{\d \theta}$, which by chain rule requires the gradient of the PDE solution with respect to the parameters, i.e. $\frac{\d u_{\theta}}{\d \theta}$, which in turn necessitates the gradient of the ML operator, i.e. $\frac{\d \G_{\theta}}{\d \theta}$. In other words, learning $\G_{\theta}$ requires end-to-end differentiability of the entire system, i.e. being able to differentiate through the PDE solution $u_{\theta}$ and through the ML components of the system to compute $\frac{\d \mathcal{L}}{\d \theta}$. How this end-to-end gradient flow is realised in practice is detailed in Section~\ref{sec:differentiable_coupling}.

Many real-world problems and engineering applications require the use of advanced numerics with state-of-the-art capabilities for PDE modelling. The simulation of complex physical systems by coupling advanced numerics for PDEs with state-of-the-art machine learning demands the composition of specialist PDE solving frameworks with industry standard machine learning tools. Hand-rolling either the PDE solver or the ML model will not cut it. Our framework introduces a generic differentiable programming interface that allows to combine the state-of-the-art Firedrake framework for PDE modelling, with different ML frameworks, including PyTorch \cite{PyTorch} and JAX \cite{jax2018github} deep learning libraries. As a result, it provides scientists and engineers with an efficient and highly productive way to learn operators combining FEM operations, e.g. solving a PDE using FEM, with ML algorithms, thanks to end-to-end  differentiability and benefiting from state-of-the-art performance of both FEM and ML libraries.
\begin{figure}[!htbp]
    \begin{center}
        \begin{tikzpicture}
        
            \def\femsolvercolor{green!30}
            \def\abstractioncolor{blue!20}
            \def\physicalmodelcolor{yellow!50}
            \def\mloperatorcolor{orange!50}
            \def\observablecolor{yellow!50}
            \def\predictioncolor{green!30}
            \def\optimisercolor{orange!50}
            \def\losscolor{green!30}
            \def\greyboxcolor{gray!30}
        
            \def\textsize{7mm}
            \def\boxwidth{10cm}  
            \def\boxheight{7cm} 
        
            \def\absspace{\textsize}
            \def\abswidth{\boxwidth - 2*\absspace}  
            \def\absheight{\boxheight - 2*\absspace} 
        
            \def\physpace{\textsize}
            \def\physwidth{\boxwidth - 2.5*\physpace}  
            \def\physheight{\boxheight - 2.5*\physpace} 

            \def\femsolverheight{0.8*\boxheight} 
            \def\physheightInner{0.8*\physheight} 

            \def\femsolverheight{0.8*\boxheight} 
            \def\physheightInner{0.8*\physheight} 
        
            \def\mlwidth{0.4*\physwidth}  
            \def\mlheight{0.8*\physheight} 
        
            \def\bottomwidth{0.25*\boxwidth}  
            \def\bottomheight{0.25*\boxheight} 
        
            \def\bottomspacing{(\boxwidth - 3*\bottomwidth)/2}
        
            \tikzset{
                thickbox/.style={draw, very thick, rounded corners},
                thickarrow/.style={->, very thick}
            }
        
            \node[thickbox, fill=\femsolvercolor, minimum width=\boxwidth, minimum height=\boxheight, anchor=center] (femsolver) 
                at (0,0) {};
            \node[above=0.2cm of femsolver.south] {\textbf{FEM Solver}};

      \node[thickbox, fill=\femsolvercolor,
            minimum width=1.25*\mlwidth, minimum height=1.5*\mlheight,
            anchor=east, align=center] (leftbox)
        at ($(0,0) + (-0.4cm - 0.5*\boxwidth,-0.3cm)$) {
                \\\textbf{Firedrake}\\
                FEM library.\\\\
                State-of-the-art\\
                numerics:\\
                $\bullet$ \textit{Advanced mesh}\\
                \textit{$\&$ geometry}\\
                \textit{support};\\
                $\bullet$ \textit{Massive parallel}\\
                \textit{scalability};\\
                $\bullet$ \textit{Custom solvers}\\
                \textit{$\&$ preconditioners};\\
                $\bullet$ \textit{Extensive FE}\\
                \textit{discretisations};\\
                $\bullet$ \textit{High-level}\\
                \textit{variational form}\\
                \textit{definition}.};

      \begin{scope}[on background layer]
      \draw[fill=\femsolvercolor, opacity=0.3]
              ($(femsolver.north west)-(0,1mm)$) --
              ($(leftbox.north east)-(0,1mm)$)     --
              ($(leftbox.south east)+(0,1mm)$)     --
              ($(femsolver.south west)+(0,1mm)$)  --
        cycle;
        \end{scope}
        
            \node[thickbox, fill=\physicalmodelcolor, minimum width=1.1*\physwidth, minimum height=\physheight, anchor=center] (physical) 
                at (0, .05*\boxheight) {};
            \node[right=0.25cm of physical.west, align=center] {\textbf{Physical Model}\\
            \textit{$F(u_{\theta}, \G_{\theta}(u_{\theta}); v) = 0$} \textit{$\forall v \in \U_{h}$;}\\\\
            
            Embedding examples:\\
            $\bullet$ \textit{Governing equation};\\
            $\bullet$ \textit{Boundary conditions};\\
            $\bullet$ \textit{Initial conditions}.};
        
            \node[thickbox, fill=\mloperatorcolor, minimum width=\mlwidth, minimum height=\mlheight, anchor=center, align=center] (mloperator) 
                at (.45*\physwidth, .3*\physheight) {\\\textbf{ML Operator}\\
                $\mathcal{G}_\theta$\\\\
                Input examples:\\
                $\bullet$ $\mathcal{G}_\theta(u_{\theta})$;\\
                $\bullet$ $\mathcal{G}_\theta(\epsilon(u_{\theta}))$;\\
                $\bullet$ $\mathcal{G}_\theta(\alpha)$.};
            \node[thickbox, fill=\mloperatorcolor,
                  minimum width=1.25*\mlwidth, minimum height=1.5*\mlheight,
                  anchor=west, align=center] (greybox)
                at ($(mloperator.east)+(.6cm,-.4cm)$) {
                \\\textbf{PyTorch}\\
                $\&$
                \textbf{JAX}\\
                deep learning\\
                libraries.\\\\
                State-of-the-art\\
                ML:\\
                $\bullet$ \textit{Extensive}\\
                \textit{set of ML}\\
                \textit{architectures};\\
                $\bullet$ \textit{Rich set of}\\
                \textit{optimisers};\\
                $\bullet$ \textit{Distributed}\\
                \textit{training support};\\
                $\bullet$ \textit{Training loop}\\
                \textit{\(SFT, RL, etc.\)}};
        
            \draw[fill=\mloperatorcolor, opacity=0.6]
              ($(mloperator.north east)-(0,1mm)$) --
              ($(greybox.north west)-(0,1mm)$)     --
              ($(greybox.south west)+(0,1mm)$)     --
              ($(mloperator.south east)+(0,1mm)$)  --
              cycle;
            \node[thickbox, fill=\observablecolor, minimum width=\bottomwidth, minimum height=\bottomheight, anchor=north, align=center] (observable) 
                at ({-\bottomwidth - \bottomspacing}, {-0.5*(\boxheight + 1.5*\bottomspacing)}) {\textbf{Observable} \\ \textbf{Data} \\ $u^{obs}$};
        
            \node[thickbox, fill=\predictioncolor, minimum width=\bottomwidth, minimum height=\bottomheight, anchor=north, align=center] (prediction) 
                at (0, {-0.5*(\boxheight + 1.5*\bottomspacing)}) {\textbf{FEM} \\ \textbf{Prediction} \\ \textbf{$u_{\theta}$}};
        
            \node[thickbox, fill=\optimisercolor, minimum width=\bottomwidth, minimum height=\bottomheight, anchor=north, align=center] (optimiser) 
                at ({\bottomwidth + \bottomspacing}, {-0.5*(\boxheight + 1.5*\bottomspacing)}) {\textbf{ML} \\ \textbf{Optimiser} \\ \textbf{$\theta^*$}};
        
            \node[thickbox, fill=\losscolor, minimum width=\bottomwidth, minimum height=\bottomheight, anchor=north, align=center] (loss) 
                at (0, {-(0.55*\boxheight + 2*\bottomspacing + .5*\bottomheight)}) {\textbf{Loss} \\ \textbf{$\mathcal{L}(u_{\theta}, u^{obs})$}};
        
            \draw[thickarrow] (observable.north) -- ($(physical.south west)!0.2!(physical.south east)$);
            \draw[thickarrow] (observable.south) -- (loss.west);
            \draw[thickarrow] (femsolver.south) -- (prediction.north);
            \draw[thickarrow] (optimiser.north) -- (mloperator.south);
            \draw[thickarrow] (prediction.south) -- (loss.north);
            \draw[thickarrow] (loss.east) -- (optimiser.south);
        
        \end{tikzpicture}
        \end{center}
        
    \caption{Schematic of the proposed framework for embedding neural networks as trainable operators within PDE systems. The framework combines finite element solvers with machine learning models to learn ML operators from observable data.}
    \label{fig:framework_schematic}
\end{figure}

\subsubsection{Operator Learning over Finite Element Spaces}


Our framework differs from the traditional operator learning literature as it embeds the learnable operator $\G_{\theta}$ into a differentiable FEM solver, thereby allowing end-to-end differentiable FEM-based operator learning. Another notable difference is that our setting entails learning over finite element spaces, i.e. $\U_{h}$ and $\V_{h}$ result from finite element discretisations. Consequently, the operator $\G_{\theta}$ can be expressed as an encode-process-decode architecture \cite{bouziani2024structure}, with a learnable processor over the spaces induced by the degrees of freedom (DoF) of $\U_{h}$ and $\V_{h}$. More precisely, $\G_{\theta}$ can be defined as
\begin{equation}
  \label{eq:Gtheta_operator_definition}
  \mathcal{G}_{\theta}(f) = \Dec \circ \Proc_{\theta} \circ \Enc (f), \quad f\in \U_h,
\end{equation}
where $\Enc$ and $\Dec$ refer to the \emph{encoder} and \emph{decoder}, while $\Proc_{\theta} \colon \mathbb{R}^{n} \mapsto \mathbb{R}^{m}$ with $n=\dim(\mathcal{U}_{h})$ and $m=\dim(\mathcal{V}_{h})$ represents a learnable model of parameters $\theta$, known as the \emph{processor}. The encoder extracts the degrees of freedom of an input function $f \in \mathcal{U}_{h}$:
\begin{equation}
  \label{eq:encoder_definition}
  \Enc(f) = (f_{1}, \ldots, f_{n}),
\end{equation}
where $f_i = \langle f,\varphi_i\rangle$ corresponds to the Galerkin projection onto the $i$-th basis function $\varphi_i$. On the other hand, the decoder maps the predicted DoFs in $\V_{h}$ to the reconstructed solution $u \in \V_{h}$ as
\begin{equation}
  \label{eq:decoder_definition}
  \Dec(u_{1}, \ldots, u_{m}) = u,
\end{equation}
with $u(x) = \sum_{i=1}^{m} u_{i} \phi_{i}(x)$, for $x \in \Omega$, and where $(\phi_{i})_{1 \le i \le m}$ a basis of $\V_h$.
Such encode-process-decode operators are referred to as \textit{structure-preserving operator networks} (SPON) \cite{bouziani2024structure} as they preserve some key mathematical and physical properties of the operator $\G$ at the discrete level and offer explicit trade-off between accuracy and efficiency. Our framework inherits the properties of SPON such as zero-shot super-resolution and theoretical bounds on the approximation error of $\G_{\theta}$.

\textbf{Zero-shot super resolution.} $\G_{\theta}$ outputs a finite element function $u \in \V_{h}$ that can be evaluated at any point $x$ in the geometrical domain $\Omega$, i.e. by simply using $\smash{u(x) = \sum_{i=1}^{m} u_{i} \phi_{i}(x)}$. This property results from the FE discretisation of $\U$ and $\V$ and holds on complex geometries and independently of the mesh and resolution $\G_{\theta}$ was trained on. This relation is crucial for transferring solutions between different meshes and spatial discretisations, such as for zero-shot super-resolution, and it enables architectures that can seamlessly operate across multiple resolutions. In practice, implementing such a property for complex geometries and/or non-trivial FE discretisations is challenging. However, our differentiable programming coupling of Firedrake and ML software allows to implement that in a single line of code for arbitrary meshes and a wide range of FE discretisations.

\textbf{Approximation error.} 
Let $\Omega$ be an open bounded domain of $\R^{n}$, $\V = H^{k}(\Omega)$ and $\U \subset H^{k}(\Omega)$. Let $\G:H^{s}(\Omega)\to \V$ be a Lipschitz continuous operator for some $0\leq s\leq k$ and $0<\epsilon<1$, and $\U_{h}$ and $\V_{h}$ be conforming finite element spaces. Under mild assumption on $\U$ and assuming $\U_{h}$ and $\V_{h}$ satisfy the standard finite element hypotheses \cite{brenner2008mathematical}, one can show that there exists a learnable operator $\mathcal{G}_\theta:\U_h\to \V_h\subset \V$ with a number of parameters bounded 
\[|\theta|< C_1\epsilon^{-C_2/{h^{k}}^{n}}(\log(1/\epsilon)+1),\] such that for all $f\in \U$,
\begin{equation}
  \label{eq:approximation_error}
    \|(\G - \mathcal{G}_\theta\circ P_\U)(f)\|_{H^{s}(\Omega)}\leq C_3 h^{k-s}\left(\|f\|_{H^{k}(\Omega)} + \|\G(f)\|_{H^{k}(\Omega)}\right)  + \epsilon(h),
\end{equation}

where $P_\U:\U\to\U_{h_1}$ is the Galerkin interpolation. We refer to \cite{bouziani2024structure} for the general approximation theorem and its proof. The first term denotes the finite element error on the input-output spaces, which can be controlled via the finite element discretisation, e.g. a higher polynomial degree $k$ results in a higher convergence rate via $h^{k - s}$. On the other hand, the second term reflects the neural network's approximation error, which can be reduced by increasing the number of parameters. The left-hand side in \cref{eq:approximation_error} can be seen as an operator aliasing error \cite{bartolucci_representation_2023}, which can be explicitly controlled by the mesh resolution and the discretisation of the input-output spaces. It is worth noting that higher-order discretisations comes with higher convergence rate but also with an additional computational cost as it increases the number of DoFs and therefore the size of the input-output of the learnable processor $\mathcal{G}_\theta$. This trade-off between accuracy and efficiency is well known in the FEM literature and is inherited by our framework.

In the following sections, we present examples of the proposed framework applied to solid mechanics and thermodynamics. Section~\ref{sec:learning_constitutive_models} focuses on learning nonlinear constitutive laws, while Section~\ref{sec:learning_thermal_properties} demonstrates the learning of nonlinear thermal properties in transient heat conduction problems.

\subsection{Learning Materials Constitutive Laws}
\label{sec:learning_constitutive_models}

In this section, we demonstrate the application of the proposed framework to solid mechanics and materials science. Different materials exhibit different deformation behaviours under applied forces. For instance, the relationship between deformation and the internal forces can varies significantly when studying rocks, metals, or other engineered metamaterials. Accurately characterising these relationships is crucial in many engineering domains, from the design of resilient infrastructure in civil engineering to the development of aircraft structures in aeronautical engineering.

We consider an unknown material, representative of a closed-cell polymeric foam, whose full elastoplastic response must be discovered from experimental data. The material exhibits two forms of nonlinearity: (i)~an elastic softening modulus that degrades under compressive volumetric strain, and (ii)~plastic hardening governed by J2 flow with a Voce-type yield law. Rather than attempting to learn both nonlinearities at once, the experiments are designed to follow a \emph{progressive discovery} strategy in which each stage isolates and learns a single piece of unknown physics, building on the knowledge acquired in the previous stage. In Section~\ref{sec:displacement_controlled}, the material is loaded within its elastic regime via a displacement-controlled uniaxial test, so that the only unknown is the nonlinear elastic response; this first experiment learns the Young's modulus $E(\kappa)$ from measured reaction forces. In Section~\ref{sec:load_controlled}, the learned elastic operator is frozen—thereby removing the elastic response from the set of unknowns—and the material is loaded beyond its elastic limit in a Brazilian disc test, so that the only remaining unknown is the plastic hardening law; this second experiment also illustrates how \emph{a priori physical knowledge} can be encoded directly into the ML architecture: here, we assume that the material hardens monotonically under continued plastic deformation—an assumption that, in a real-world setting, might stem from preliminary test data, domain expertise, or theoretical considerations. A monotone neural network is used to parameterise the yield-stress evolution $\sigma_y(p)$, guaranteeing this property by construction rather than relying on a soft penalty or post-hoc projection. The learned yield-stress evolution is obtained from full-field displacement measurements. In Section~\ref{sec:torque_holed_plate}, the two pretrained operators are combined into a foundation constitutive model and deployed zero-shot on a three-dimensional cylindrical rod with a transverse keyhole-shaped through-slot subjected to torsion—a problem that differs from the training setup in geometry, dimensionality, and loading.

The four preceding experiments use synthetic data, which provides an exact ground-truth operator against which the learned model can be quantitatively verified. As a demonstration of real-world applicability, Section~\ref{sec:shear_coupon} applies the framework to \emph{real} experimental data from a benchmark shear-coupon test on a titanium alloy, where the constitutive law is genuinely unknown. This example serves two roles: it extends the discovered description beyond elasticity and hardening to \emph{ductile damage}---a second learned operator that captures post-peak softening within a finite-strain formulation---and it demonstrates that the framework operates on real laboratory measurements, not only synthetic ones. The two settings are complementary: the synthetic studies verify the method against known ground truth, while the real-data study demonstrates its practical applicability where no ground truth exists.

The formulation of the solid mechanics problem with the embedded ML constitutive model is detailed in Section~\ref{sec:appendix_solid_mechanics_problem_formulation}.

\subsubsection{Learning Nonlinear Elastic Laws from Loads in Displacement-Controlled Experiments}
\label{sec:displacement_controlled}

This first experiment targets the elastic regime of the material, where the sole unknown is the nonlinear relationship between strain and stress in the absence of permanent deformation. By restricting the loading to remain below the yield stress, plastic effects are excluded and the learning problem reduces to discovering the elastic constitutive law alone. The elastic operator learned here will serve as a fixed, known component in the subsequent plasticity experiment (Section~\ref{sec:load_controlled}).

We focus on displacement-controlled uniaxial tests—standard experiments in materials science, rock mechanics, and civil engineering to assess the compressibility and strength of materials.

\begin{figure}[!htbp]
    \centering
    \begin{subfigure}[b]{0.30\textwidth}
        \centering
        \begin{tikzpicture}[xscale=0.9, yscale=0.99]

            \fill[gray!20] (0, 0) rectangle (3, 5); 
            \draw[thick] (0, 0) rectangle (3, 5);

            \node at (1.5, -0.5) {\textbf{\( \boldsymbol{u} = \boldsymbol{0}\)}};
            \foreach \x in {0.25, 0.75, 1.25, 1.75, 2.25, 2.75} {
                \draw[thick] (\x+0.2, -0.2) -- (\x, 0.0);
            }

            \node at (1.5, 6.2) {\textbf{\( \boldsymbol{u} = (0, \bar{u}_i) \)}};
            \foreach \x in {0.25, 0.75, 1.25, 1.75, 2.25, 2.75} {
                \draw[thick,<-] (\x, 5.1) -- (\x, 5.8);
            }

            \draw[thick,<->] (3.75, 0) -- (3.75, 5) node[midway, right] {10 cm};
            \draw[thick,<->] (0, -1) -- (3, -1) node[midway, below] {5 cm};

        \end{tikzpicture}
        \subcaption{}
        \label{fig:uniaxial_test_schematic}
    \end{subfigure}
    \raisebox{0.8cm}{%
      \begin{subfigure}[b]{0.22\textwidth}
        \centering
        \includegraphics[height=5.20cm]{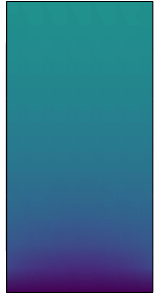}
        \subcaption{}
        \label{fig:appendix_learning_from_forces_displacements_0}
      \end{subfigure}
    }
    \raisebox{0.8cm}{%
      \begin{subfigure}[b]{0.31\textwidth}
        \centering
        \includegraphics[height=4.95cm]{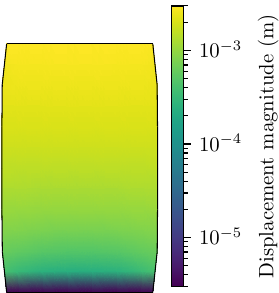}
        \subcaption{}
        \label{fig:appendix_learning_from_forces_displacements_5}
      \end{subfigure}
    }
    \caption{Displacement-controlled uniaxial test: (a) schematic of the 2D test setup with free vertical boundaries, a fixed bottom, and a prescribed displacement $\bar{u}_i$ at the top; displacement magnitude at the (b) first and (c) last loading increment of the training data. The deformation has been amplified by five times for visualisation.}
    \label{fig:uniaxial_test}

\end{figure}

We idealise the sample as a 2D rectangular domain, as depicted in Figure~\ref{fig:uniaxial_test_schematic}. The two vertical sides of the sample are free, while the bottom and top surfaces have imposed displacements. The bottom surface is fixed throughout the simulation. A prescribed displacement $\bar{u_i}$ is applied to the top surface, where $i$ is an integer from 1 to $N$ that represents the sequence of displacements applied during the experiment. The loading is kept within the elastic regime of the material (i.e.\ the maximum stress remains below the yield stress $\sigma_{y0}$). This configuration simulates the quasi-static loading condition in which time increments correspond to sequential deformation steps rather than physical time. Figure~\ref{fig:framework_schematic_displacement_controlled} illustrates the mathematical definition of the problem and the proposed framework for learning constitutive laws from displacement-controlled experiments.
The loss function is computed as:
\begin{equation}
    \mathcal{L} = \frac{1}{N} \sum_{i}^{N} \frac{\left| F_i^\text{fem} - F_i^\text{obs} \right|}{\left| F_i^\text{obs} \right|}
\end{equation}

where \( F_i^\text{fem} \) is the applied load predicted by the FEM model with the ML-based constitutive model at time step $i$, and \( F_i^\text{obs} \) is the corresponding synthetic experimental load.
It is important to note that the Poisson effects are not taken into account in this definition of the loss function and therefore the constitutive law is not univocally defined. In this example, the Poisson's ratio is assumed known ($\nu = 0.3$). This simplification is adopted because the uniaxial test provides no lateral displacement information from which $\nu$ could be inferred. In principle, $\nu$ could be learned jointly with $E$ by augmenting the loss with horizontal displacement measurements—for example from a digital image correlation (DIC) system—but we omit this additional data channel here to keep the example focused on the elastic softening law. A more general architecture that can learn both Lam\'e parameters from full-field displacements is used in Section~\ref{sec:load_controlled}. More details about the ML constitutive model architecture, the symmetries enforced by the architecture, and the physical constraints it does and does not guarantee are described in Section~\ref{sec:appendix_load_controlled}.

The synthetic experimental data is generated by solving the same PDE system with a known constitutive law: a nonlinear elastic softening law representative of a closed-cell polymeric foam (see Section~\ref{sec:appendix_displacement_controlled} for the explicit form). The training dataset consists of six force-displacement pairs, with four additional pairs used for model validation. To represent real data, a 1\% noise is added to the scalar force values in the synthetic experimental data.

\begin{figure}[!htbp]
    \centering
    \includegraphics[width=0.8\textwidth]{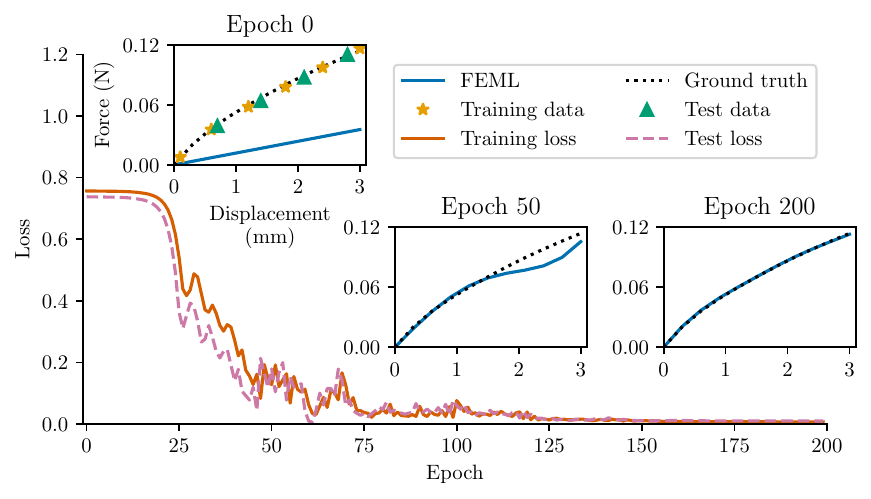}
    \caption{Training loss curve and force-displacement response at different training stages. The figure presents the training and test loss curve alongside the force-displacement response predicted by the FEM solver incorporating the ML-based constitutive model at training epochs 0, 50, and 200. In the plot corresponding to epoch 0, star markers denote the six force-displacement data points used for training, and triangle markers denote the 4 values used for model validation (test). The dotted line represents the reference (ground truth) force-displacement curve, while the blue line corresponds to the response predicted by the FEM solver employing the learned constitutive model. Owing to the universal approximation properties of MLs, the model initially approximates the material behaviour as an almost linear elastic response (epoch 0 and 50). After 200 epochs, the ML-based constitutive model accurately captures the softening behaviour of the material, achieving a close match to the reference curve despite being trained on only six data points.}
    \label{fig:displacement_controlled_plot}
\end{figure}

Figure~\ref{fig:displacement_controlled_plot} shows the training and validation loss curves together with the evolution of the force–displacement response predicted by the FEM solver incorporating the ML-based constitutive model at successive training stages (epochs 0, 50, and 200). The dotted line denotes the reference force–displacement curve, while the blue line indicates the model prediction. At initialisation (epoch 0) and after 50 epochs, the ML-based model captures an almost linear elastic response. After 200 epochs, it accurately reproduces the material’s softening behaviour, demonstrating its ability to learn general nonlinear constitutive responses from a minimal training dataset.

\begin{figure}[!htbp]
    \begin{center}
        \begin{tikzpicture}
        
            \def\textsize{5mm}
            \def\boxwidth{10cm}
            \def\boxheight{7cm}
            \def\femsolverheight{0.9*\boxheight}
        
            \def\femsolvercolor{green!30}
            \def\physicalmodelcolor{yellow!50}
            \def\mloperatorcolor{orange!50}
            \def\observablecolor{yellow!50}
            \def\predictioncolor{green!30}
            \def\optimisercolor{orange!50}
            \def\losscolor{green!30}
        
            \def\physpace{\textsize}
            \def\physwidth{\boxwidth - 2.5*\physpace}
            \def\physheight{\boxheight - 2.5*\physpace}
            \def\physheightInner{0.88*\physheight}
        
            \def\mlwidth{0.20*\physwidth}
            \def\mlheight{0.8*\physheight}
            \def\mlheightInner{0.8*\mlheight}
        
            \def\bottomwidth{2.6cm}
            \def\bottomheight{1.3cm}
            \def\lossboxheight{1.3cm}
        
            \def\bottomspacing{(\boxwidth - 3*\bottomwidth)/4}
            \def\rowgap{0.5cm}           
            \def\lossgap{0.6cm}           
        
            \tikzset{
                thickbox/.style={draw, very thick, rounded corners},
                thickarrow/.style={->, very thick}
            }
        
            \node[thickbox, fill=\femsolvercolor, minimum width=\boxwidth, minimum height=\femsolverheight, anchor=center] (femsolver) 
                at (0,0) {};
            \node[above=0.15cm of femsolver.south] {\textbf{FEM Solver}};
        
            \node[thickbox, fill=\physicalmodelcolor, minimum width=\physwidth, minimum height=\physheightInner, anchor=center] (physical) 
            at (0, .05*\boxheight) {};
            \node[right=0.4cm of physical.west, align=center] {
                \textbf{Geometry:} \\
                Rectangle \\\\
                \textbf{Governing Equations:} \\
                $\nabla \cdot \boldsymbol{\sigma} + \boldsymbol{f} = 0$ \\
                $\boldsymbol{\varepsilon} = \frac{1}{2} \left( \nabla \boldsymbol{u} + (\nabla \boldsymbol{u})^\text{T} \right)$ \\\\
                \textbf{Boundary Conditions:} \\
                $\boldsymbol{u} = \boldsymbol{0}$ on $\Gamma_b$ \\
                $\boldsymbol{u} = (0, \bar{u}_i)$ on $\Gamma_t$};
        
            \node[thickbox, fill=\mloperatorcolor, minimum width=\mlwidth, minimum height=\mlheightInner, anchor=center, align=center] (mloperator) 
            at ($(physical.center)+(0.4*\physwidth,0cm)$) {\textbf{ML} \\ \textbf{Elastic} \\ \textbf{Constitutive}\\ \textbf{Model} \\[6pt]
            {\footnotesize $\boldsymbol{\sigma} = \mathbb{C}\bigl(\mathcal{N}_\theta(I_1)\bigr) : \boldsymbol{\varepsilon}$}};
        
            \node[thickbox, fill=\observablecolor, minimum width=\bottomwidth, minimum height=\bottomheight, anchor=north, align=center] (observable) 
                at ({-\bottomwidth - \bottomspacing}, {-0.5*\femsolverheight - \rowgap}) {\textbf{Compression} \\ \textbf{Experiment} \\ $(\bar{u}_i, F^{obs}_i$)};
        
            \node[thickbox, fill=\predictioncolor, minimum width=\bottomwidth, minimum height=\bottomheight, anchor=north, align=center] (prediction) 
                at (0, {-0.5*\femsolverheight - \rowgap}) {\textbf{Applied} \\ \textbf{Loads} \\ $(F^{fem}_i)$};
        
            \node[thickbox, fill=\optimisercolor, minimum width=\bottomwidth, minimum height=\bottomheight, anchor=north, align=center] (optimiser) 
                at ({\bottomwidth + \bottomspacing}, {-0.5*\femsolverheight - \rowgap}) {\textbf{Optimiser} \\ $\theta^{C} = \underset{\theta}{\operatorname{arg\,min}} \, \mathcal{L}$};
        
            \node[thickbox, fill=\losscolor, minimum width=\bottomwidth, minimum height=\lossboxheight, anchor=north, align=center] (loss) 
                at (0, {-0.5*\femsolverheight - \rowgap - \bottomheight - \lossgap}) {\textbf{Loss} \\ $\mathcal{L} = \frac{1}{N} \sum_{i}^{N} \frac{\left| F_i^\text{fem} - F_i^\text{obs} \right|}{\left| F_i^\text{obs} \right|}$};
        
            \draw[thickarrow] (observable.north) -- ($(physical.south west)!0.2!(physical.south east)$);
            \draw[thickarrow] (observable.south) -- (loss.west);
            \draw[thickarrow] (femsolver.south) -- (prediction.north);
            \draw[thickarrow] (optimiser.north) -- (mloperator.south);
            \draw[thickarrow] (prediction.south) -- (loss.north);
            \draw[thickarrow] (loss.east) -- (optimiser.south);
        
        \end{tikzpicture}
    \end{center}
    \caption{Schematic of the mathematical definition of the problem and the proposed framework for learning constitutive laws from displacement-controlled experiments.}
    \label{fig:framework_schematic_displacement_controlled}
\end{figure}

\subsubsection{Learning Plastic Hardening Laws from Displacements in Load-Controlled Experiments}
\label{sec:load_controlled}

Having established the elastic constitutive law in the previous experiment, we now proceed to the second stage of the progressive discovery strategy: learning what happens when the material is loaded beyond its elastic limit and begins to deform plastically. The elastic operator $E(\kappa)$ learned in Section~\ref{sec:displacement_controlled} is frozen—its parameters are no longer updated—so that the only remaining unknown is the plastic hardening law. This separation is the key advantage of the progressive approach: by building on previously acquired knowledge, each new experiment can focus on a single unknown, reducing ambiguity and improving the identifiability of the learned operator.

Beyond progressive discovery, this experiment also demonstrates how a priori knowledge of the expected physical behaviour can be embedded directly in the ML architecture. In many practical situations, one may have qualitative prior knowledge about the unknown physical law—for example, from preliminary experiments, theoretical arguments, or domain expertise. Here, we assume that the yield stress increases monotonically with accumulated plastic strain (i.e.\ the material exhibits strict isotropic hardening); in other applications, analogous priors might take different forms, such as convexity of a free-energy potential or positivity of a transport coefficient. Rather than treating this as an unconstrained regression problem and hoping the network discovers monotonicity from data, we encode this prior by parameterising $\sigma_y(p)$ with a monotone neural network whose architecture \emph{guarantees} non-decreasing output by construction. This design choice reduces the hypothesis space to physically admissible hardening laws, improves data efficiency, and ensures thermodynamic consistency of the learned plastic response.

We consider the \textit{Brazilian disc test}, a standard method widely used in engineering to indirectly measure the tensile strength of materials. During the test, the displacement field is typically recorded using digital image correlation~\cite{farsi2017}. The experiment is performed under force control, where the applied load~$\bar{F_i}$~is prescribed rather than the displacement. The same unknown foam is now loaded beyond its elastic limit so that the disc undergoes J2 elastoplastic deformation. Because J2 plastic flow is isochoric ($\operatorname{tr}(\boldsymbol{\varepsilon}^p)=0$, where $\boldsymbol{\varepsilon}^p$ is the plastic strain; the elastoplastic kinematics are defined in Section~\ref{sec:appendix_load_controlled}), the frozen elastic operator $E(\kappa)$ continues to receive the \emph{total} volumetric strain $\kappa=\langle -\operatorname{tr}(\boldsymbol{\varepsilon})\rangle$ as input, avoiding a circular dependence on the elastic–plastic strain decomposition.

\begin{figure}[!htbp]
    \centering
    \begin{subfigure}[b]{0.28\textwidth}
        \centering
        \begin{tikzpicture}[scale=0.8]

            \fill[gray!20] (0, 0) circle [radius=2]; 
            \draw[thick] (0, 0) circle [radius=2];

            \foreach \x in {-0.5, 0, 0.5}
            {\draw[thick] (\x+0.2, -2.2) -- (\x, -2);}
            \node at (0, -2.5) {\(\boldsymbol{u} = \boldsymbol{0}\)};

            \draw[thick,->] (0, 3.2) -- (0, 2.25) node[midway, right] {\(\boldsymbol{F_i}\)};

            \draw[thick,<->] (2.5, -2) -- (2.5, 2) node[midway, right] {10 cm};

        \end{tikzpicture}
        \subcaption{}
        \label{fig:brazilian_disc_test_schematic}
    \end{subfigure}\hfill
    \begin{subfigure}[b]{0.33\textwidth}
        \centering
        \includegraphics[width=\textwidth]{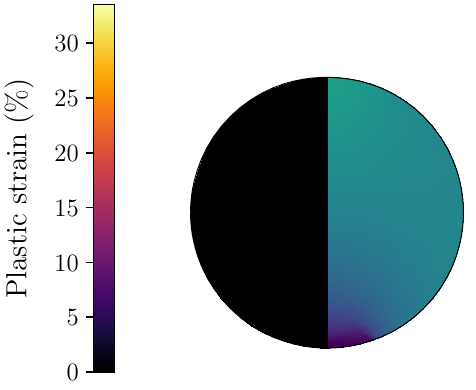}
        \subcaption{}
        \label{fig:learning_from_displacements_b}
    \end{subfigure}\hfill
    \begin{subfigure}[b]{0.33\textwidth}
        \centering
        \includegraphics[width=\textwidth]{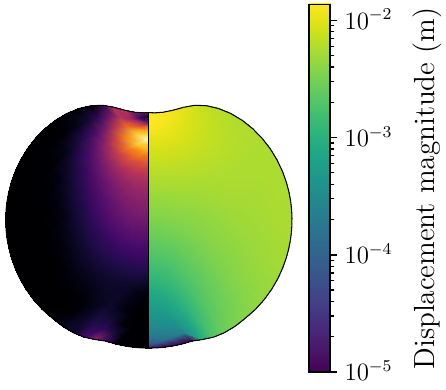}
        \subcaption{}
        \label{fig:learning_from_displacements_c}
    \end{subfigure}
    \caption{Load-controlled Brazilian disc test: (a) schematic of the test setup with a fixed bottom and a prescribed force $\boldsymbol{F_i}$ at the top; (b) first and (c) last loading increment of the training data. The left half of each disc shows the accumulated plastic strain~$p$ and the right half shows the displacement magnitude.}
    \label{fig:brazilian_disc_test}
\end{figure}

In this example, the sample is idealised as a 2D circular disc, as shown in Figure~\ref{fig:brazilian_disc_test_schematic}. The bottom of the disc is fixed throughout the simulation, while a force $\bar{F_i}$ is applied at the top of the sample, where $i$ is an integer from 1 to $N$ that represents the sequence of forces applied during the test. Similarly to the previous example, this configuration simulates the quasi-static loading condition in which time increments correspond to sequential loading steps rather than physical time.

Figure~\ref{fig:framework_schematic_load_controlled} illustrates the mathematical definition of the problem and the proposed framework for learning constitutive laws from load-controlled experiments. The only trainable operator is the yield-stress function $\sigma_y(p)$; all other components—including the frozen elastic modulus $E(\kappa)$—are fixed during training. The parameters of the neural network-based constitutive model are learned by minimising the following loss function:

\begin{equation}
    \mathcal{L} = \frac{1}{N} \sum_{i}^{N} \frac{\left| \boldsymbol{u}_i^\text{fem} - \boldsymbol{u}_i^\text{obs} \right|}{\left| \boldsymbol{u}_i^\text{obs} \right|},
\end{equation}

where \( \boldsymbol{u}_i^\text{fem} \) represents the displacement field predicted by the FEM solver with the ML-based constitutive model at loading step \( i \), and \( \boldsymbol{u}_i^\text{obs} \) is the corresponding synthetic experimental displacement field.

The synthetic experimental data is generated by solving the same PDE system with the known foam constitutive law: the elastic softening law of Section~\ref{sec:displacement_controlled} combined with a Voce isotropic hardening law for the yield stress (see Section~\ref{sec:appendix_load_controlled} for the explicit forms). Because full-field displacement data are available, the Poisson's ratio is in principle identifiable; however, we retain $\nu = 0.3$ as a known constant for consistency with the elastic-regime experiment. The training dataset consists of 15 force-displacement field pairs, with 8 additional pairs used for model validation. To represent real data, a 1\% noise is added to each nodal value of the displacement fields in the synthetic experimental data, providing a substantially richer perturbation than the scalar noise of the previous example. The architecture of the monotone neural network used to parameterise $\sigma_y(p)$ is described in Section~\ref{sec:appendix_load_controlled}.

\begin{figure}[!htbp]
    \centering
    \includegraphics[width=0.8\textwidth]{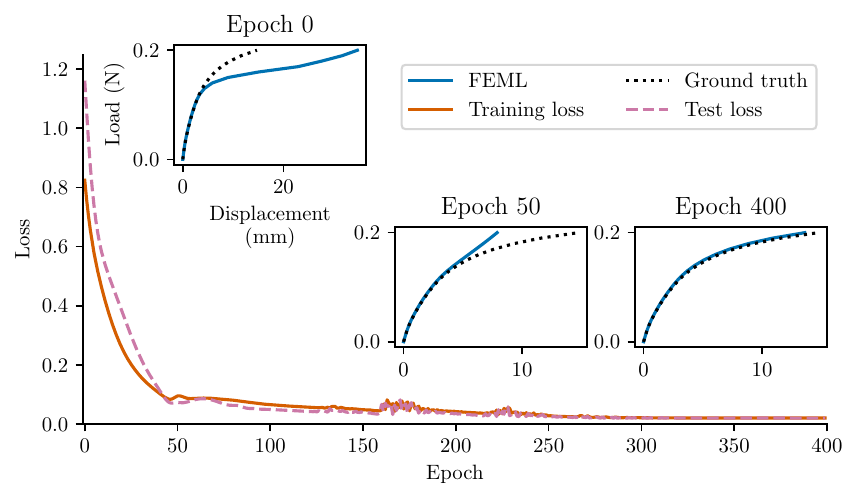}
    \caption{Training loss curve and load versus maximum displacement magnitude response at different training stages. The figure presents the training and test loss curves alongside the load--displacement response predicted by the FEM solver incorporating the ML-based constitutive model at training epochs 0, 50, and 400. The dotted line represents the reference (ground truth) load--displacement curve, while the blue line corresponds to the response predicted by the FEM solver employing the learned constitutive model. Initially, the untrained model (epoch~0) overestimates deformations under applied loads. As training progresses, the model improves its accuracy; after 400 epochs, the ML-based constitutive model closely matches the reference response.}
    \label{fig:plasticity_loss_curves}
\end{figure}

Figure~\ref{fig:plasticity_loss_curves} shows the training and test loss curves together with the evolution of the load versus maximum displacement magnitude response predicted by the FEM solver incorporating the ML-based constitutive model at successive training stages (epochs 0, 50, and 400). The dotted line denotes the reference load--displacement curve, while the blue line indicates the model prediction. Both the training loss and test error decrease steadily over 400 epochs, indicating convergence of the learned hardening law. Initially, the untrained model (epoch~0) overestimates deformations under applied loads, progressively improving its accuracy. By epoch 400, the ML-based constitutive model closely matches the reference response, benefiting from the richer training data comprising nodal displacement fields at 15 load increments. Compared to the previous example, each node undergoes a distinct strain-stress state, enriching the training dataset and enhancing the constraints on the ML-based constitutive model.

\begin{figure}[!htbp]
    \begin{center}
        \begin{tikzpicture}
        
            \def\textsize{5mm}
            \def\boxwidth{10cm}
            \def\boxheight{8cm}
            \def\femsolverheight{0.85*\boxheight}
        
            \def\femsolvercolor{green!30}
            \def\physicalmodelcolor{yellow!50}
            \def\mloperatorcolor{orange!50}
            \def\observablecolor{yellow!50}
            \def\predictioncolor{green!30}
            \def\optimisercolor{orange!50}
            \def\losscolor{green!30}
        
            \def\physpace{\textsize}
            \def\physwidth{\boxwidth - 2.5*\physpace}
            \def\physheight{\boxheight - 2.5*\physpace}
            \def\physheightInner{0.82*\physheight}
        
            \def\mlwidth{0.20*\physwidth}
            \def\mlheight{0.8*\physheight}
            \def\mlheightInner{0.8*\mlheight}
        
            \def\bottomwidth{2.6cm}
            \def\bottomheight{1.3cm}
            \def\lossboxheight{1.3cm}
        
            \def\bottomspacing{(\boxwidth - 3*\bottomwidth)/4}
            \def\rowgap{0.5cm}           
            \def\lossgap{0.6cm}           
        
            \tikzset{
                thickbox/.style={draw, very thick, rounded corners},
                thickarrow/.style={->, very thick}
            }
        
            \node[thickbox, fill=\femsolvercolor, minimum width=\boxwidth, minimum height=\femsolverheight, anchor=center] (femsolver) 
                at (0,0) {};
            \node[above=0.15cm of femsolver.south] {\textbf{FEM Solver}};
        
            \node[thickbox, fill=\physicalmodelcolor, minimum width=\physwidth, minimum height=\physheightInner, anchor=center] (physical) 
            at (0, .05*\boxheight) {};
            \node[right=0.4cm of physical.west, align=center] {
                \textbf{Geometry:} \\
                Disc \\\\
                \textbf{Governing Equations:} \\
                $\nabla \cdot \boldsymbol{\sigma} + \boldsymbol{f} = 0$ \\
                $\boldsymbol{\varepsilon} = \frac{1}{2} \left( \nabla \boldsymbol{u} + (\nabla \boldsymbol{u})^\text{T} \right)$ \\\\
                \textbf{Boundary Conditions:} \\
                $\boldsymbol{u} = \boldsymbol{0}$ on $\Gamma_b$ \\
                $\boldsymbol{u} \cdot \boldsymbol{t} = 0$ on $\Gamma_t$ \\
                $\boldsymbol{\sigma} \cdot \boldsymbol{n} = \boldsymbol{F_i} /S_{\Gamma_t} $ on $\Gamma_t$};
        
            \node[thickbox, fill=\mloperatorcolor, minimum width=\mlwidth, minimum height=\mlheightInner, anchor=center, align=center] (mloperator) 
            at ($(physical.center)+(0.35*\physwidth,0cm)$) {\textbf{ML} \\ \textbf{Elastoplastic} \\ \textbf{Constitutive}\\ \textbf{Model} \\[6pt]
            {\footnotesize $\boldsymbol{\sigma} = \mathbb{C}\bigl(\mathcal{N}_{\theta^{\ast}}(I_1)\bigr) : \bigl(\boldsymbol{\varepsilon} - \boldsymbol{\varepsilon}^p\bigr)$} \\[4pt]
            {\footnotesize $f = \sigma_{\mathrm{eq}} - \sigma_y(p) \le 0$} \\
            {\footnotesize $\sigma_y(p) = \mathcal{M}_{\theta}(p)$}};
        
            \node[thickbox, fill=\observablecolor, minimum width=\bottomwidth, minimum height=\bottomheight, anchor=north, align=center] (observable) 
                at ({-\bottomwidth - \bottomspacing}, {-0.5*\femsolverheight - \rowgap}) {\textbf{Compression} \\ \textbf{Experiment} \\ $(\boldsymbol{F}_i, \boldsymbol{u}^{obs}_i$)};
        
            \node[thickbox, fill=\predictioncolor, minimum width=\bottomwidth, minimum height=\bottomheight, anchor=north, align=center] (prediction) 
                at (0, {-0.5*\femsolverheight - \rowgap}) {\textbf{Displacement} \\ \textbf{Fields} \\ $(\boldsymbol{u}^{fem}_i)$};
        
            \node[thickbox, fill=\optimisercolor, minimum width=\bottomwidth, minimum height=\bottomheight, anchor=north, align=center] (optimiser) 
                at ({\bottomwidth + \bottomspacing}, {-0.5*\femsolverheight - \rowgap}) {\textbf{Optimiser} \\ $\theta^{C} = \underset{\theta}{\operatorname{arg\,min}} \, \mathcal{L}$};
        
            \node[thickbox, fill=\losscolor, minimum width=\bottomwidth, minimum height=\lossboxheight, anchor=north, align=center] (loss) 
                at (0, {-0.5*\femsolverheight - \rowgap - \bottomheight - \lossgap}) {\textbf{Loss} \\ $\mathcal{L} = \frac{1}{N} \sum_{i}^{N} \frac{\left| \boldsymbol{u}_i^\text{fem} - \boldsymbol{u}_i^\text{obs} \right|}{\left| \boldsymbol{u}_i^\text{obs} \right|}$};
        
            \draw[thickarrow] (observable.north) -- ($(physical.south west)!0.2!(physical.south east)$);
            \draw[thickarrow] (observable.south) -- (loss.west);
            \draw[thickarrow] (femsolver.south) -- (prediction.north);
            \draw[thickarrow] (optimiser.north) -- (mloperator.south);
            \draw[thickarrow] (prediction.south) -- (loss.north);
            \draw[thickarrow] (loss.east) -- (optimiser.south);
        
        \end{tikzpicture}
    \end{center}
    \caption{Schematic of the mathematical definition of the problem and the proposed framework for learning constitutive laws from load-controlled experiments.}
    \label{fig:framework_schematic_load_controlled}
\end{figure}

\subsubsection{Zero-Shot Inference with an Elastoplastic Foundation Model}
\label{sec:torque_holed_plate}

We demonstrate the portability of our pretrained constitutive model by deploying it zero-shot on a problem that differs from the training setup. We treat the constitutive operator as a part of a foundation model: a reusable, data-driven component pretrained once and then inserted into new simulations without retraining or architectural changes, while allowing modifications to the surrounding governing PDEs, geometry, and/or boundary/initial conditions.

\begin{figure}[!htbp]
    \centering
    \def\torqueResultSubfigureRaise{2cm} 
    \begin{subfigure}[b]{0.22\textwidth}
        \centering
        \begin{tikzpicture}[scale=0.90, every node/.style={font=\normalsize}]
            \def\L{6.0}      
            \def\R{1.0}      
            \def\E{0.32}     
            \def\sc{0.5}     
            \def\sh{0.25}    

            \fill[gray!20] (-\R, -\L/2) rectangle (\R, \L/2);

            \fill[gray!10] (0, \L/2) ellipse [x radius=\R, y radius=\E];
            \draw[thick] (0, \L/2) ellipse [x radius=\R, y radius=\E];

            \fill[gray!10] (0, -\L/2) ellipse [x radius=\R, y radius=\E];
            \draw[thick] (-\R, -\L/2) arc[start angle=180, end angle=360, x radius=\R, y radius=\E];
            \draw[thick, dashed] (\R, -\L/2) arc[start angle=0, end angle=180, x radius=\R, y radius=\E];

            \draw[thick] (-\R, -\L/2) -- (-\R, \L/2);
            \draw[thick] (\R, -\L/2) -- (\R, \L/2);

            \fill[white] (-\sc, -\sh) rectangle (\sc, \sh);
            \fill[white] (-\sc, 0) circle (\sh);
            \fill[white] (\sc, 0) circle (\sh);

            \def\dx{-0.05}   
            \def\dy{0.04}    
            \def\sho{0.20}   
            \def\sco{0.50}   
            \draw[thin, gray!70] (-0.692, -\sho + \dy) -- (\sco + \dx, -\sho + \dy);
            \draw[thin, gray!70] (\sco + \dx, -\sho + \dy)
                arc[start angle=-90, end angle=90, radius=\sho];
            \draw[thin, gray!70] (\sc, -\sh) -- (\sco + \dx, -\sho + \dy);
            \draw[thin, gray!70] (\sc, \sh) -- (\sco + \dx, \sho + \dy);

            \draw[thick] (-\sc, \sh) -- (\sc, \sh);
            \draw[thick] (-\sc, -\sh) -- (\sc, -\sh);
            \draw[thick] (-\sc, \sh) arc[start angle=90, end angle=270, radius=\sh];
            \draw[thick] (\sc, -\sh) arc[start angle=-90, end angle=90, radius=\sh];

            \draw[thick, dash dot] (0, \L/2) -- (0, \L/2 + 1.15);
            \draw[thick, dash dot] (0, -\L/2) -- (0, -\L/2 - 1.15);

            \draw[->, >=stealth, thick]
                ($(0, \L/2 + 0.75) + (200:0.55 and 0.18)$)
                arc[start angle=200, end angle=-20, x radius=0.55, y radius=0.18];
            \node[above] at (0, \L/2 + 1.30) {$\boldsymbol{u} = \bar{\boldsymbol{u}}(+\Theta/2)$};

            \draw[->, >=stealth, thick]
                ($(0, -\L/2 - 0.75) + (-20:0.55 and 0.18)$)
                arc[start angle=-20, end angle=-200, x radius=0.55, y radius=0.18];
            \node[below] at (0, -\L/2 - 1.30) {$\boldsymbol{u} = \bar{\boldsymbol{u}}(-\Theta/2)$};

            \draw[thick, <->] (\R + 0.55, -\L/2) -- (\R + 0.55, \L/2)
                node[midway, right] {60 cm};
            \draw[thick, <->] (-\R, -\L/2 - 2.30) -- (\R, -\L/2 - 2.30)
                node[midway, below] {20 cm};
            \draw[thick, <->] (-\sc - \sh, \sh + 0.30) -- (\sc + \sh, \sh + 0.30)
                node[midway, above] {15 cm};
            \draw[thick, <->] (-\R - 0.30, -\sh) -- (-\R - 0.30, \sh)
                node[midway, left] {5 cm};
        \end{tikzpicture}
        \subcaption{}
        \label{fig:torque_holed_plate_schematic}
    \end{subfigure}%
    \hspace{0.04\textwidth}%
    \begin{subfigure}[b]{0.24\textwidth}
        \centering
        \vspace*{-\torqueResultSubfigureRaise}%
        \begin{tikzpicture}[inner sep=0, every node/.style={font=\normalsize}]
            \node[anchor=south west] (dimg) at (0.35\linewidth,0)
                {\includegraphics[width=0.53865\linewidth]{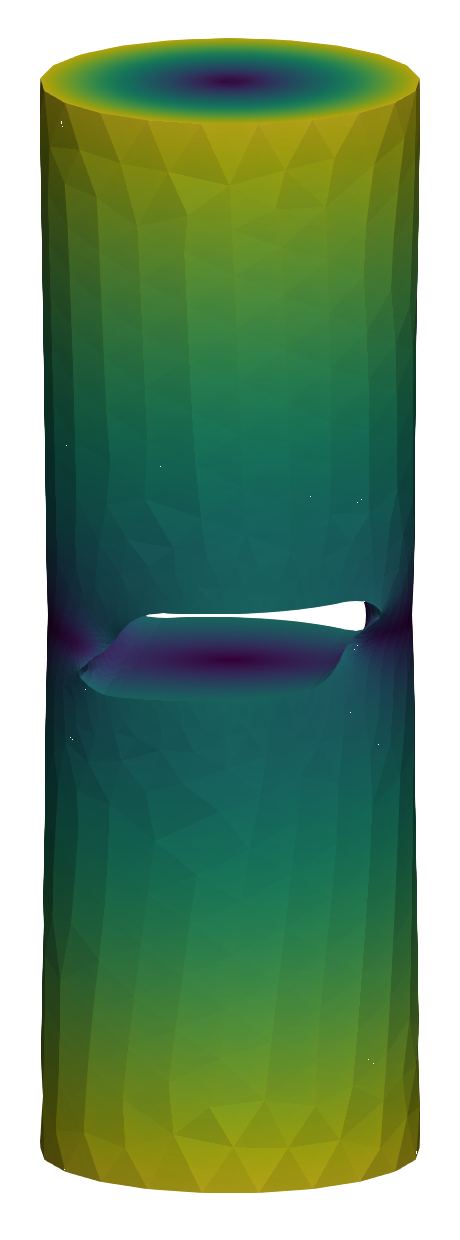}};

            \node[anchor=south west] (dcb) at (0.225\linewidth,0.324\linewidth)
                {\includegraphics[height=1.05\linewidth]{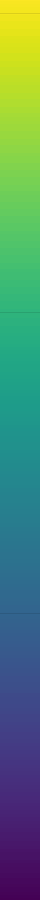}};
            \draw (0.225\linewidth,0.324\linewidth) rectangle (0.2717\linewidth,1.374\linewidth);
            \foreach \yy/\lab in {0.324/0,0.668/1,1.012/2,1.357/3}{%
                \draw (0.225\linewidth,\yy\linewidth) -- (0.210\linewidth,\yy\linewidth);
                \node[anchor=east] at (0.200\linewidth,\yy\linewidth) {\lab};
            }
            \node[rotate=90, anchor=center]
                at (0.105\linewidth,0.849\linewidth) {Displacement (mm)};
        \end{tikzpicture}
        \subcaption{}
        \label{fig:torque_holed_plate_gt_disp}
        \par\vspace*{\torqueResultSubfigureRaise}%
    \end{subfigure}%
    \hspace{-0.04\textwidth}%
    \begin{subfigure}[b]{0.53\textwidth}
        \centering
        \vspace*{-\torqueResultSubfigureRaise}%
        \begin{tikzpicture}[inner sep=0, every node/.style={font=\normalsize}]
            \def\vmsX{0.425}   
            \def\vmsW{0.243675}  
            \def\vmsH{0.65742} 
            \def\zoomLx{0.16690}
            \def\zoomRx{0.41000}
            \def\zoomBy{0.26391}  
            \def\zoomTy{0.50500}
            \def\pImgW{1.10500}
            \def\pImgX{-0.16460}
            \def\pImgY{-0.42923}

            \node[anchor=south west] (vimg) at (\vmsX\linewidth,0)
                {\includegraphics[width=\vmsW\linewidth]{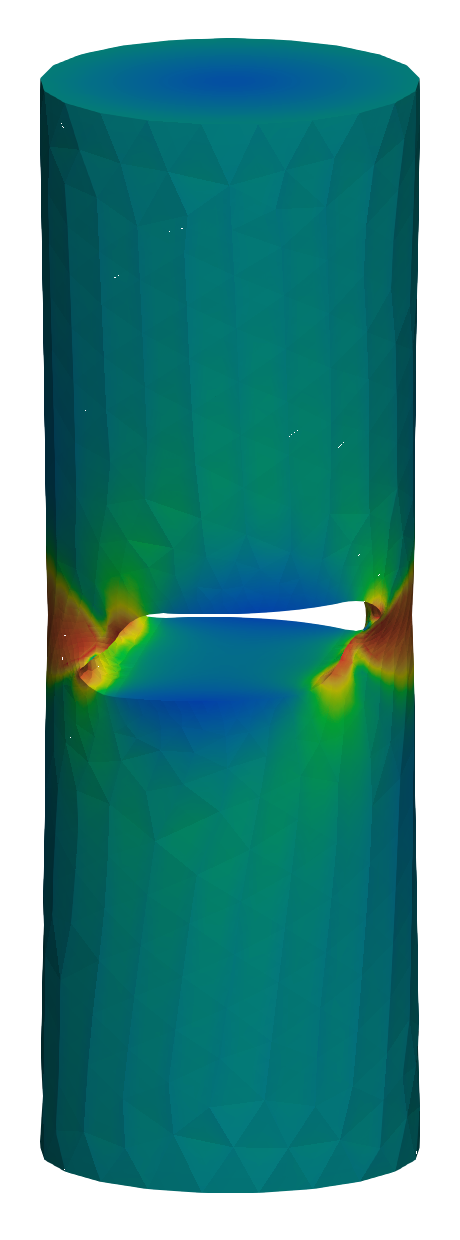}};

            \begin{scope}[x={(vimg.south east)},y={(vimg.north west)}]
                \coordinate (rTL) at (0.308,0.556);
                \draw[black, dashed, line width=1.0pt] (rTL) rectangle (0.5544,0.396);
            \end{scope}

            \begin{scope}
                \path[clip] (\zoomLx\linewidth,\zoomBy\linewidth)
                    rectangle (\zoomRx\linewidth,\zoomTy\linewidth);
                \node[anchor=south west]
                    at (\pImgX\linewidth,\pImgY\linewidth)
                    {\includegraphics[width=\pImgW\linewidth]{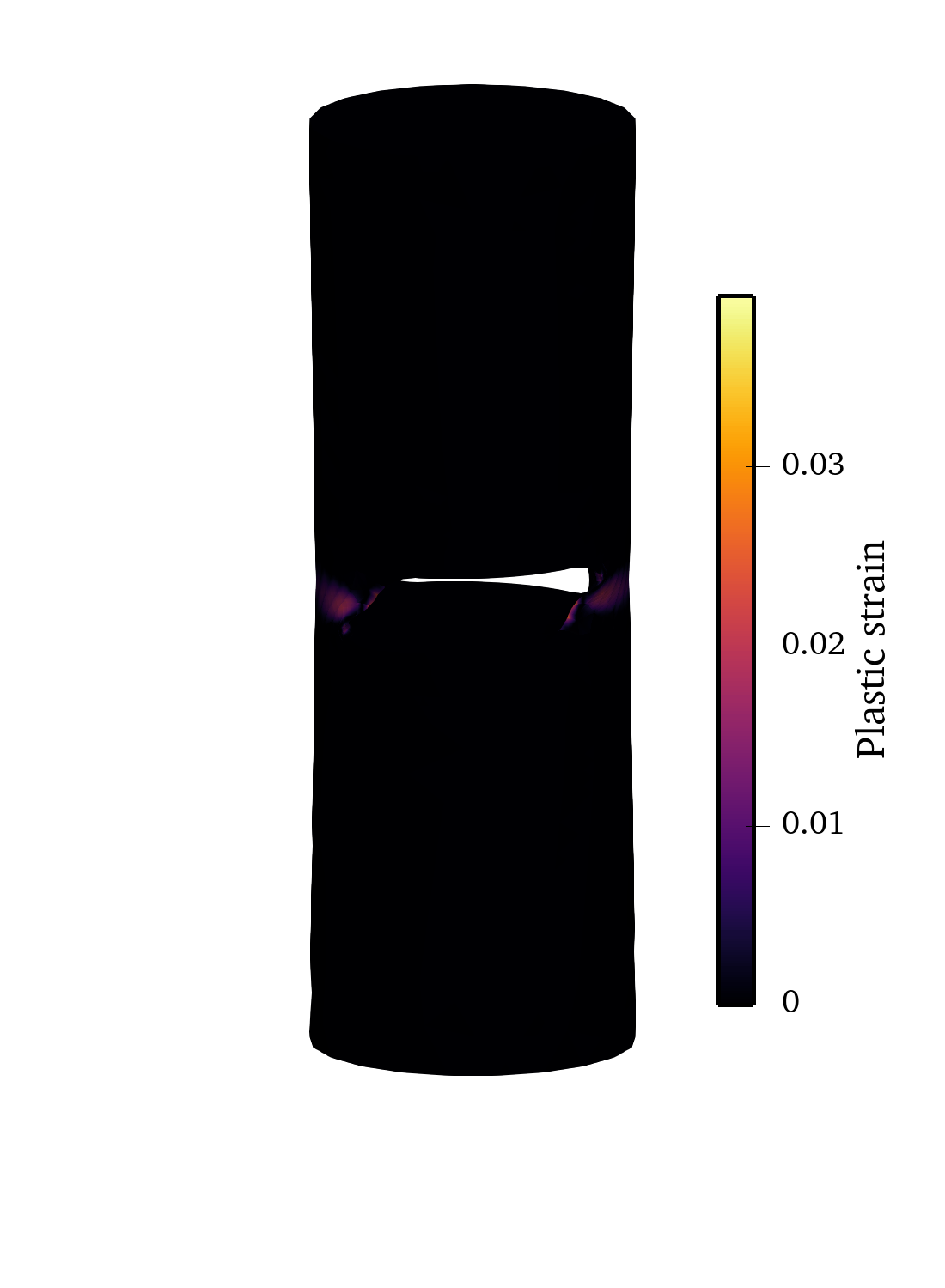}};
            \end{scope}
            \draw[black, dashed, line width=1.0pt]
                (\zoomLx\linewidth,\zoomBy\linewidth)
                rectangle (\zoomRx\linewidth,\zoomTy\linewidth);

            \draw[black, line width=0.6pt]
                (rTL) -- (\zoomRx\linewidth,\zoomTy\linewidth);

            \node[anchor=south west] (pcb) at (0.130\linewidth,0.10946\linewidth)
                {\includegraphics[height=0.55\linewidth]{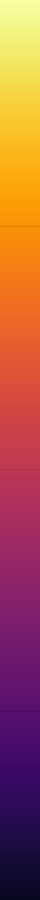}};
            \draw (0.130\linewidth,0.10946\linewidth) rectangle (0.1544\linewidth,0.65946\linewidth);
            \foreach \yy/\lab in {0.10946/0,0.26660/1,0.42374/2,0.58088/3}{%
                \draw (0.130\linewidth,\yy\linewidth) -- (0.115\linewidth,\yy\linewidth);
                \node[anchor=east] at (0.105\linewidth,\yy\linewidth) {\lab};
            }
            \node[rotate=90, anchor=center]
                at (0.055\linewidth,0.38446\linewidth) {Plastic strain (\%)};

            \node[anchor=south west] (vcb) at (0.730\linewidth,0.10946\linewidth)
                {\includegraphics[height=0.55\linewidth]{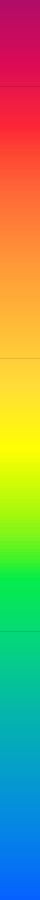}};
            \draw (0.730\linewidth,0.10946\linewidth) rectangle (0.7544\linewidth,0.65946\linewidth);
            \foreach \yy/\lab in {0.10946/0,0.27462/1,0.43979/2,0.60495/3}{%
                \draw (0.7544\linewidth,\yy\linewidth) -- (0.7694\linewidth,\yy\linewidth);
                \node[anchor=west] at (0.779\linewidth,\yy\linewidth) {\lab};
            }
            \node[rotate=90, anchor=center]
                at (0.845\linewidth,0.38446\linewidth) {Von Mises stress (MPa)};
        \end{tikzpicture}
        \subcaption{}
        \label{fig:torque_holed_plate_gt_vms}
        \par\vspace*{\torqueResultSubfigureRaise}%
    \end{subfigure}
    \caption{Cylindrical rod with a transverse keyhole-shaped through-slot subjected to torsion: (a)~schematic of the rod geometry, viewed from the front so that the stadium-shaped through-slot is visible at mid-span. Rigid rotations of equal magnitude and opposite sign, $\pm\Theta/2$, are prescribed on the two circular end faces, producing a relative twist of $\theta = 3.5^\circ$ between the ends. Deformed configuration from the ground-truth constitutive model, coloured by (b)~the displacement magnitude, and (c)~the von Mises stress on the full cylinder together with an enlargement (left) of the slot region---highlighted by the red rectangle---coloured by the accumulated plastic strain~$p$, where plastic flow concentrates. The deformation has been amplified by a factor of ten for visualisation.}
    \label{fig:torque_holed_plate}
\end{figure}

 The foundation constitutive model combines the two operators learned from the material: (i)~the elastic modulus $E(\kappa)$ from Section~\ref{sec:displacement_controlled} and (ii)~the hardening law $\sigma_y(p)$ from Section~\ref{sec:load_controlled}. Both are frozen and embedded inside a three-dimensional J2 elastoplastic solver for a cylindrical rod with a transverse keyhole-shaped through-slot at its mid-span, subjected to torsion. The smooth circular cross-section of the rod has zero Saint-Venant warping function, so prescribing rigid rotations on the end faces introduces no warping stresses, and plastic flow is confined to the neighbourhood of the slot rather than the boundary surfaces. As shown in Figure~\ref{fig:torque_holed_plate_schematic}, the two end faces are rotated in opposite directions by $\pm\theta/2$ with $\theta = 3.5^\circ$, yielding a relative twist of $\theta$ between the ends and producing a strongly heterogeneous stress field around the slot. The applied twist is deliberately calibrated so that the maximum accumulated plastic strain remains within the range covered by the training data of Section~\ref{sec:load_controlled} (peak $p \approx 4$\%), thereby exercising the learned constitutive operator strictly inside its training envelope. Plastic flow develops in approximately 5.9\% of the domain. Figures~\ref{fig:torque_holed_plate_gt_disp} and~\ref{fig:torque_holed_plate_gt_vms} show, respectively, the displacement magnitude and the von Mises stress on the deformed configuration obtained with the ground-truth constitutive model; the latter also includes an enlargement of the slot region coloured by the accumulated plastic strain~$p$, where plastic flow concentrates.

The problem formulation is summarized in Figure~\ref{fig:framework_schematic_holed_plate}. Standard FEM provides the kinematics, balance laws, and boundary/loading conditions. The two pretrained operators supply the elastic and plastic responses from the computed strain and history fields, thereby closing the system without additional calibration.

\begin{figure}[!htbp]
    \centering
    \begin{subfigure}[b]{0.43\textwidth}
        \centering
        \includegraphics[width=\textwidth]{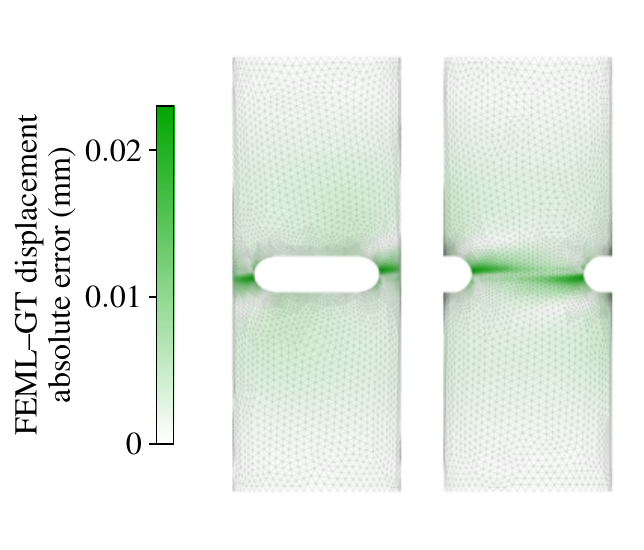}
        \caption{}
        \label{fig:torque_holed_plate_absdiff_disp}
    \end{subfigure}
    \begin{subfigure}[b]{0.43\textwidth}
        \centering
        \includegraphics[width=\textwidth]{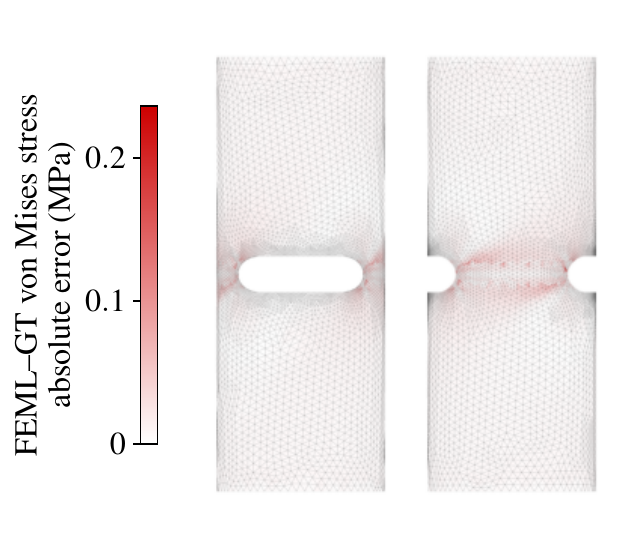}
        \caption{}
        \label{fig:torque_holed_plate_absdiff_vms}
        
    \end{subfigure}
    \\
    \vspace{0.6em}
    \begin{subfigure}[b]{0.43\textwidth}
        \centering
        \includegraphics[width=\textwidth]{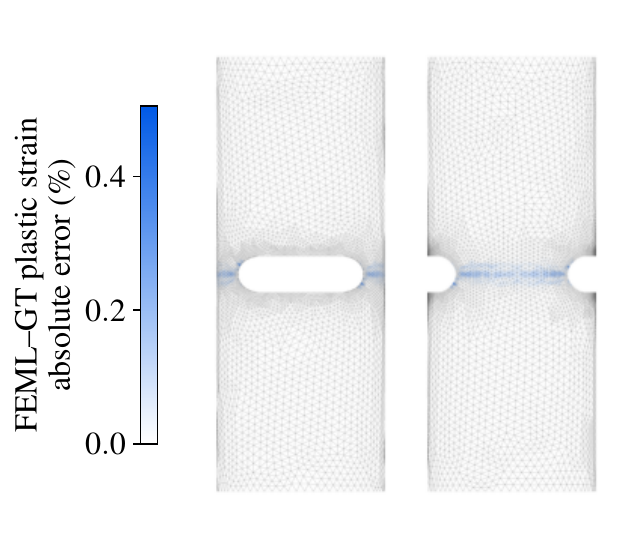}
        \caption{}
        \label{fig:torque_holed_plate_absdiff_p}
        
    \end{subfigure}

    \caption{Absolute error fields at the end of the simulation, shown with two views (front and side): (a)~absolute displacement difference, (b)~absolute von Mises stress difference, and (c)~absolute accumulated plastic-strain difference in \%.}
    \label{fig:mean_stress_comparison}
\end{figure}

Figure~\ref{fig:mean_stress_comparison} reports the absolute error fields at the final simulation step between the foundation-model prediction and the synthetic ground-truth solution, viewed from the front and from the side, with black mesh edges overlaid consistently across all three panels. The maximum absolute errors are $2.30\times10^{-2}$~mm for displacement, 0.23~MPa for von Mises stress, and $0.5\%$ for accumulated plastic strain. These margins are particularly compelling given that inference is performed in a different loading regime on a geometry with strong stress-localising features (the keyhole slot), while the constitutive operators were trained on noise-corrupted data. Together, these results indicate that the learned operators transfer robustly to a new geometry, three-dimensional setting, and loading regime—including plastic deformation—in a zero-shot manner. This portability highlights the utility of embedding pretrained operators within PDE solvers to enable efficient simulation across varied engineering scenarios.

\begin{figure}[!htbp]
    \begin{center}
        \begin{tikzpicture}
        
            \def\textsize{5mm}
            \def\boxwidth{10cm}
            \def\boxheight{8cm}
            \def\femsolverheight{0.85*\boxheight}
        
            \def\femsolvercolor{green!30}
            \def\physicalmodelcolor{yellow!50}
            \def\mloperatorcolor{orange!50}
            \def\predictioncolor{green!30}
        
            \def\physpace{\textsize}
            \def\physwidth{\boxwidth - 2.5*\physpace}
            \def\physheight{\boxheight - 2.5*\physpace}
            \def\physheightInner{0.82*\physheight}
        
            \def\mlwidth{0.20*\physwidth}
            \def\mlheight{0.8*\physheight}
            \def\mlheightInner{0.8*\mlheight}
        
            \def\bottomwidth{2.6cm}
            \def\bottomheight{1.3cm}
        
            \def\rowgap{0.5cm}           
        
            \tikzset{
                thickbox/.style={draw, very thick, rounded corners},
                thickarrow/.style={->, very thick}
            }
        
            \node[thickbox, fill=\femsolvercolor, minimum width=\boxwidth, minimum height=\femsolverheight, anchor=center] (femsolver) 
                at (0,0) {};
            \node[above=0.15cm of femsolver.south] {\textbf{FEM Solver}};
        
            \node[thickbox, fill=\physicalmodelcolor, minimum width=\physwidth, minimum height=\physheightInner, anchor=center] (physical) 
            at (0, .05*\boxheight) {};
            \node[right=0.4cm of physical.west, align=center] {
                \textbf{Geometry:} \\
                Cylindrical rod \\ with a through-slot \\\\
                \textbf{Governing Equations:} \\
                $\nabla \cdot \boldsymbol{\sigma} + \boldsymbol{f} = 0$ \\
                $\boldsymbol{\varepsilon} = \frac{1}{2} \left( \nabla \boldsymbol{u} + (\nabla \boldsymbol{u})^\text{T} \right)$ \\\\
                \textbf{Boundary Conditions:} \\
                $\boldsymbol{u} = \bar{\boldsymbol{u}}(-\Theta/2)$ on $\Gamma_b$ \\
                $\boldsymbol{u} = \bar{\boldsymbol{u}}(+\Theta/2)$ on $\Gamma_t$};
        
            \node[thickbox, fill=\mloperatorcolor, minimum width=\mlwidth, minimum height=\mlheightInner, anchor=center, align=center] (mloperator) 
            at ($(physical.center)+(0.35*\physwidth,0cm)$) {\textbf{ML} \\ \textbf{Elastoplastic} \\ \textbf{Constitutive}\\ \textbf{Model} \\[6pt]
            {\footnotesize $\boldsymbol{\sigma} = \mathbb{C}\bigl(\mathcal{N}_{\theta^{\ast}}(I_1)\bigr) : \bigl(\boldsymbol{\varepsilon} - \boldsymbol{\varepsilon}^p\bigr)$} \\[4pt]
            {\footnotesize $f = \sigma_{\mathrm{eq}} - \sigma_y(p) \le 0$} \\
            {\footnotesize $\sigma_y(p) = \mathcal{M}_{\theta^{\ast}}(p)$}};
        
            \node[thickbox, fill=\predictioncolor, minimum width=\bottomwidth, minimum height=\bottomheight, anchor=north, align=center] (prediction) 
                at (0, {-0.5*\femsolverheight - \rowgap}) {\textbf{Displacements} \\ \textbf{Stresses, and} \\ \textbf{Plastic Strain} \\ $(\boldsymbol{u}_\Theta, \boldsymbol{\sigma}_\Theta, p_\Theta)$};
        
            \draw[thickarrow] (femsolver.south) -- (prediction.north);

        \end{tikzpicture}
    \end{center}
    \caption{Schematic of the foundation model with the pretrained constitutive operator for zero-shot transfer to a three-dimensional cylindrical rod with a transverse through-slot under torsional loading.}
    \label{fig:framework_schematic_holed_plate}
\end{figure}

\subsubsection{Learning Plasticity and Damage from Real Shear-Coupon Experiments}
\label{sec:shear_coupon}

The preceding experiments established and validated the framework on \emph{synthetic} data, where an exact ground-truth operator is available for quantitative verification. We now apply FEML to \emph{real} experimental measurements, where the underlying constitutive law is genuinely unknown. Beyond demonstrating real-world applicability, this example is the most demanding in the constitutive-modelling sequence in two respects. First, it extends the discovered material description beyond elasticity and plastic hardening to \emph{ductile damage}---the progressive loss of load-carrying capacity that produces post-peak softening and eventual failure---which is represented here by a \emph{second} learned operator embedded in the solver alongside the hardening law. Second, it does so within a \emph{finite-strain} formulation, departing from the small-strain setting of the previous solid-mechanics examples. Because no ground-truth law exists for real data, success is assessed by whether the model reproduces the measured response to within the experimental specimen-to-specimen scatter.

\begin{figure}[!htbp]
    \centering
    \begin{subfigure}[b]{0.483\textwidth}
        \centering
        \resizebox{\linewidth}{!}{%
        \begin{tikzpicture}[every node/.append style={font=\fontsize{20}{24}\selectfont}]
            \def\W{6.35}\def\H{2.794}\def\no{1.35}\def\nt{0.743}
            \draw[thick,fill=gray!20]
                (-\W,-\H) -- (-\W,\H) -- (-\no,\H) -- (0,\nt) -- (\no,\H)
                -- (\W,\H) -- (\W,-\H) -- (\no,-\H) -- (0,-\nt) -- (-\no,-\H) -- cycle;
            \foreach \y in {-2.4,-1.6,...,2.5}{\draw[thick] (-\W,\y) -- (-\W-0.32,\y+0.32);}
            \foreach \x in {-5.8,-5.0,-4.2}{%
                \draw[thick] (\x,\H) -- (\x-0.32,\H+0.32);
                \draw[thick] (\x,-\H) -- (\x-0.32,-\H-0.32);}
            \node[anchor=south west] at (-\W,\H+1.25) {\textbf{\(\boldsymbol{u}=\boldsymbol{0}\)}};
            \foreach \y in {-1.8,-0.6,0.6,1.8}{\draw[thick,->] (\W+0.55,\y-0.425) -- (\W+0.55,\y+0.425);}
            \foreach \x in {4.3,5.1,5.9}{%
                \draw[thick,->] (\x,\H+0.30) -- (\x,\H+1.15);
                \draw[thick,->] (\x,-\H-1.15) -- (\x,-\H-0.30);}
            \node[anchor=south east] at (\W,\H+1.25) {\textbf{\(\boldsymbol{u}=(0,\bar{u}_i)\)}};
            \draw[thick,<->] (-\W,-\H-1.5) -- (\W,-\H-1.5) node[midway,below]{12.7 cm};
            \draw[thick,<->] (-\W-1.2,-\H) -- (-\W-1.2,\H) node[midway,left]{5.59 cm};
        \end{tikzpicture}}
        \subcaption{}
        \label{fig:shear_coupon_test_schematic}
    \end{subfigure}\hfill
    \begin{subfigure}[b]{0.477\textwidth}
        \centering
        \includegraphics[width=\linewidth]{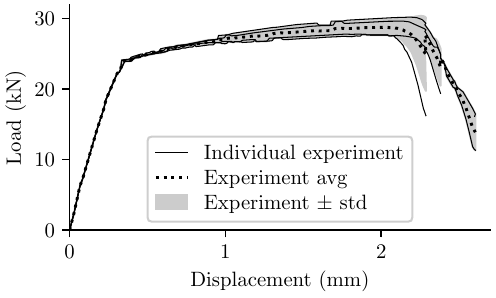}
        \subcaption{}
        \label{fig:shear_coupon_data}
    \end{subfigure}
    \caption{Shear-coupon experiment (SFC2): (a) notched-coupon geometry and boundary conditions---the left grip is fixed ($\boldsymbol{u}=\boldsymbol{0}$) and the right grip is displaced by $\boldsymbol{u}=(0,\bar{u}_i)$, localising shear in the ligament between the two notches; (b) measured shear load--displacement curves for the four specimens, with the specimen mean and the $\pm$ one-standard-deviation scatter band.}
    \label{fig:shear_coupon_test}
\end{figure}

We use the shear-coupon experiment from the Second Sandia Fracture Challenge (SFC2)~\cite{boyce2016second}, a community benchmark designed to test the predictive limits of computational fracture mechanics for a titanium alloy (Ti--6Al--4V). The specimen is a notched coupon loaded in shear by two grips: the left grip is held fixed while the right grip is displaced parallel to the notch ligament by a prescribed amount~$\bar{u}_i$, where $i$ indexes the sequence of imposed grip displacements. The two opposing notches localise the deformation into the ligament between them, producing a near-pure-shear stress state and driving the specimen through yielding, plastic flow, and eventual softening as damage accumulates (Figure~\ref{fig:shear_coupon_test_schematic}). The training target is the measured shear load--displacement response of four nominally identical specimens (VA1, VA2, VP2, VP6); we target the specimen-mean force and weight the loss by the specimen-to-specimen scatter. The mean response reaches a peak of approximately $28.7$~kN before softening (Figure~\ref{fig:shear_coupon_data}).

Both unknown material functions---the plastic hardening law $\sigma_y(p)$ and the ductile-damage law---are represented by learnable neural operators embedded in the finite element solver and trained end-to-end through the adjoint, reusing the progressive-discovery strategy of the previous examples. In the first stage, the elastic constants are fixed (the Young's modulus $E = 115$~GPa is taken from an independent tensile gauge) and the hardening law $\sigma_y(p)$ is learned on the rising branch of the curve, with damage disabled; as in Section~\ref{sec:load_controlled}, $\sigma_y(p)$ is parameterised by a monotone neural network that guarantees a non-decreasing yield stress by construction. In the second stage, the hardening operator is frozen and the only remaining unknown is the damage law, represented by a \emph{rectified integral network} (Section~\ref{sec:appendix_shear_coupon}) whose architecture enforces the qualitative properties expected of a ductile-damage hazard---a dormant plateau below a strain onset, then monotone, accelerating growth---by construction. The damage is regularised by a nonlocal (implicit-gradient) length scale to prevent spurious mesh-dependent localisation, and is learned on the full curve, including the post-peak softening branch. Because a shear coupon samples essentially a single stress state along each material point's path, the triaxiality dependence of a general damage model is not identifiable from this experiment; the reduced scalar form adopted here, driven by the accumulated plastic strain~$p$, is therefore both sufficient and well-posed. The finite-strain formulation and both network architectures are detailed in Section~\ref{sec:appendix_shear_coupon}.

Figure~\ref{fig:framework_schematic_shear} summarises the problem and the framework. Both stages minimise the same scatter-normalised mean-squared residual between the simulated and measured shear force,
\begin{equation}
    \mathcal{L} = \frac{1}{N} \sum_{i}^{N} \left( \frac{F_i^\text{fem} - F_i^\text{obs}}{\sigma_{\mathrm{eff},i}} \right)^{2},
    \label{eq:shear_loss}
\end{equation}
where $F_i^\text{fem}$ is the reaction force predicted by the FEM solver at grip displacement~$\bar{u}_i$, $F_i^\text{obs}$ is the experimental mean, and $\sigma_{\mathrm{eff},i} = \sqrt{\mathrm{std}_i^2 + (0.1\,F_{\mathrm{peak}})^2}$ normalises each residual by the measured specimen-to-specimen scatter---the standard deviation $\mathrm{std}_i$ across specimens at step~$i$, with a floor of $10\%$ of the peak force so that low-scatter points near zero do not dominate the fit. A residual of unity thus corresponds to a prediction within one experimental standard deviation.

\begin{figure}[!htbp]
    \begin{center}
        \begin{tikzpicture}
        \def\textsize{5mm}\def\boxwidth{10cm}\def\boxheight{8cm}\def\femsolverheight{0.85*\boxheight}
        \def\femsolvercolor{green!30}\def\physicalmodelcolor{yellow!50}\def\mloperatorcolor{orange!50}
        \def\observablecolor{yellow!50}\def\predictioncolor{green!30}\def\optimisercolor{orange!50}\def\losscolor{green!30}
        \def\physpace{\textsize}\def\physwidth{\boxwidth - 2.5*\physpace}\def\physheight{\boxheight - 2.5*\physpace}
        \def\physheightInner{0.82*\physheight}\def\mlwidth{0.34*\physwidth}\def\mlheight{0.9*\physheight}\def\mlheightInner{0.83*\mlheight}
        \def\bottomwidth{3.0cm}\def\bottomheight{2.0cm}\def\lossboxheight{1.4cm}
        \def\bottomspacing{(\boxwidth - 3*\bottomwidth)/4}\def\rowgap{0.5cm}\def\lossgap{0.6cm}
        \tikzset{thickbox/.style={draw, very thick, rounded corners}, thickarrow/.style={->, very thick}}
        \node[thickbox, fill=\femsolvercolor, minimum width=\boxwidth, minimum height=\femsolverheight, anchor=center] (femsolver) at (0,0) {};
        \node[above=0.15cm of femsolver.south] {\textbf{FEM Solver}};
        \node[thickbox, fill=\physicalmodelcolor, minimum width=\physwidth, minimum height=\physheightInner, anchor=center] (physical) at (0, .05*\boxheight) {};
        \node[right=0.4cm of physical.west, align=center] {
            \textbf{Geometry:} \\ Notched shear coupon \\\\
            \textbf{Governing Equations:} \\
            $\nabla \cdot \boldsymbol{P} + \boldsymbol{f} = 0$ \\ $\boldsymbol{P} = \boldsymbol{F}\boldsymbol{S}$ \\
            $\boldsymbol{F} = \boldsymbol{I} + \nabla \boldsymbol{u}$ \\
            $\boldsymbol{E} = \tfrac{1}{2}\!\left(\boldsymbol{F}^\text{T}\boldsymbol{F} - \boldsymbol{I}\right)$ \\\\
            \textbf{Boundary Conditions:} \\
            $\boldsymbol{u} = \boldsymbol{0}$ on $\Gamma_{\ell}$ \\ $\boldsymbol{u} = (0,\, \bar{u}_i)$ on $\Gamma_{r}$};
        \node[thickbox, fill=\mloperatorcolor, minimum width=\mlwidth, minimum height=\mlheightInner, anchor=center, align=center] (mloperator)
            at ($(physical.center)+(0.32*\physwidth,0cm)$) {\textbf{ML Elastoplastic} \\ \textbf{Damage} \\ \textbf{Constitutive Model} \\[5pt]
            {\footnotesize $\boldsymbol{S} = g(\bar{D})\,\mathbb{C} : (\boldsymbol{E} - \boldsymbol{E}^p)$} \\[3pt]
            {\footnotesize $f = \sigma_{\mathrm{eq}} - \sigma_y(p) \le 0$} \\[3pt]
            {\footnotesize $\sigma_y(p) = \mathcal{M}_{\theta_1}(p)$} \\[3pt]
            {\footnotesize $\Phi_{\mathrm{loc}}(p) = \mathcal{H}_{\theta_2}(p)$} \\[3pt]
            {\footnotesize $\bar{\Phi} - \ell^2 \nabla^2 \bar{\Phi} = \Phi_{\mathrm{loc}}(p)$} \\[3pt]
            {\footnotesize $\bar{D} = 1 - e^{-\bar{\Phi}},\ \ g = (1-\bar{D})^2$}};
        \node[thickbox, fill=\observablecolor, minimum width=\bottomwidth, minimum height=\bottomheight, anchor=north, align=center] (observable)
            at ({-\bottomwidth - \bottomspacing}, {-0.5*\femsolverheight - \rowgap}) {\textbf{Shear} \\ \textbf{Experiment} \\ $(\bar{u}_i,\, F^{obs}_i)$};
        \node[thickbox, fill=\predictioncolor, minimum width=\bottomwidth, minimum height=\bottomheight, anchor=north, align=center] (prediction)
            at (0, {-0.5*\femsolverheight - \rowgap}) {\textbf{Reaction} \\ \textbf{Force} \\ $(F^{fem}_i)$};
        \node[thickbox, fill=\optimisercolor, minimum width=\bottomwidth, minimum height=\bottomheight, anchor=north, align=center] (optimiser)
            at ({\bottomwidth + \bottomspacing}, {-0.5*\femsolverheight - \rowgap}) {\textbf{Optimiser} \\[3pt]
            {\footnotesize \textbf{Stage 1:} $\theta_1^{\ast} = \underset{\theta_1}{\operatorname{arg\,min}}\,\mathcal{L}$} \\[4pt]
            {\footnotesize \textbf{Stage 2:} $\theta_2^{\ast} = \underset{\theta_2}{\operatorname{arg\,min}}\,\mathcal{L}$}};
        \node[thickbox, fill=\losscolor, minimum width=\bottomwidth, minimum height=\lossboxheight, anchor=north, align=center] (loss)
            at (0, {-0.5*\femsolverheight - \rowgap - \bottomheight - \lossgap}) {\textbf{Loss} \\ $\mathcal{L} = \frac{1}{N} \sum_{i}^{N} \left( \dfrac{F_i^\text{fem} - F_i^\text{obs}}{\sigma_{\mathrm{eff},i}} \right)^{2}$};
        \draw[thickarrow] (observable.north) -- ($(physical.south west)!0.2!(physical.south east)$);
        \draw[thickarrow] (observable.south) -- (loss.west);
        \draw[thickarrow] (femsolver.south) -- (prediction.north);
        \draw[thickarrow] (optimiser.north) -- (mloperator.south);
        \draw[thickarrow] (prediction.south) -- (loss.north);
        \draw[thickarrow] (loss.east) -- (optimiser.south);
        \end{tikzpicture}
    \end{center}
    \caption{Mathematical definition of the problem and the FEML framework for the shear coupon, in two stages: the hardening law $\sigma_y(p)=\mathcal{M}_{\theta_1}(p)$ and the nonlocal ductile-damage hazard $\Phi_{\mathrm{loc}}(p)=\mathcal{H}_{\theta_2}(p)$, the latter a rectified integral network.}
    \label{fig:framework_schematic_shear}
\end{figure}

\begin{figure}[!htbp]
    \centering
    \begin{subfigure}{\textwidth}
        \centering
        \includegraphics[width=0.92\textwidth]{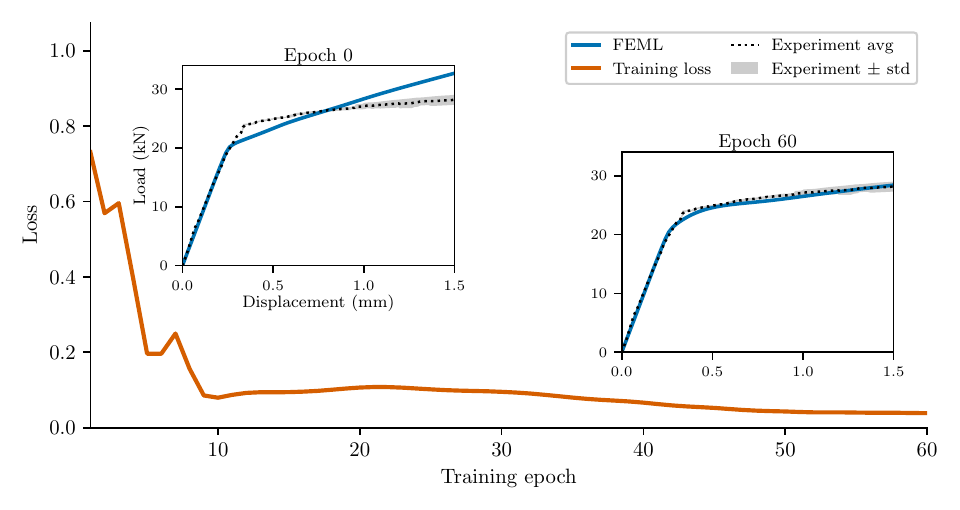}
        \subcaption{}
        \label{fig:shear_training_hardening}
    \end{subfigure}

    \vspace{1ex}
    \begin{subfigure}{\textwidth}
        \centering
        \includegraphics[width=0.92\textwidth]{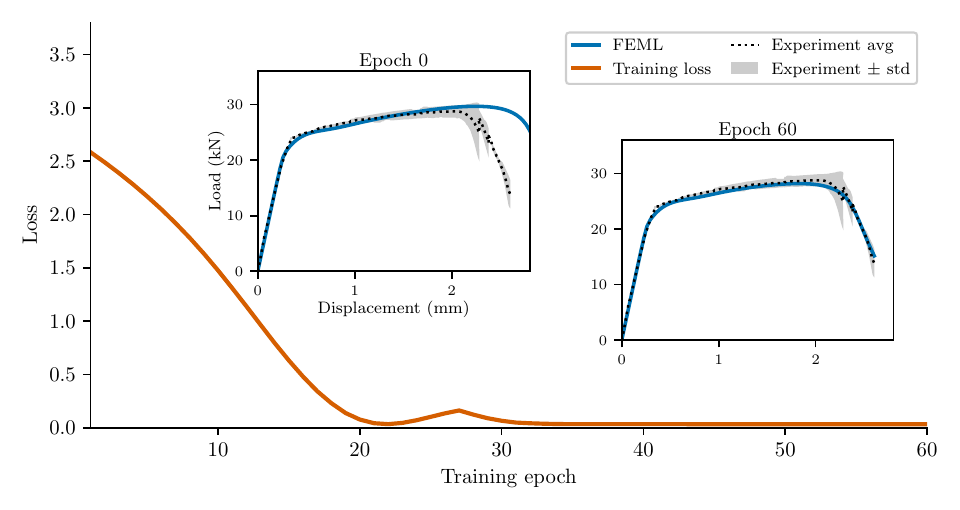}
        \subcaption{}
        \label{fig:shear_training_damage}
    \end{subfigure}
    \caption{Training the shear-coupon constitutive model. (a)~Stage~1, plastic hardening $\sigma_y(p)$ on the rising branch; (b)~Stage~2, the rectified-integral-network damage law on the full curve with the hardening operator frozen. Each panel shows the training loss and, in the boxed insets, the predicted (blue) versus measured (dotted) load--displacement response at the initial and final epochs.}
    \label{fig:shear_training}
\end{figure}

Figure~\ref{fig:shear_training} reports both training stages. Figure~\ref{fig:shear_training_hardening} shows the hardening stage: the loss decreases steadily and the predicted load--displacement curve converges to the measured rising branch, with the boxed insets at the initial and final epochs showing the elastic knee and hardening shoulder being captured. Figure~\ref{fig:shear_training_damage} shows the damage stage on the full curve: at initialisation the model does not yet soften and overshoots the experimental peak, whereas the trained rectified integral network reproduces the peak and the post-peak softening tail. The final model tracks the experiment within the specimen-scatter band across the entire curve, achieving a root-mean-square error of $0.54$~kN, approximately $1.9\%$ of the peak force.

\begin{figure}[!htbp]
    \centering
    \begin{subfigure}[b]{0.42\textwidth}
        \centering
        \includegraphics[width=\textwidth]{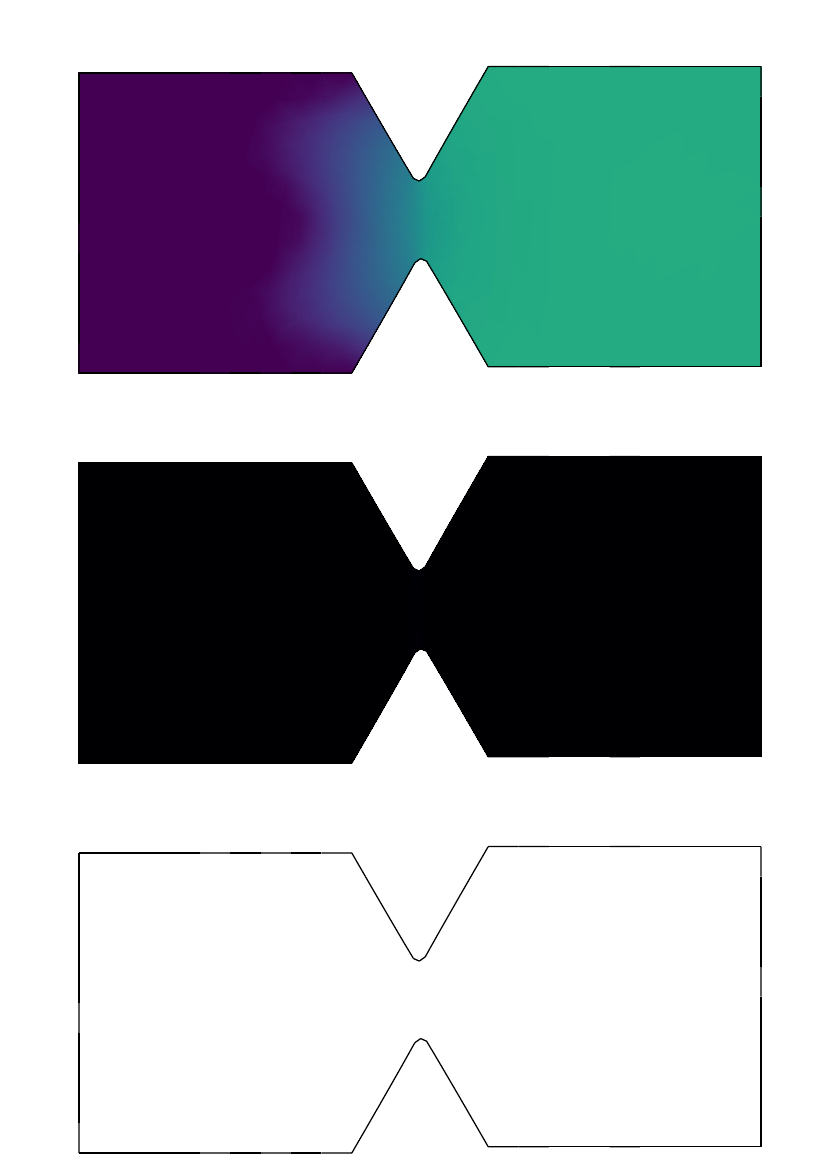}
        \subcaption{}
        \label{fig:shear_fields_a}
    \end{subfigure}\hfill
    \begin{subfigure}[b]{0.525\textwidth}
        \centering
        \includegraphics[width=\textwidth]{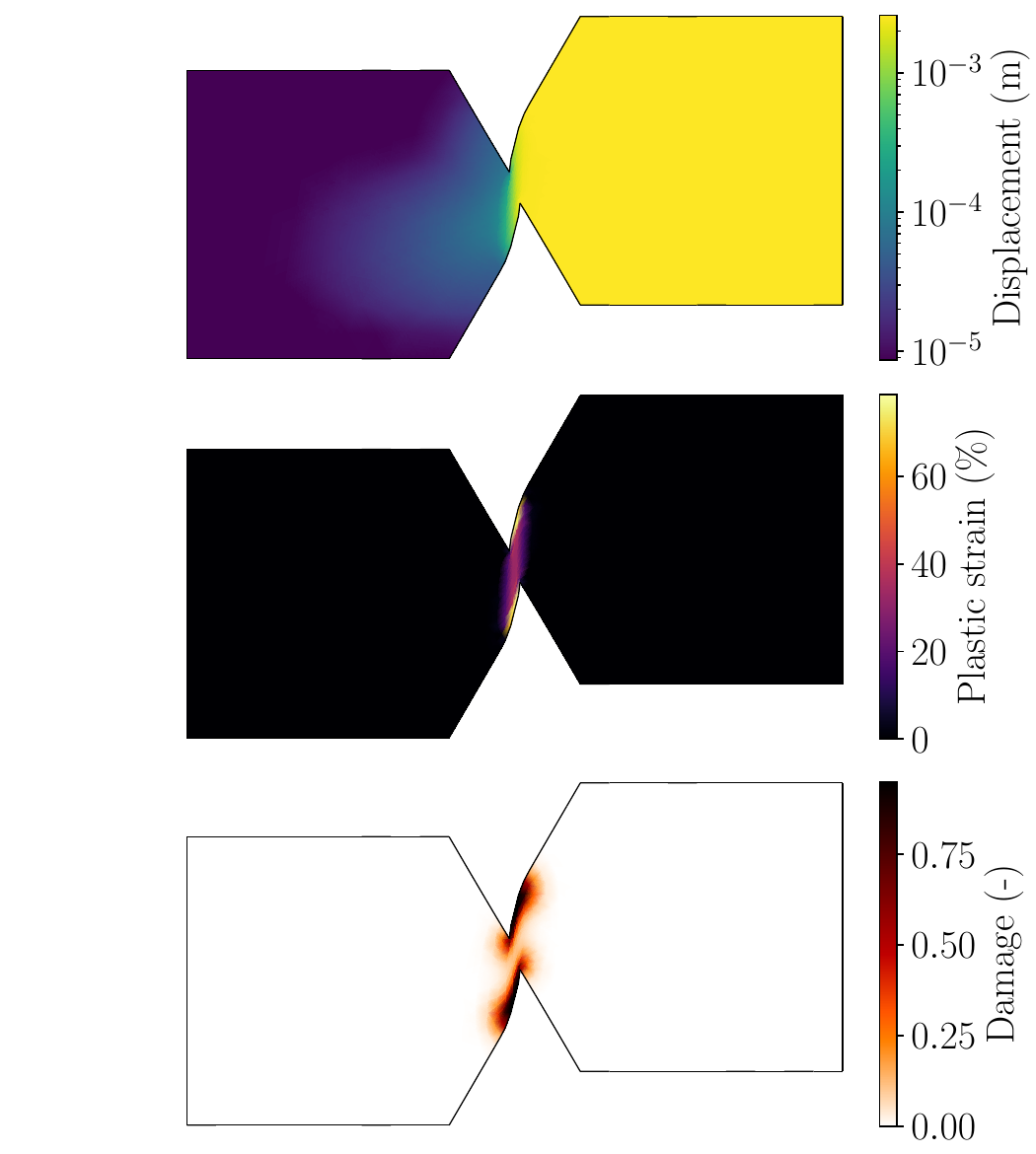}
        \subcaption{}
        \label{fig:shear_fields_b}
    \end{subfigure}
    \caption{Predicted fields (a)~before and (b)~after the load peak, shown on the deformed configuration. Top to bottom in each column: displacement magnitude, accumulated plastic strain, and nonlocal damage. Plastic strain and damage localise into the shear band across the ligament between the notches. The deformation has been amplified by four times for visualisation.}
    \label{fig:shear_fields}
\end{figure}

Figure~\ref{fig:shear_fields} shows the fields predicted by the trained model before and after the load peak. As deformation accumulates, the plastic strain and damage concentrate in the ligament between the two notches, forming the localised shear band that drives the macroscopic softening; the nonlocal length scale gives this band a finite, mesh-objective width. Together, these results demonstrate that FEML can recover a complete elastoplastic--damage description---two learned constitutive operators, including post-peak softening---directly from real laboratory measurements, extending the framework beyond the ground-truth-verified synthetic studies to a genuine experimental benchmark.

\subsection{Learning in Transient Thermodynamics Problems}
\label{sec:learning_thermal_properties}

In this section, we showcase the application of the proposed framework to transient thermodynamics problems, focusing on the learning of nonlinear thermal properties. Many materials exhibit temperature-dependent thermal behaviour, which significantly influences heat transfer processes. For instance, rocks, and ceramics respond differently to temperature variations, affecting their thermal performance in geothermal applications, civil engineering applications like bridge thermal expansion analysis, and ceramics, including high-temperature industrial reactors. Capturing these nonlinear thermal properties is essential for accurate simulations and predictive modelling.

The formulation of the transient thermodynamics problem is detailed in Section~\ref{sec:transient_thermodynamics_problem_formulation}. In the following sections, we show how the proposed framework can be used to learn the nonlinear thermal properties of the square plate from temperature measurements.

\subsubsection{Learning Thermal Properties from Temperature Measurements}
\label{sec:thermal_properties}

\begin{figure}[!htbp]
    \centering
    \subfloat[]{%
        \begin{tikzpicture}
        \begin{scope}[shift={(-3,0)}, scale=0.75]
            \pgfmathsetmacro{\shiftval}{(0.4*0.6)/sqrt(2)}
            \coordinate (shift) at (-\shiftval,-\shiftval);
            
            \fill[gray!5] (0,0) circle (2.5);
            \draw[thick] (0,0) circle (2.5);
            
            \fill[gray!20] ($(0,0)+(shift)$) circle (2.5);
            \draw[thick] ($(0,0)+(shift)$) circle (2.5);
            
            \begin{scope}
                \fill[white] ($(0,0)+(shift)$) circle (1.25);
                \draw[thick] ($(0,0)+(shift)$) circle (1.25);
            \end{scope}
            
            \draw[dashed] ($(-1.25,-1.25)+(shift)$) rectangle ($(1.25,1.25)+(shift)$);
            \draw[dashed] ($(-1.25,-1.25)+2.2*(shift)$) rectangle ($(1.25,1.25)+2.2*(shift)$);
            \draw[dashed] ($(-1.25,-1.25)+(shift)$) -- ($(-1.25,-1.25)+2.2*(shift)$);
            \draw[dashed] ($(-1.25,1.25)+(shift)$) -- ($(-1.25,1.25)+2.2*(shift)$);
            \draw[dashed] ($(1.25,-1.25)+(shift)$) -- ($(1.25,-1.25)+2.2*(shift)$);
            \draw[dashed] ($(1.25,1.25)+(shift)$) -- ($(1.25,1.25)+2.2*(shift)$);
            
            \begin{scope}
                \def\startAngle{145}
                \def\endAngle{305}
                \def\radius{1.25cm}
                \coordinate (O) at (0,0);
                \coordinate (A) at ($(O)+(\startAngle:\radius)$);
                \coordinate (B) at ($(O)+(\endAngle:\radius)$);
                \draw[thick] (A) arc (\startAngle:\endAngle:\radius);
            \end{scope}
            
            \begin{scope}
                \def\startAngle{145}
                \def\endAngle{215}
                \def\radius{2.49}
                \coordinate (O) at ($(0,0)+(shift)$);
                \coordinate (A) at ($(O)+(\startAngle:\radius)$);
                \coordinate (B) at ($(O)+(\endAngle:\radius)$);
                \fill[gray!50] (A) arc (\startAngle:\endAngle:\radius) -- cycle;
            \end{scope}
            
            \draw[thick, <->] (-2.5,2.8) -- (2.5,2.8) node[midway, above] {10 cm};
            
            \draw[thick, <->] (-2.5,2.5) -- ($(-2.5,2.5)+3*(shift)$)
                node[midway, left, xshift=-0.3cm] {0.4 cm};
            
            \node[overlay] at (-4.25,-0.25) {\textbf{\( T = \bar{T}_i\)}};
            
            \coordinate (disc_dashed_right) at ($ (1.25,0)+(shift) $);
        \end{scope}
        
        \begin{scope}[shift={(0.3,-0.5)}, scale=0.35]
            \coordinate (shift) at (0.45,0.45);
            \fill[gray!20] (0,0) rectangle (5,5);
            \draw[thick] (0,0) rectangle (5,5);
            \draw[thick] (0,5) -- ($(0,5)+(shift)$) -- ($(5,5)+(shift)$) -- (5,5) -- cycle;
            \fill[gray!20] (5,0) -- ($(5,0)+(shift)$) -- ($(5,5)+(shift)$) -- (5,5) -- cycle;
            \draw[thick] (5,0) -- ($(5,0)+(shift)$);
            \draw[thick] (5,5) -- ($(5,5)+(shift)$);
            \draw[black, thick] ($(5,5)+(shift)$) -- ($(5,0)+(shift)$);
            \draw[thick, <->] ($(6.75,0)+(shift)$) -- ($(6.75,5)+(shift)$) node[midway, right] {5 cm};
            \draw[thick, <->] ($(0,6.5)+(shift)$) -- ($(5,6.5)+(shift)$) node[midway, above] {5 cm};
            \draw[thick, <->] (-1.25,5) -- (-0.25,6.2) node[midway,left] {0.3 cm \qquad};
            \coordinate (sq_left) at ($(0,2.5)$);
        \end{scope}
        
        \draw[very thick,->] ($(sq_left)-(.25,0)$) -- ($(disc_dashed_right)+(.25,0)$);
        \end{tikzpicture}
    }
    \quad
    \subfloat[]{%
        \includegraphics[width=0.6\linewidth]{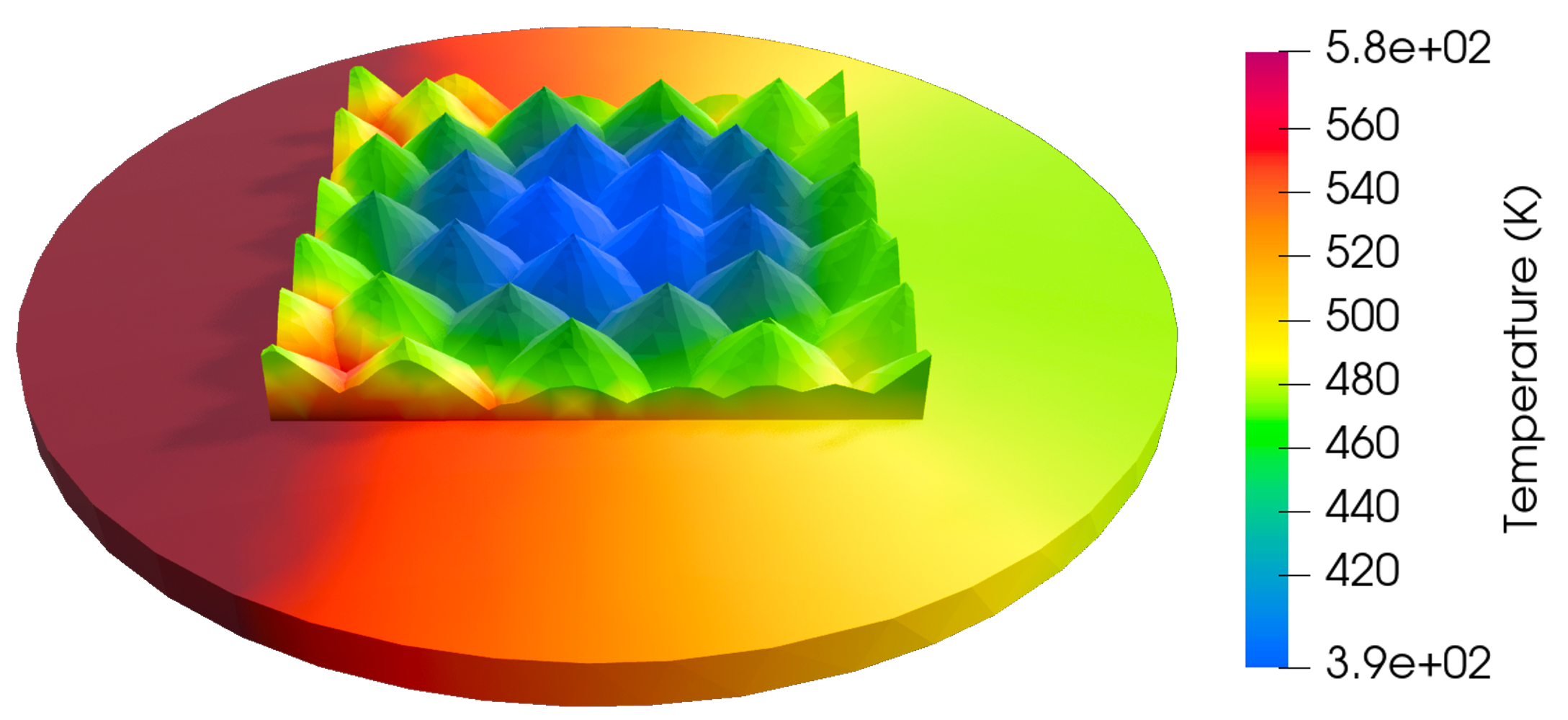}
    }
    \caption{(a) Schematic of the problem showing a disc-shaped copper plate (diameter 10 cm, thickness 0.4 cm, central hole diameter 5 cm) with a square plate of edge 5 cm with an irregular rughness and average thickness 0.3 cm. (b) displays the corresponding temperature distribution.}
    \label{fig:3d_heat_flow_experiment}
\end{figure}

In this example, we consider a transient heat conduction problem on two bodies: a copper disc and a square plate, as shown in Figure~\ref{fig:3d_heat_flow_experiment}(a). The copper disc is assumed to have a known constant thermal conductivity, while the square plate has an unknown thermal conductivity that is a nonlinear function of temperature which is parametrised with an ML model.

The two bodies start at room temperature, and then a heat source is applied to the left side of the copper disc, highlighted in dark gray. The heat source generates a fluctuating temperature boundary condition $\bar{T}_i$ that increaes in amplitude over time. The temperature field of the square plate is then measured at different times $i$. Figure~\ref{fig:3d_heat_flow_experiment}(b) shows the temperature field at the end of the experiment. The mathematical definition of the problem is illustrated in Figure~\ref{fig:framework_schematic_3d_heat_flow}.

The loss function used to train the ML model is defined as:

\begin{equation}
    \mathcal{L} = \frac{1}{N} \sum_{i}^{N} \frac{\left| T_i^\text{fem} - T_i^\text{obs} \right|}{\left| T_i^\text{obs} \right|},
\end{equation}

where $T_i^\text{fem}$ represents the temperature field predicted by the FEM solver with the nonlinear ML thermal conductivity model at time step $i$, and $T_i^\text{obs}$ is the corresponding synthetic experimental temperature field.
The synthetic experimental data is generated by solving the same PDE system with a known nonlinear thermal conductivity model. Both the training and test datasets consist of temperature fields at 12 time steps, but the two datasets are generated from two different synthetic experiments with different temperature boundary conditions. A 2\% noise is added to each nodal value of the temperature fields in the synthetic experimental data to represent real data. This noise level is higher than that used in the solid-mechanics examples (1\%), and is applied to full-field nodal snapshots at every time step, reflecting a more demanding perturbation.

\begin{figure}[!htbp]
    \centering
    \includegraphics[width=0.8\textwidth]{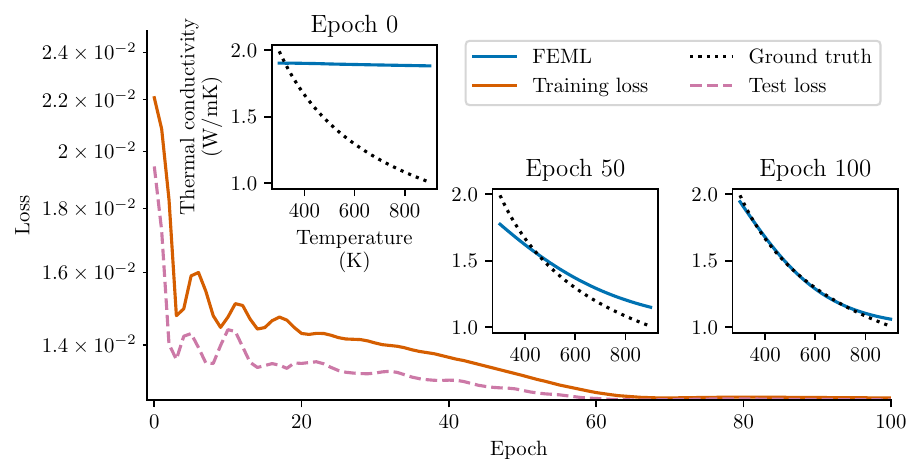}
    \caption{Training and test loss curves are shown alongside the thermal conductivity as a function of temperature for the machine learning model integrated within the FEM solver. Results are presented for training epochs 0, 50 and 100. The dotted line indicates the reference thermal conductivity model (ground truth), while the blue line represents the relationship learned by the ML model. By epoch 100 the model accurately captures the nonlinear dependence of thermal conductivity on temperature observed in the reference model.}
    \label{fig:3d_heat_conduction_plot}
\end{figure}

Figure~\ref{fig:3d_heat_conduction_plot} shows the evolution of the training and test loss, as well as the thermal conductivity profile predicted by the machine learning model embedded in the FEM solver, evaluated at epochs 0, 50 and 100. The dotted line represents the reference thermal conductivity model, which exhibits a nonlinear dependence on temperature, while the blue line shows the corresponding predictions from the ML model. At initialisation (epoch 0), the model fails to capture the correct trend and deviates significantly from the ground truth. By epoch 50, the learned conductivity shows partial agreement, although noticeable discrepancies remain. After 100 epochs, the ML-based model successfully reproduces the nonlinear thermal response, closely matching the reference profile across the entire temperature range. This highlights the model's ability to progressively learn the underlying physical law through training.

\begin{figure}[!htbp]
    \begin{center}
        \begin{tikzpicture}
        
            \def\textsize{5mm}
            \def\boxwidth{10cm}
            \def\boxheight{8cm}
            \def\femsolverheight{0.88*\boxheight}
            \def\physheightInner{0.85*\physheight}
            \def\mlheightInner{0.8*\mlheight}
        
            \def\femsolvercolor{green!30}
            \def\physicalmodelcolor{yellow!50}
            \def\mloperatorcolor{orange!50}
            \def\observablecolor{yellow!50}
            \def\predictioncolor{green!30}
            \def\optimisercolor{orange!50}
            \def\losscolor{green!30}
        
            \def\physpace{\textsize}
            \def\physwidth{\boxwidth - 2.5*\physpace}
            \def\physheight{\boxheight - 2.5*\physpace}
        
            \def\mlwidth{0.20*\physwidth}
            \def\mlheight{0.8*\physheight}
        
            \def\bottomwidth{2.6cm}
            \def\bottomheight{1.3cm}
            \def\lossboxheight{1.3cm}
        
            \def\bottomspacing{(\boxwidth - 3*\bottomwidth)/4}
            \def\rowgap{0.5cm}           
            \def\lossgap{0.6cm}           
        
            \tikzset{
                thickbox/.style={draw, very thick, rounded corners},
                thickarrow/.style={->, very thick}
            }
        
            \node[thickbox, fill=\femsolvercolor, minimum width=\boxwidth, minimum height=\femsolverheight, anchor=center] (femsolver) 
                at (0,0) {};
            \node[above=0.15cm of femsolver.south] {\textbf{FEM Solver}};
        
            \node[thickbox, fill=\physicalmodelcolor, minimum width=\physwidth, minimum height=\physheightInner, anchor=center] (physical) 
            at (0, .05*\boxheight) {};
            \node[right=0.4cm of physical.west, align=center] {
                \textbf{Geometry:} \\
                Disc-shaped plate\\
                with square sample\\\\
                \textbf{Governing Equations:} \\
                $\rho c_p \frac{\partial T}{\partial t} = \nabla \cdot (k \nabla T)$ \\\\
                \textbf{Initial Conditions:} \\
                $T = 300 K$ at $t=0$\\\\
                \textbf{Boundary Conditions:} \\
                $T = \bar{T}_i$ on $\Gamma_l$};
        
            \node[thickbox, fill=\mloperatorcolor, minimum width=\mlwidth, minimum height=\mlheightInner, anchor=center, align=center] (mloperator) 
            at ($(physical.center)+(0.35*\physwidth,0cm)$) {\textbf{ML} \\ \textbf{Heat}\\ \textbf{Conductivity}\\\textbf{Model} \\\\ $k = \mathcal{G}_\theta(T)$};
        
            \node[thickbox, fill=\observablecolor, minimum width=\bottomwidth, minimum height=\bottomheight, anchor=north, align=center] (observable) 
                at ({-\bottomwidth - \bottomspacing}, {-0.5*\femsolverheight - \rowgap}) {\textbf{Thermal} \\ \textbf{Experiment} \\ $(\bar{T}_i, T^{obs}_i$)};
        
            \node[thickbox, fill=\predictioncolor, minimum width=\bottomwidth, minimum height=\bottomheight, anchor=north, align=center] (prediction) 
                at (0, {-0.5*\femsolverheight - \rowgap}) {\textbf{Temperature} \\ \textbf{Fields} \\ $(T^{fem}_i)$};
        
            \node[thickbox, fill=\optimisercolor, minimum width=\bottomwidth, minimum height=\bottomheight, anchor=north, align=center] (optimiser) 
                at ({\bottomwidth + \bottomspacing}, {-0.5*\femsolverheight - \rowgap}) {\textbf{Optimiser} \\ $\theta^{C} = \underset{\theta}{\operatorname{arg\,min}} \, \mathcal{L}$};
        
            \node[thickbox, fill=\losscolor, minimum width=\bottomwidth, minimum height=\lossboxheight, anchor=north, align=center] (loss) 
                at (0, {-0.5*\femsolverheight - \rowgap - \bottomheight - \lossgap}) {\textbf{Loss} \\ $\mathcal{L} = \frac{1}{N} \sum_{i}^{N} \frac{\left| T_i^\text{fem} - T_i^\text{obs} \right|}{\left| T_i^\text{obs} \right|}$};
        
            \draw[thickarrow] (observable.north) -- ($(physical.south west)!0.2!(physical.south east)$);
            \draw[thickarrow] (observable.south) -- (loss.west);
            \draw[thickarrow] (femsolver.south) -- (prediction.north);
            \draw[thickarrow] (optimiser.north) -- (mloperator.south);
            \draw[thickarrow] (prediction.south) -- (loss.north);
            \draw[thickarrow] (loss.east) -- (optimiser.south);
        
        \end{tikzpicture}
    \end{center}
    \caption{Schematic of the mathematical formulation of the transient heat conduction problem and the proposed framework for learning thermal properties from temperature measurements.}
    \label{fig:framework_schematic_3d_heat_flow}
\end{figure}

\subsubsection{Learning a Closed-Form Thermal Conductivity Law via Symbolic Regression}
\label{sec:symbolic_regression_thermal}

Having learned the thermal conductivity operator as a neural network, we now extract a closed-form expression from it using symbolic regression~\cite{schmidt2009distilling}. This post-processing step converts the trained model into a mathematical formula that can be inspected and interpreted by domain specialists.

We evaluate the trained operator on 500 uniformly spaced temperatures in $[400, 700]$~K---the regime reliably covered during training (see Section~\ref{sec:thermal_properties})---and fit symbolic expressions to the resulting $(T, k_{\text{ML}})$ data using PySR~\cite{cranmer2023pysr}. Because thermal conductivity is strictly positive, we augment the mean-squared-error loss with a large penalty for negative predictions, which steers the evolutionary search away from unphysical candidates without constraining the functional form. The resulting Pareto front is then post-filtered: every candidate is evaluated on a dense grid over $[1, 1500]$~K and any expression that produces a negative or non-finite value is discarded. This two-stage strategy---penalised loss during search, global positivity check afterwards---follows standard practice in constrained symbolic regression~\cite{kronberger2022shape} and guarantees $k > 0$ without biasing the discovered expression towards a specific structure such as $\exp(\cdot)$.

\begin{figure}[!htbp]
    \centering
    \includegraphics[width=0.8\textwidth]{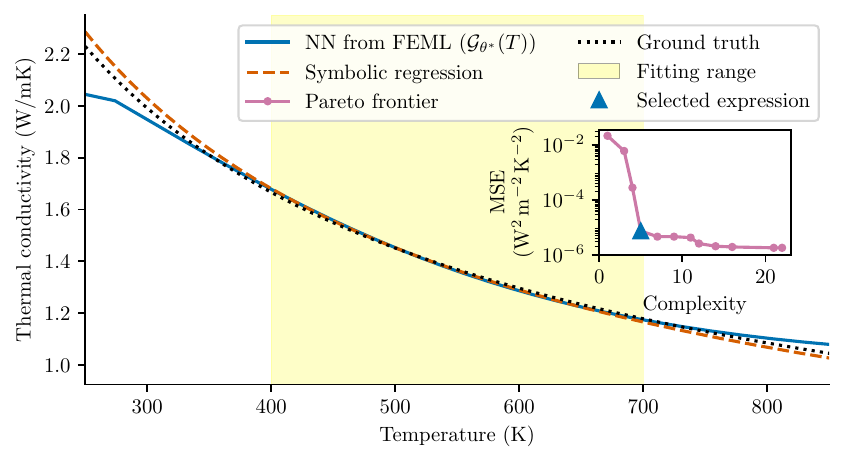}
    \caption{Symbolic regression of the learned thermal conductivity operator. A penalised loss steers the search away from negative predictions and a post-hoc positivity filter over $[1, 1500]$~K ensures $k > 0$ globally. The main axes compare the thermal conductivity as a function of temperature: ground truth (dotted black), NN from FEML $\mathcal{G}_{\theta^*}(T)$ (solid blue), and the selected symbolic regression expression (dashed vermillion). The yellow band marks the fitting range $[400, 700]$~K. The inset shows the Pareto front of candidate expressions (mean squared error versus complexity); the blue triangle marks the selected expression (complexity~5), chosen as the highest-scoring candidate among those that satisfy the positivity constraint.}
    \label{fig:symbolic_regression}
\end{figure}

Figure~\ref{fig:symbolic_regression} compares the ground-truth conductivity, the neural network prediction, and the symbolic expression selected from the filtered Pareto front over $[250, 900]$~K, which includes the extrapolation regime beyond 700~K (the yellow band marks the fitting range). From the candidates that pass the positivity filter we select the one with the highest PySR \emph{score}, defined as the log-loss improvement per unit of added complexity, which balances approximation accuracy against parsimony. The selected expression, of complexity~5, reads:
\begin{equation}\label{eq:sr_k}
    k_{\text{SR}}(T) = \left(\frac{a}{T}\right)^{\!b},
\end{equation}
with $a = 884.2$ and $b = 0.6547$ (temperature in K, conductivity in W\,m$^{-1}$\,K$^{-1}$). This pure power-law form is noteworthy because the ground truth rewrites as $k = (911.6/T)^{0.62}$: the physical signal captured by FEML was rich enough for symbolic regression to recover, without any prior knowledge of the underlying law, essentially the same $(C/T)^{\alpha}$ structure---with $a$ within 3\% of the exact constant and $b$ within 6\% of the exact exponent. The expression achieves $R^{2} = 0.997$ against the ground truth on the fitting range $[400, 700]$~K and $R^{2} = 0.886$ in the extrapolation region beyond 700~K. The inset of Figure~\ref{fig:symbolic_regression} displays the Pareto front of all candidate expressions returned by the symbolic regression search, plotting mean squared error on a logarithmic scale against expression complexity. The MSE decreases by roughly three orders of magnitude between complexities~3 and~5, after which the front plateaus: expressions of complexity~7 through~20 offer only marginal reductions in error. The blue triangle marks the selected expression (complexity~5), which sits at the knee of the Pareto front where accuracy gains per additional unit of complexity become negligible.

\section{Discussion and Conclusions}
\label{sec:conclusions}

We have introduced FEML, a fully differentiable finite element–based machine learning framework for discovering missing physics in systems governed by partial differential equations. The key idea is to retain the known governing equations in their standard finite element (FEM) form, while representing the unknown physical relationships as trainable operators embedded within the variational formulation. By differentiating end-to-end through both the FEM solver and the machine learning components, FEML enables the identification of internal operators—such as constitutive or transport laws—from indirect, experimentally accessible quantities. In contrast to surrogate models that learn configuration-specific PDE solutions, the learned operator in FEML is defined on function spaces and can be reused across geometries, boundary conditions, dimensions, and discretisations.

A central ingredient of the framework is the use of structure-preserving operator networks (SPONs) to parameterise the missing-physics operator. By operating on the degrees of freedom of finite element spaces, SPONs preserve key properties of the continuous operators at the discrete level, and come with theoretical guarantees. SPON models offer zero-shot  cross-discretisation capabilities, which we demonstrated on our numerical experiments: once trained, the operator can be evaluated on meshes and finite element spaces different from those used during training, without additional retraining.

We assessed FEML on a sequence of solid mechanics problems in which the hidden operator is a nonlinear constitutive relation for an unknown material. The constitutive law was discovered progressively in two stages. In the first stage (displacement-controlled uniaxial test), FEML recovered a nonlinear elastic softening law from a noisy and extremely sparse dataset: just six force–displacement values, perturbed with \(1\%\) noise. The loading was kept within the elastic regime, so only the strain-dependent Young's modulus $E(\kappa)$ was active. Despite having access only to global load measurements, and no direct stress information, the learned operator reconstructed the full elastic response closely matching the reference behaviour. This illustrates both the data efficiency of the framework and the advantage of exploiting equilibrium constraints through the embedded FEM solver.

In the second stage (load-controlled Brazilian disc test), the learned elastic operator was frozen and the framework leveraged full-field displacement data to identify the plastic hardening law $\sigma_y(p)$ of the same material. The learned yield-stress evolution reproduced complex displacement fields under indirect loading when trained on noisy nodal displacement fields at 15 load levels. This sequential discovery strategy—learning elastic, then plastic, behaviour from progressively richer data—demonstrates that FEML can compose frozen and trainable operators within a single differentiable framework to incrementally build up a complete constitutive description.

We note that the sequential strategy adopted here is not the only option available within FEML. Because the framework supports end-to-end differentiability through arbitrary compositions of PDE solvers and ML operators, one could alternatively define a joint loss that combines data from both experiments—the force residuals from the uniaxial test and the displacement-field residuals from the Brazilian disc test—and train the elastic and plastic operators concurrently. Joint training may be advantageous when the phenomena of interest are strongly coupled or when data from a single experiment type are insufficient to uniquely identify either operator in isolation. The sequential approach was preferred here because it decouples the learning tasks, simplifies the optimisation landscape, and mirrors the staged experimental workflow commonly adopted in practice; however, the joint formulation remains a natural extension enabled by the framework's differentiable architecture.

We then combined both pretrained operators—$E(\kappa)$ and $\sigma_y(p)$—into a foundation elastoplastic constitutive model and applied it in a zero-shot fashion to a three-dimensional cylindrical rod with a transverse keyhole-shaped through-slot under torsion. Without any retraining, the learned operators produced displacement and stress fields in close agreement with the reference solution. This level of accuracy is noteworthy given that the operators were trained on noisy data from different two-dimensional mechanical tests and then transferred to a 3D geometry, with different loading and boundary conditions. This result illustrates a main advantage of operator-level learning over solution-level surrogates: once the hidden physics has been identified, the same operators can be deployed as reusable components in new scenarios, rather than being restricted to the configuration on which they were trained.

As a demonstration of real-world applicability, we moved from synthetic to \emph{real} experimental data, applying the framework to a benchmark shear-coupon test on a titanium alloy. This example extends the discovered description beyond elasticity and hardening to ductile damage: a plastic-hardening law and a ductile-damage hazard are represented by two learned neural operators---both monotone networks built from the integral of a non-negative activation, with the activation matched to each law (a smooth rise for hardening, a hard onset followed by accelerating growth for damage)---and trained end-to-end through the adjoint within a finite-strain formulation. Because the underlying constitutive law is unknown for real data, the damage operator is regularised by a nonlocal length scale that renders the softening band mesh-objective, and success is judged by agreement with the measured response. The trained model reproduces the full shear load--displacement curve---elastic loading, yield, plateau, peak, and post-peak softening---to within the experimental specimen-to-specimen scatter, with the plastic strain and damage localising into a shear band across the notched ligament. This demonstrates that FEML composes multiple learned constitutive operators and recovers a complete elastoplastic--damage description directly from real laboratory measurements, not only from synthetic data.

To demonstrate FEML applications beyond quasi-static solid mechanics, we apply the framework to a transient heat-conduction problem in which the missing physics is an unknown temperature-dependent thermal conductivity. The operator was trained from noisy temperature fields (with \(2\%\) perturbations) measured on a heterogeneous, coupled 3D geometry comprising a copper disc and an irregular square plate. The learned thermal law accurately reproduced the underlying nonlinear conductivity profile and generalised to a test experiment with different temperature boundary conditions. This example demonstrates that the framework extends naturally to time-dependent problems and heterogeneous materials, and that it can accommodate realistic levels of measurement noise in complex three-dimensional configurations. Furthermore, we demonstrated that a compact closed-form expression can be learned from the neural operator via symbolic regression (Section~\ref{sec:symbolic_regression_thermal}). The selected expression achieves $R^{2} > 0.99$ on the fitting range and, notably, recovers the same pure power-law structure $(C/T)^{\alpha}$ as the ground truth, with the fitted constants within 3\% and 6\% of their exact values---even though no prior knowledge of the underlying functional form was provided to the search. This indicates that the physical signal encoded in the FEML-trained neural network is faithful enough for the true analytical law to be extracted automatically. Taken together, these results demonstrate that the combined FEML--symbolic-regression framework can recover closed-form physical laws from indirect, noisy measurements without presupposing the functional form of the underlying relation.

The present work also suggests several opportunities and limitations that warrant further investigation. First, although end-to-end differentiability through a mature FEM stack enables flexible model discovery, it comes at a computational cost: each training epoch requires both a forward FEM solve and an adjoint solve for gradient computation, and this cost is repeated for every load or time step in the training dataset. Across all examples, end-to-end training takes from a few minutes for the simplest case to a few hours for the most demanding, on a single laptop---a MacBook Air with a 10-core Apple M4 chip and 32\,GB of unified memory---with the cost set mainly by the mesh resolution, the number of load or time steps evaluated per epoch, and the complexity of the constitutive update.

The displacement-controlled example uses a coarse $6\times6$ rectangular mesh with cubic elements and a single ML model, resulting in the shortest training time. The load-controlled Brazilian disc example is more expensive owing to a finer disc mesh, the J2 elastoplastic radial-return algorithm, and the evaluation of full-field displacement errors at each epoch. The transient heat conduction example operates on a 3D tetrahedral mesh with 12 time steps per epoch, which increases the per-epoch cost. We note that the computational overhead of the ML operator during forward solves is negligible: performance benchmarks show that the difference in solve time between a standard FEM and an ML-augmented FEM solve is less than 1.3\% across all mesh sizes tested; the training cost is dominated by the adjoint solves required for backpropagation. Efficient solvers, reduced-order models, and mixed-precision or multi-fidelity training strategies will be important to scale FEML to field-scale applications. A summary of all ML hyper-parameters and a sensitivity analysis with respect to training epochs are provided in Section~\ref{sec:appendix_sensitivity}; the analysis shows that the learned operators stabilise well before final convergence and that training and test losses track each other closely, indicating low sensitivity to the specific hyper-parameter choices. Second, identifiability issues can arise when multiple operators produce similar observables. In the FEML setting, three features mitigate this risk: (i)~the operator is embedded inside a PDE whose governing equations and boundary conditions must be satisfied at every spatial point simultaneously and across multiple load or time steps (e.g., momentum balance in the solid-mechanics examples, energy conservation in the thermal example), imposing far more constraints than learning from paired input--output data alone; (ii)~the specific neural-network parameterisation within the SPON structure restricts the hypothesis class to physically admissible mappings; in the constitutive examples, the invariant-based inputs, isotropic tensor basis, and positive activation functions (e.g., SoftPlus) enforce isotropy, frame indifference, and positivity of moduli by construction, eliminating non-physical parametrisations; and (iii)~spatially heterogeneous loading (e.g., the Brazilian disc test) activates diverse local strain states from a single experiment, densely sampling the operator's input domain. The zero-shot generalisation result, where constitutive operators trained on 2D tests are deployed on a three-dimensional cylindrical rod with a keyhole through-slot under torsion, provides empirical evidence that the learned operators capture the true material law rather than merely fitting the training configuration. Nevertheless, formal identifiability guarantees are not established in this work, and in general, addressing non-uniqueness will require careful experiment design, regularisation informed by prior knowledge, and possibly multi-modal data.

All experiments except the shear-coupon study (Section~\ref{sec:shear_coupon}) use synthetic data generated from the same finite element solver (Firedrake). This choice is deliberate: synthetic data provides an exact ground-truth operator against which the learned model can be quantitatively validated, something that is not possible with real experimental data, where the true constitutive or transport law is unknown. We note, however, that several structural features of the framework mitigate the risk of overfitting to solver-specific numerical artefacts. The ML model learns a local, pointwise material law (e.g., strain invariants to Lam\'e parameters, or temperature to conductivity) and has no access to the mesh topology, element shape functions, or linear-algebra internals of the solver. Furthermore, the synthetic examples presented span fundamentally different governing equations, quasi-static momentum balance (an elliptic vector system) and transient heat conduction (a parabolic scalar equation), with different discretisations, element types, and time-stepping schemes; there is no common ``numerical behaviour'' across these problems to which the ML model could overfit. The zero-shot generalisation experiment provides additional evidence: constitutive operators trained on 2D meshes under uniaxial compression and diametral loading were deployed on a 3D tetrahedral mesh under torsion. Had the model memorised solver-specific artefacts, this cross-dimension, cross-geometry, cross-loading transfer would have failed. We have also taken a first step towards validation on real measurements: the shear-coupon study (Section~\ref{sec:shear_coupon}) learns both a plastic-hardening and a ductile-damage operator directly from a benchmark experiment, reproducing the measured response---including post-peak softening---to within the experimental specimen-to-specimen scatter. Broader validation across materials, geometries, and loading regimes remains a natural next step.

We note that a direct quantitative comparison between FEML and solution-level surrogate methods such as FNO or PINO is not applicable, because the two classes of methods address fundamentally different learning problems and produce different types of output. Solution-level surrogates learn the full mapping from PDE inputs (e.g., forcing, boundary conditions) to PDE solutions; they are designed for fast evaluation within a training distribution but do not extract a reusable, configuration-independent operator. FEML, by contrast, learns the hidden operator itself, which can then be ported across geometries, boundary conditions, dimensions, and discretisations---as demonstrated by the zero-shot 2D$\to$3D transfer experiment in Section~\ref{sec:torque_holed_plate}. Similarly, while inverse-PINN formulations~\cite{raissi_physics-informed_2019} can in principle be set up for operator-discovery problems, they enforce PDE residuals at collocation points (strong form) rather than through the variational formulation used by FEML, and do not employ structure-preserving operator networks---the key ingredient enabling zero-shot cross-discretisation generalisation. These methods are thus best understood as complementary rather than competing: surrogates deliver fast inference for parameterised PDE families, while FEML discovers the underlying physical law.

Future work that goes beyond direct physics discovery, will include the use of FEML in a surrogate-modelling, where the embedded ML operator emulates expensive high-order coupling terms in a full-order PDE model. In such a workflow, high-fidelity simulations provide training data for the operator, which then replaces the costly components in subsequent simulations. This strategy has the potential to deliver reduced-order models that retain the fidelity of the original FEM formulation while significantly lowering computational cost, with applications ranging from nonlinear constitutive modelling and fracture to multiphase flow and thermomechanical coupling.

Although the current implementation is built on a finite element solver (Firedrake), the conceptual framework is not restricted to FEM. The underlying mathematical formulation, i.e. minimising a loss where the forward state satisfies a discrete residual containing a trainable operator, applies equally to finite-volume, finite-difference, or other discretisation methods, provided that (i)~the solver supports adjoint-based or algorithmic differentiation for computing gradients through the forward solve, and (ii)~its differentiation system can be composed with the computational graph of the ML library to enable end-to-end gradient flow. As discussed in the Introduction, differentiable non-FEM solvers already exist; however, certain features of the current framework, in particular, structure-preserving operator networks (SPONs) and zero-shot super-resolution, exploit the finite element function-space structure and would require adaptation for alternative discretisations. Importantly, the learned operator itself is solver-independent: since the ML model learns a local, pointwise material law, it can in principle be deployed inside any compatible solver at inference time.

In summary, FEML provides a general and flexible route to combine physics-based finite element solvers with data-driven operator learning. By preserving known physics, learning only the missing relations, and respecting the underlying function-space structure, the framework yields data-efficient and interpretable operators that can be ported across problems, geometries, and discretisations. The solid-mechanics examples presented here demonstrate sequential discovery of elastic and plastic behaviour for the same material, with the learned operators composed into a foundation model that generalises zero-shot to a three-dimensional torsion problem. On real laboratory data, the framework composed two learned operators---a plastic-hardening law and a ductile-damage hazard---and reproduced the measured shear-coupon response, including post-peak softening, to within the experimental specimen-to-specimen scatter. The thermal conductivity example further shows that closed-form expressions can be learned from the neural operators via symbolic regression, completing the framework from indirect measurements to human-interpretable physical laws. We expect this combination of differentiable PDE solvers and structure-preserving operator networks to underpin a new class of predictive tools for computational solid mechanics, thermodynamics, and, more broadly, for scientific and engineering applications in which incomplete physical knowledge must be reconciled with limited but informative data.

\section{Methods}

\subsection{Differentiable Coupling of Firedrake and ML Frameworks}
\label{sec:differentiable_coupling}

Learning the operator $\G_{\theta}$ from indirect observations requires end-to-end gradient flow through the coupled FEM--ML system. This section describes how that gradient flow is realised; full details are given in~\cite{bouziani_diffprogram_2024}.

\paragraph{Computational graph composition.}
The key idea is to view the coupled FEM--ML simulation as a single directed acyclic graph (DAG). The coupling strategy introduced in~\cite{bouziani_diffprogram_2024} partitions this DAG into two sub-graphs: Firedrake-based operations are differentiated by \textit{pyadjoint}, whereas ML operations are handled by the native AD engine of the ML library (\texttt{autograd} for PyTorch~\cite{PyTorch}, or \texttt{grad} for JAX~\cite{jax2018github}). This partitioning retains the full capabilities and performance of each framework.

\begin{figure}[htbp]
    \centering
    \begin{subfigure}[b]{0.48\textwidth}
        \centering
        \includegraphics[width=\textwidth]{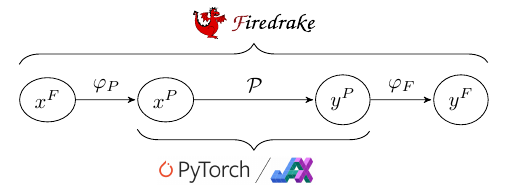}
        \caption{Embedding ML models into the Firedrake computational graph. $\mathcal{P}$ denotes ML operations; $P$ and $F$ denote PyTorch/JAX and Firedrake variables, respectively. Adapted from~\cite{bouziani_diffprogram_2024}.}
        \label{fig:coupling_ml_in_firedrake}
    \end{subfigure}
    \hfill
    \begin{subfigure}[b]{0.48\textwidth}
        \centering
        \includegraphics[width=\textwidth]{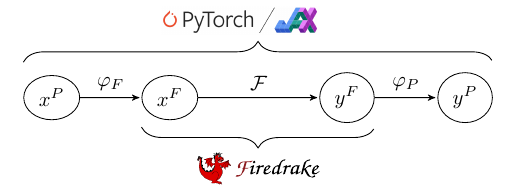}
        \caption{Embedding Firedrake operations into the PyTorch/JAX computational graph. $\mathcal{F}$ denotes Firedrake operations. Adapted from~\cite{bouziani_diffprogram_2024}.}
        \label{fig:coupling_firedrake_in_ml}
    \end{subfigure}
    \caption{Differentiable coupling between Firedrake and ML frameworks. In both directions, the casting operations $\varphi_{F}$ and $\varphi_{P}$ convert between PyTorch/JAX tensors and Firedrake functions. Adapted from~\cite{bouziani_diffprogram_2024}.}
    \label{fig:coupling_diagrams}
\end{figure}

The coupling operates in two directions (Figures~\ref{fig:coupling_ml_in_firedrake} and~\ref{fig:coupling_firedrake_in_ml}):

\begin{enumerate}
    \item \emph{Embedding ML models into PDE systems} (Figure~\ref{fig:coupling_ml_in_firedrake}). An ML model, such as a PyTorch neural network, is exposed as a symbolic operator within UFL~\cite{UFL_2014}, the domain-specific language that Firedrake uses to express variational forms. A dedicated \texttt{ml\_operator} interface represents the ML model as a UFL linear form with $k$ coefficient operands. Symbolic UFL operations such as \texttt{derivative}, \texttt{action}, and \texttt{adjoint} are then transparently dispatched to the ML framework's AD engine. In practice, this means that the derivatives $\frac{\partial N}{\partial u}$ (required for Newton solves) and $\frac{\partial N}{\partial \theta}$ (required for training) are computed by PyTorch or JAX without any manual implementation of derivative expressions.

    \item \emph{Embedding PDE solvers into ML frameworks} (Figure~\ref{fig:coupling_firedrake_in_ml}). A Firedrake solve is registered as a custom differentiable operator in $\operatorname{torch.autograd}$ (or equivalently in JAX). During the forward pass, the PDE is solved by Firedrake; during the backward pass, \textit{pyadjoint} evaluates the adjoint model~\eqref{eq:adjoint_model}--\eqref{eq:adjoint_equation}. Lightweight casting operations $\varphi_{F}$ and $\varphi_{P}$ convert between ML tensors and Firedrake functions, handling the different data representations of the two frameworks~\cite{bouziani_diffprogram_2024}. The resulting composite operator makes the PDE solution, together with its derivative information, a first-class node in the ML computational graph, so that standard optimisers such as Adam can update $\theta$ through end-to-end backpropagation.
\end{enumerate}

\paragraph{Adjoint-based differentiation through the PDE solve.}
Consider the following concrete illustration of how backpropagation is realised when a variational problem appears as a node in the coupled DAG. The PDE solution $u_{\theta}$ depends implicitly on the ML parameters $\theta$ through $\G_{\theta}$, and computing $\frac{\d u_{\theta}}{\d \theta}$ by differentiating through every Newton iteration would be prohibitively expensive. Instead, the gradient is obtained via the \emph{adjoint method}. Let $V$ and $M$ be Hilbert spaces, and let $u \in V$ be the solution of the parametrised PDE
\begin{equation}
    \label{eq:adjoint_F_residual}
    F(u, m; v) = 0 \quad \forall v \in V,
\end{equation}
where $m \in M$ denotes the control (in our context, the ML parameters $\theta$ or any other quantity that $\G_{\theta}$ depends on). Under the assumptions that $F$ is continuously Fr\'echet differentiable and that the linearised operator $\frac{\partial F}{\partial u}$ is non-singular, the implicit function theorem ensures existence and differentiability of the solution map $u(\cdot) \colon M \to V$~\cite{hinze_optimization_2009}. Implicit differentiation of $F(u(m), m; v) = 0$ with respect to $m$ then gives
\begin{equation}
    \label{eq:adjoint_dudm}
    \frac{\d u}{\d m} = -\left(\frac{\partial F}{\partial u}\right)^{-1} \frac{\partial F}{\partial m}.
\end{equation}
Backpropagation through the PDE solve is realised by the adjoint model of $u(m)$, which takes the form
\begin{equation}
    \label{eq:adjoint_model}
    \mathcal{J}^{*}_{u,m}(w) = -\frac{\partial F}{\partial m}^{*} \lambda, \quad \forall w \in V^{*},
\end{equation}
where $\lambda \in V$ is the solution of the \emph{adjoint equation}
\begin{equation}
    \label{eq:adjoint_equation}
    \frac{\partial F}{\partial u}^{*} \lambda = w.
\end{equation}
In FEML, this adjoint solve is performed automatically by the \textit{pyadjoint} package~\cite{dolfin-adjoint_2019}, which tapes the forward Firedrake computation and derives the corresponding adjoint system without user intervention.

\paragraph{End-to-end gradient flow for training.}
As a natural consequence of combining both embedding directions, the complete gradient chain required for training in FEML is assembled automatically. Given a loss $\mathcal{L}(u_{\theta}, u^{\text{obs}})$, the gradient with respect to the ML parameters $\theta$ is computed as follows: (i)~the ML framework differentiates $\mathcal{L}$ with respect to $u_{\theta}$; (ii)~\textit{pyadjoint} solves the adjoint equation~\eqref{eq:adjoint_equation} to propagate gradients through the PDE solve; and (iii)~the ML framework differentiates through $\G_{\theta}$ to obtain $\frac{\d \mathcal{L}}{\d \theta}$. This three-stage chain requires no manual derivation of adjoint equations or gradient expressions. The implementation is built on Firedrake~\cite{FiredrakeUserManual}, which is publicly available, and uses its native \textit{pyadjoint} integration for adjoint computations. All examples presented in this work can be reproduced using Firedrake.

\subsection{Solid mechanics problem formulation}
\label{sec:appendix_solid_mechanics_problem_formulation}
All solid-mechanics examples are quasi-static, but they span two kinematic regimes. The displacement-controlled (Section~\ref{sec:appendix_displacement_controlled}), load-controlled (Section~\ref{sec:appendix_load_controlled}), and zero-shot torsion (Section~\ref{sec:appendix_torque_holed_plate}) examples undergo small strains and rotations and are modelled with infinitesimal, small-displacement kinematics; the shear-coupon example (Section~\ref{sec:appendix_shear_coupon}) develops large rotations within its ligament and is modelled with a finite-strain, total-Lagrangian formulation. The momentum balance and the strain and stress measures differ between the two regimes, but the way a learnable operator is embedded in the constitutive law does not.

In the small-displacement regime, equilibrium is expressed in the (undeformed) configuration through the balance of linear momentum,

\begin{equation}
    \nabla \cdot \boldsymbol{\sigma} + \boldsymbol{f} = 0,
    \label{eq:momentum}
\end{equation}

where \( \boldsymbol{\sigma} \) (N/m\(^2\)) is the Cauchy stress tensor and \( \boldsymbol{f} \) (N/m\(^3\)) denotes body forces, taken to be zero in all examples presented here.

The infinitesimal strain tensor is defined by the symmetric gradient of the displacement field:

\begin{equation}
    \boldsymbol{\varepsilon} = \frac{1}{2} \left( \nabla \boldsymbol{u} + (\nabla \boldsymbol{u})^\text{T} \right),
    \label{eq:strain}
\end{equation}

where \( \boldsymbol{u} \) (m) is the displacement field and \( \boldsymbol{\varepsilon} \) is dimensionless.

When strains or rotations are no longer small, this linearisation is inadequate and a finite-strain, total-Lagrangian description is used instead. Equilibrium is then posed in the reference (undeformed) configuration,

\begin{equation}
    \nabla_{\!0} \cdot \boldsymbol{P} + \boldsymbol{f} = 0,
    \label{eq:momentum_ref}
\end{equation}

where \( \boldsymbol{P} \) is the first Piola--Kirchhoff stress and \( \nabla_{\!0} \) the gradient taken with respect to the reference coordinates. Deformation is measured by the deformation gradient and the work-conjugate Green--Lagrange strain,

\begin{equation}
    \boldsymbol{F} = \boldsymbol{I} + \nabla \boldsymbol{u}, \qquad
    \boldsymbol{E} = \frac{1}{2} \left( \boldsymbol{F}^\text{T} \boldsymbol{F} - \boldsymbol{I} \right),
    \label{eq:finite_strain_kinematics}
\end{equation}

and the constitutive law relates the second Piola--Kirchhoff stress \( \boldsymbol{S} \) to \( \boldsymbol{E} \), with \( \boldsymbol{P} = \boldsymbol{F} \boldsymbol{S} \). The small-displacement equations are the limiting case of this description: as strains and rotations vanish \( \boldsymbol{F} \to \boldsymbol{I} \), so that \( \boldsymbol{E} \to \boldsymbol{\varepsilon} \) and the stress measures coincide, \( \boldsymbol{P} \to \boldsymbol{S} \to \boldsymbol{\sigma} \), recovering Eqs.~\eqref{eq:momentum}--\eqref{eq:strain}.

In either regime the system is closed by a constitutive relation between a strain measure and its work-conjugate stress. In a classical setting this relation is fully specified by known material laws; here, one or more material functions are unknown and are replaced by learnable operators. In its most general form the ML constitutive model is written as

    \begin{equation}
        \boldsymbol{\sigma} = \mathcal{G}_\theta(\boldsymbol{\varepsilon}) \quad \text{(small strain)}, \qquad
        \boldsymbol{S} = \mathcal{G}_\theta(\boldsymbol{E}) \quad \text{(finite strain)},
        \label{eq:constitutive_ml}
    \end{equation}

where \( \mathcal{G}_\theta \) denotes the constitutive operator with learnable parameters \( \theta \), acting on the strain measure appropriate to the regime. Crucially, $\mathcal{G}_\theta$ is not a black-box mapping from strain to stress; it retains the known physical structure of the constitutive law and embeds a learnable neural network only in place of the specific material function that is unknown. This structure-preserving embedding is identical in both regimes, so the same learning strategy applies whether the kinematics are linearised or finite.

The following subsections detail the problem setup, ground-truth material law, and ML model architecture for each experiment: in the displacement-controlled experiment (Section~\ref{sec:appendix_displacement_controlled}), the sole unknown is the nonlinear elastic modulus, which is expressed as a neural network learning the exponential softening behavior of the ground-truth law. In the load-controlled experiment (Section~\ref{sec:appendix_load_controlled}), the learned elastic operator is frozen and coupled with a second neural network that parameterises the yield-stress function $\sigma_y(p)$ to learn the ground-truth Voce hardening law. The two pretrained operators are then composed into a foundation constitutive model and deployed zero-shot on a three-dimensional torsion problem (Section~\ref{sec:appendix_torque_holed_plate}). Finally, the shear-coupon experiment (Section~\ref{sec:appendix_shear_coupon}) learns a plastic-hardening and a ductile-damage operator from real data within a finite-strain formulation; as there is no ground-truth law, that section describes the forward model and the two network architectures rather than a reference material law.

\subsubsection{Simulating Displacement-Controlled Uniaxial Experiments}
\label{sec:appendix_displacement_controlled}

\begin{figure}[!htbp]
  \centering
  \includegraphics[width=0.27\textwidth]{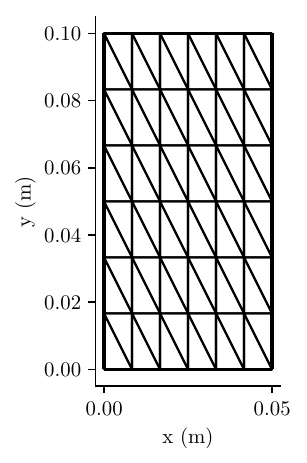}
  \caption{Computational mesh for the displacement-controlled uniaxial test.}
  \label{fig:appendix_learning_from_forces}
  \label{fig:appendix_learning_from_forces_mesh}
\end{figure}

The specimen simulated in Section~\ref{sec:displacement_controlled} is a discretised with a triangurlar mesh, as shown in Figure~\ref{fig:appendix_learning_from_forces_mesh}. The displacement field is approximated with cubic polynomial basis functions.

\paragraph{Ground Truth}
The ground truth constitutive law describes a closed-cell polymeric foam: the Poisson's ratio is fixed to a value of \( \nu = 0.3 \), while the Young's modulus softens under compressive volumetric strain:

\begin{equation}
E(\kappa) = E_{\infty} + (E_{0} - E_{\infty})\,e^{-\kappa / \kappa_0},
\label{eq:young_modulus}
\end{equation}

where \( \kappa = \langle -\operatorname{tr}(\boldsymbol{\varepsilon}) \rangle \) is the Macaulay bracket of the compressive volumetric strain, \( E_0 = 145 \)~MPa is the initial (undeformed) modulus, \( E_{\infty} = 57.3 \)~MPa is the fully softened modulus, and \( \kappa_0 = 0.008 \) controls the rate of softening.

\paragraph{ML Model Architecture}
In this example only the Young's modulus is unknown; the Poisson's ratio is treated as a known constant. The constitutive operator of Eq.~\eqref{eq:constitutive_ml} therefore takes the form:
\begin{equation}
\boldsymbol{\sigma} = \mathcal{G}_\theta(\boldsymbol{\varepsilon})
\;=\; \lambda\!\bigl(\mathcal{N}_\theta(I_1),\,\nu\bigr)\,\mathrm{tr}(\boldsymbol{\varepsilon})\,\mathbf{I}
\;+\; 2\,\mu\!\bigl(\mathcal{N}_\theta(I_1),\,\nu\bigr)\,\boldsymbol{\varepsilon},
\label{eq:displacement_constitutive}
\end{equation}
where \(\mathcal{N}_\theta(I_1)\) is a single MLP that takes the first strain invariant \(I_1 = \mathrm{tr}(\boldsymbol{\varepsilon})\) as input and outputs a positive scalar replacing the Young's modulus, \(\nu = 0.3\) is the known Poisson's ratio, and the Lam\'e coefficients are computed from these quantities via the isotropic linear elastic constitutive relations:
\begin{equation}
\lambda = \frac{\mathcal{N}_\theta(I_1)\,\nu}{(1+\nu)(1-2\nu)}, \qquad
\mu = \frac{\mathcal{N}_\theta(I_1)}{2(1+\nu)}.
\label{eq:lame_from_E}
\end{equation}

The MLP has two hidden layers, each containing thirty neurons, with SiLU and SoftPlus activation functions. A single MLP outputs $E$ from $I_1$, and the Lam\'e parameters are computed via the standard isotropic relations, guaranteeing frame indifference, isotropy, and positivity of the elastic moduli.


\subsubsection{Simulating Load-Controlled Brazilian Disc Experiments}
\label{sec:appendix_load_controlled}

\begin{figure}[!htbp]
  \centering
  \includegraphics[width=0.45\textwidth]{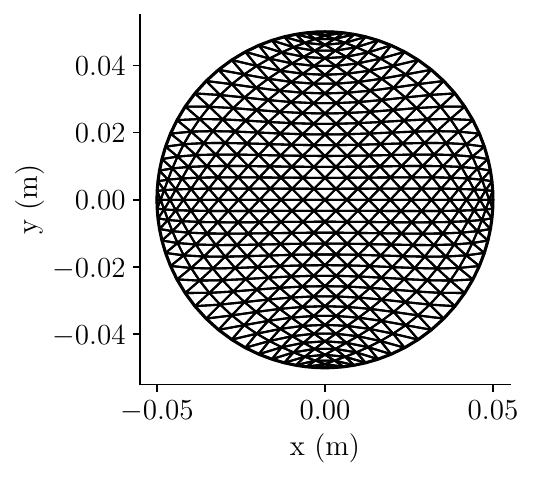}
  \caption{Computational mesh for the load-controlled Brazilian disc test.}
  \label{fig:appendix_learning_from_displacements}
  \label{fig:learning_from_displacements_mesh}
\end{figure}

The specimen simulated in Section~\ref{sec:load_controlled} is a discretised with a triangurlar mesh, as shown in Figure~\ref{fig:learning_from_displacements_mesh}. The displacement field is approximated with linear polynomial basis functions.

\paragraph{Ground Truth}

The ground truth constitutive law combines the elastic softening modulus of Eq.~\eqref{eq:young_modulus} with J2 elastoplasticity and Voce isotropic hardening. The Poisson's ratio is fixed at $\nu = 0.3$ and the elastic modulus $E(\kappa)$ is the same foam law described in Section~\ref{sec:appendix_displacement_controlled}. The Lam\'e parameters are computed from $E$ and $\nu$ via the isotropic linear elastic constitutive relations. The yield stress evolves according to Voce hardening:
\begin{equation}
\sigma_y(p) = \sigma_{y0} + R_\infty\!\left(1 - e^{-b\,p}\right),
\label{eq:voce_hardening}
\end{equation}
where $p$ is the accumulated plastic strain, $\sigma_{y0} = 2.354$~MPa is the initial yield stress, $R_\infty = 0.9$~MPa is the saturation hardening, and $b = 25$ controls the rate of hardening. Given the total strain tensor $\boldsymbol{\varepsilon}$, the stress is computed from the elastic strain as
\begin{equation}
\boldsymbol{\sigma} = \mathbb{C}(\kappa) : \bigl(\boldsymbol{\varepsilon} - \boldsymbol{\varepsilon}^p\bigr),
\qquad
f(\boldsymbol{\sigma},p) = \sigma_{\mathrm{eq}}(\boldsymbol{\sigma}) - \sigma_y(p) \le 0,
\end{equation}
where $\boldsymbol{\varepsilon} = \boldsymbol{\varepsilon}^e + \boldsymbol{\varepsilon}^p$ is the additive decomposition of the (infinitesimal) total strain into elastic and plastic parts, $\sigma_{\mathrm{eq}}$ is the von~Mises equivalent stress, and $f$ is the yield function, so that $\sigma_y(p)$ sets the current admissible stress. The plastic strain $\boldsymbol{\varepsilon}^p$ obeys the associative flow rule, and the accumulated (equivalent) plastic strain $p$---the scalar internal variable that drives hardening---is the time integral of its rate,
\begin{equation}
\dot{\boldsymbol{\varepsilon}}^p = \dot{p}\,\frac{\partial f}{\partial \boldsymbol{\sigma}}, \qquad
\dot{p} = \sqrt{\tfrac{2}{3}\,\dot{\boldsymbol{\varepsilon}}^p \!:\! \dot{\boldsymbol{\varepsilon}}^p}, \qquad
p(t) = \int_0^{t} \dot{p}\,\mathrm{d}t',
\label{eq:accumulated_plastic_strain}
\end{equation}
subject to the Karush--Kuhn--Tucker loading/unloading conditions $\dot{p} \ge 0$, $f \le 0$, $\dot{p}\,f = 0$. The J2 radial-return mapping enforces these conditions at each quadrature point.

\paragraph{ML Model Architecture}

The pretrained elastic operator $E(\kappa)$ from Section~\ref{sec:displacement_controlled} is embedded in the solver with its weights frozen. Because J2 plastic flow is isochoric ($\operatorname{tr}(\boldsymbol{\varepsilon}^p) = 0$), the total volumetric strain equals the elastic volumetric strain, so $\kappa = \langle -\operatorname{tr}(\boldsymbol{\varepsilon})\rangle$ can be evaluated directly from the displacement field without requiring the elastic--plastic decomposition. The only trainable component is the yield-stress function $\sigma_y(p)$, parameterised by the integral monotone neural network described below.

The yield-stress function $\sigma_y(p)$ must be non-decreasing with respect to the accumulated plastic strain $p$ to ensure thermodynamic consistency. A standard approach would be to train a neural network to output $\sigma_y(p)$ directly and add a penalty term to the loss to discourage monotonicity violations. However, a penalty only \emph{encourages} the desired property; it does not guarantee it. Instead, we encode monotonicity directly in the network architecture via an integral formulation that \emph{enforces} it by construction.

The key idea is to let the network learn not the yield stress itself, but the local \emph{hardening rate}—how fast the yield stress grows with plastic strain—and then integrate that rate to obtain the yield stress:
\begin{equation}
\sigma_y(p) \;=\; \sigma_{y0} + \int_0^{p}\!\operatorname{softplus}\!\bigl(\mathrm{MLP}(s)\bigr)\,\mathrm{d}s,
\label{eq:integral_monotone}
\end{equation}
where $\sigma_{y0}$ is a trainable scalar representing the initial yield stress at $p=0$, and $\mathrm{MLP}(s)$ is a multi-layer perceptron with a single hidden layer. Because $\operatorname{softplus}(x) = \log(1 + e^x) > 0$ for all $x$, the integrand is strictly positive everywhere, and therefore the accumulated integral can only increase with $p$. Formally:
\begin{equation}
\frac{\mathrm{d}\sigma_y}{\mathrm{d}p} = \operatorname{softplus}\!\bigl(\mathrm{MLP}(p)\bigr) > 0,
\label{eq:hardening_tangent}
\end{equation}
so $\sigma_y(p)$ is monotonically increasing by construction, regardless of the learned network weights. This is stronger than a soft penalty: the hypothesis space is restricted to physically admissible hardening laws.

Because the integrand contains a neural network, the integral in Eq.~\eqref{eq:integral_monotone} generally has no closed form. We approximate it numerically using Gauss--Legendre quadrature:
\begin{equation}
\int_0^{p}\!\operatorname{softplus}\!\bigl(\mathrm{MLP}(s)\bigr)\,\mathrm{d}s
\;\approx\; \sum_{i=1}^{n_{\mathrm{quad}}} w_i \;\operatorname{softplus}\!\bigl(\mathrm{MLP}(s_i)\bigr),
\end{equation}
where $\{s_i, w_i\}$ are the quadrature nodes and weights. Standard Gauss--Legendre nodes are defined on the reference interval $[-1,\,1]$; at each evaluation they are affinely mapped to the current interval $[0,\,p]$, which varies from one material point and load step to another. Because the integrand is smooth, Gauss--Legendre quadrature converges rapidly; we use $n_{\mathrm{quad}} = 16$ points, which provides sufficient accuracy. Each evaluation of $\sigma_y(p)$ therefore requires $n_{\mathrm{quad}}$ forward passes through the (small) MLP—a negligible cost compared with the PDE solve.

In J2 plasticity, the radial-return algorithm used in the plastic corrector step requires not only $\sigma_y(p)$ but also its derivative with respect to $p$, the hardening tangent $H(p) \coloneqq \mathrm{d}\sigma_y / \mathrm{d}p$. A key advantage of the integral formulation is that, by the fundamental theorem of calculus, this tangent is simply the integrand evaluated at the upper limit:
\begin{equation}
H(p) = \operatorname{softplus}\!\bigl(\mathrm{MLP}(p)\bigr).
\end{equation}
This requires only a single MLP evaluation at $p$—no additional integration, finite differencing, or automatic differentiation through the quadrature routine is needed. The resulting tangent is both exact (it follows from the analytic definition of the model) and inexpensive, providing a smooth and accurate input to the radial-return solver and thereby contributing to the stability of the local plasticity update.

\subsubsection{Simulating Torsional Behaviour in a Slotted Cylindrical Rod}
\label{sec:appendix_torque_holed_plate}

\begin{figure}[!htbp]
    \centering
    \includegraphics[width=0.6\textwidth]{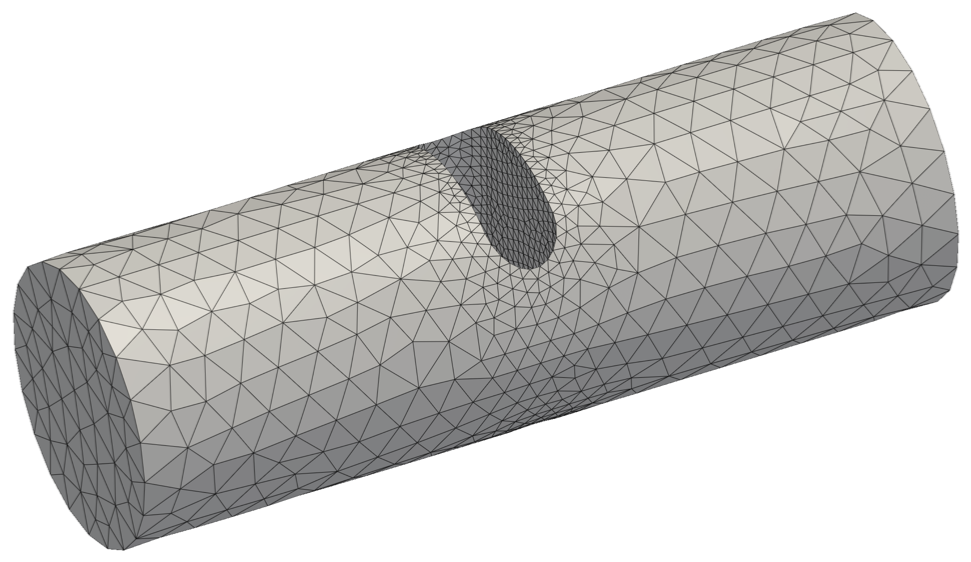}
    \caption{Computational mesh used in the torsion experiment on the slotted cylindrical rod.}
    \label{fig:torque_holed_plate_mesh}
\end{figure}

The cylindrical rod simulated in Section~\ref{sec:torque_holed_plate} is discretised with the tetrahedral mesh shown in Figure~\ref{fig:torque_holed_plate_mesh}, with refinement concentrated in the neighbourhood of the keyhole slot. The displacement field is approximated with cubic polynomial basis functions. The ground truth constitutive law is the foam elastoplastic model: the elastic softening modulus $E(\kappa)$ of Eq.~\eqref{eq:young_modulus} combined with J2 plasticity and the Voce hardening law of Eq.~\eqref{eq:voce_hardening}—the same material used for the training in Sections~\ref{sec:displacement_controlled} and~\ref{sec:load_controlled}.

\subsubsection{Simulating Shear-Coupon Experiments}
\label{sec:appendix_shear_coupon}

\begin{figure}[!htbp]
  \centering
  \includegraphics[width=0.6\textwidth]{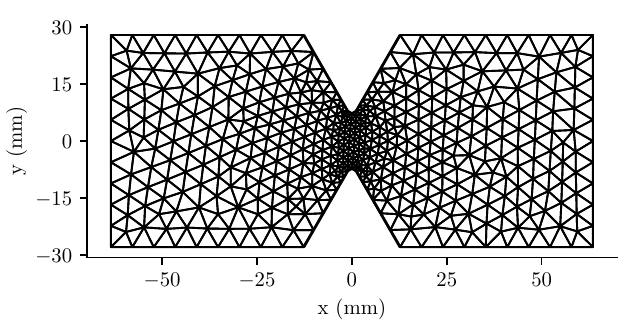}
  \caption{Computational mesh for the shear-coupon experiment.}
  \label{fig:shear_coupon_mesh}
\end{figure}

The specimen simulated in Section~\ref{sec:shear_coupon} is discretised with a triangular mesh, as shown in Figure~\ref{fig:shear_coupon_mesh}. Plane stress is modelled with a mixed formulation: the in-plane displacement is approximated with linear (CG1) basis functions, and the out-of-plane normal strain $E_{33}$ is carried as an additional unknown so that the plane-stress condition $S_{33}=0$ is enforced weakly.

Unlike the previous solid-mechanics examples, the shear coupon develops large rotations within the ligament and is therefore modelled with the finite-strain, total-Lagrangian formulation of Section~\ref{sec:appendix_solid_mechanics_problem_formulation} (Eqs.~\eqref{eq:momentum_ref}--\eqref{eq:finite_strain_kinematics}), with the additive elastic--plastic split $\boldsymbol{E} = \boldsymbol{E}^e + \boldsymbol{E}^p$ of the Green--Lagrange strain. The second Piola--Kirchhoff stress follows the St.~Venant--Kirchhoff law on the elastic strain, degraded by the scalar damage,
\begin{equation}
    \boldsymbol{S} = g(\bar{D})\,\mathbb{C} : \bigl(\boldsymbol{E} - \boldsymbol{E}^p\bigr),
    \label{eq:svk_damaged}
\end{equation}
and the weak form pairs the first Piola--Kirchhoff stress $\boldsymbol{P} = \boldsymbol{F}\boldsymbol{S}$ with the gradient of the test function. Plasticity is governed by a J2 (von Mises) yield criterion with associative flow, $f = \sigma_{\mathrm{eq}} - \sigma_y(p) \le 0$, solved by a radial return on the trial deviatoric stress at each loading step; here $p$ is the accumulated plastic strain (Eq.~\eqref{eq:accumulated_plastic_strain}) and $\sigma_y(p)$ the learned hardening law. The elastic constants are fixed throughout ($E = 115$~GPa from an independent tensile gauge, $\nu = 0.3$).

The scalar damage $\bar{D} \in [0,1]$ degrades the stress through $g(\bar{D}) = (1-\bar{D})^2$. It is driven by the accumulated plastic strain through a local hazard $\Phi_{\mathrm{loc}}(p)$ (the learned damage operator, below). A local damage law of this kind localises into a single element under mesh refinement; we regularise it with an implicit-gradient (nonlocal) filter, smoothing the hazard by a Helmholtz equation with length scale~$\ell$,
\begin{equation}
    \bar{\Phi} - \ell^2 \nabla^2 \bar{\Phi} = \Phi_{\mathrm{loc}}(p),
    \label{eq:helmholtz_damage}
\end{equation}
with natural (Neumann) boundary conditions, and define the nonlocal damage as $\bar{D} = 1 - e^{-\bar{\Phi}}$ (capped at $0.95$). The length scale~$\ell$ sets the width of the softening band, making the localisation mesh-objective rather than collapsing onto a single element. Damage is advanced explicitly (staggered): equilibrium at each step is solved with the damage from the previous step, after which $\bar{D}$ is updated from the new plastic state.

\paragraph{ML Model Architecture}
Two operators are learned, both as monotone neural networks built from the integral of a non-negative activation, so that the relevant physical monotonicity holds by construction; the activation is matched to the qualitative behaviour of each law. The hardening law $\sigma_y(p)$ uses the same integral monotone network as the Brazilian disc test (Section~\ref{sec:appendix_load_controlled}, Eq.~\eqref{eq:integral_monotone}): a strictly positive hardening rate obtained from a SoftPlus output is integrated by Gauss--Legendre quadrature, giving a smooth yield stress that rises monotonically from $p=0$.

The damage hazard $\Phi_{\mathrm{loc}}(p)$ is parameterised by a \emph{rectified integral network}: a single hidden layer of $H$ rectified-quadratic units with non-negative output weights, whose integral over plastic strain gives the cumulative hazard,
\begin{equation}
    r(p) = \sum_{j=1}^{H} s_j\,\langle p - b_j \rangle_{+}^{2}, \quad s_j \ge 0,
    \qquad
    \Phi_{\mathrm{loc}}(p) = \int_0^p r(s)\,\mathrm{d}s = \sum_{j=1}^{H} \frac{s_j}{3}\,\langle p - b_j \rangle_{+}^{3},
    \label{eq:rectified_integral_network}
\end{equation}
where $\langle\,\cdot\,\rangle_{+}$ denotes the positive part (rectifier), $\{b_j\}$ are learnable breakpoints and $\{s_j\}$ learnable non-negative slopes (we use $H=8$, i.e.\ $16$ parameters). This architecture enforces, by construction, the qualitative properties expected of a ductile-damage hazard: $\Phi_{\mathrm{loc}}(0)=0$; monotone non-decreasing accumulation ($s_j\ge 0 \Rightarrow r\ge 0$); a \emph{hard onset}, since each rectified-quadratic unit is exactly zero below its breakpoint, so $r(p)=0$ for $p<\min_j b_j$ (a dormant plateau, with no leakage below onset); and a smooth ($C^1$-continuous), convex, accelerating growth above the onset. The contrast with the hardening operator is deliberate: hardening integrates a \emph{SoftPlus} activation (a smooth rise from $p=0$), whereas damage integrates a \emph{rectified-quadratic} activation (a dormant plateau, a hard onset, then smooth accelerating growth), each matched to the physics of the law it represents. Squared rectifiers are used in place of plain ReLUs precisely so that the learned hazard rate is smooth rather than piecewise-linear. Unlike a fixed closed-form rate, the network learns the shape of the hazard directly from the experimental response.

\subsection{Transient thermodynamics problem formulation}
\label{sec:transient_thermodynamics_problem_formulation}
We consider a transient heat conduction problem on two bodies, where the thermal conductivity of one of the two bodies is a nonlinear function of temperature. This problem is governed by the time-dependent heat equation, expressed as:

\begin{equation}
\rho c_p \frac{\partial T}{\partial t} = \nabla \cdot (k \nabla T),
\end{equation}

where $T$ represents the temperature field in Kelvin, $\rho$ denotes the material density measured in $kg/m^3$, and $c_p$ is the specific heat capacity, expressed in $J\cdot kg^{-1}\cdot K^{-1}$.
The expression $\nabla \cdot (k \nabla T)$ describes the heat flux divergence, while $\frac{\partial T}{\partial t}$ corresponds to the transient variation of temperature.
The term $k$ is the thermal conductivity, measured in $W\cdot m^{-1}\cdot K^{-1}$, which can be a nonlinear function of temperature and parametrised with a neural network:

\begin{equation}
k = \mathcal{G}_\theta(T).
\end{equation}

\subsubsection{Simulating Transient Heat Conduction Experiments}

\begin{figure}[!htbp]
    \centering
    \includegraphics[width=0.6\textwidth]{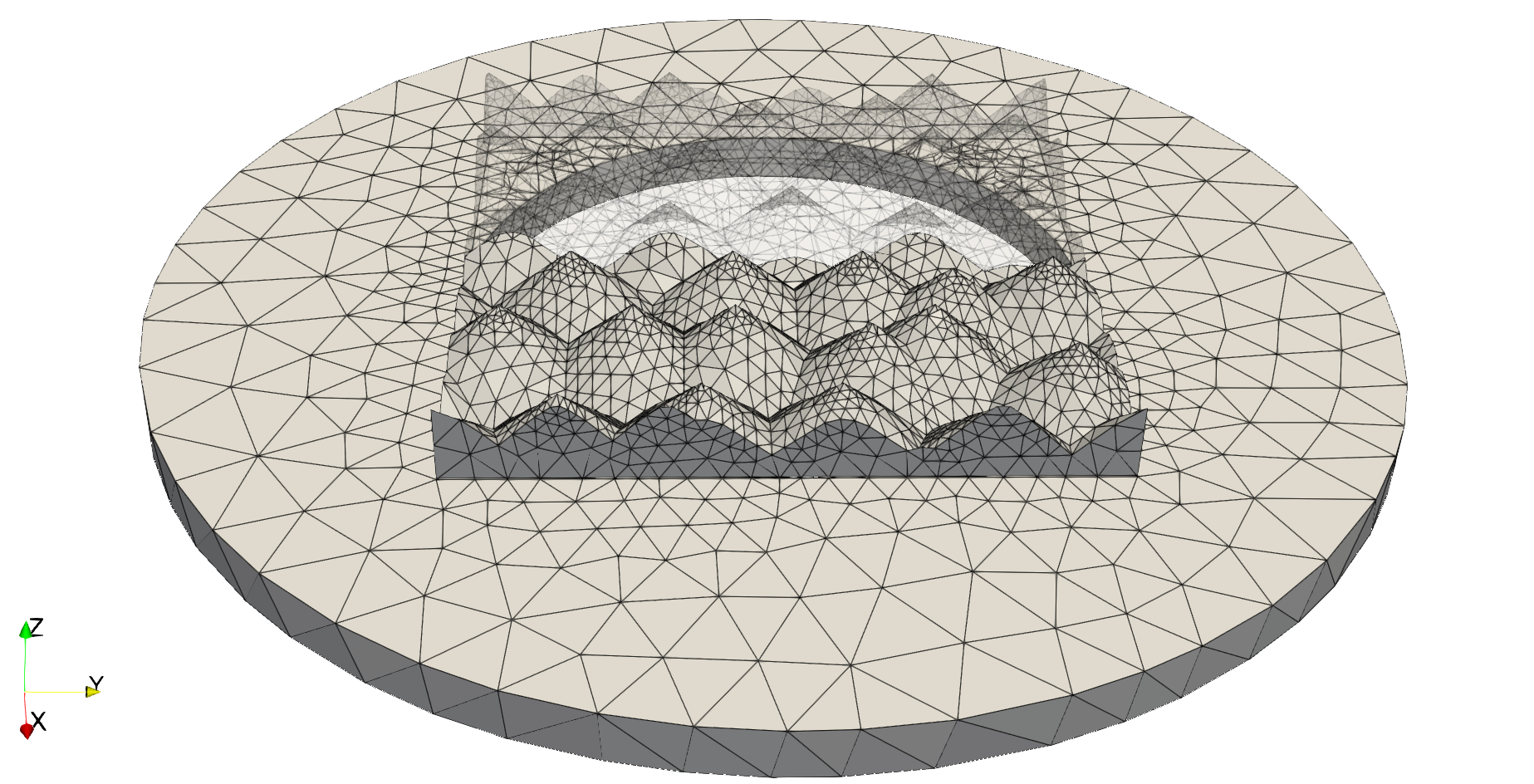}
    \caption{Computational mesh used in the thermal conduction experiment.}
    \label{fig:learning_from_temperatures_mesh}
\end{figure}

The two bodies simulated in Section~\ref{sec:learning_thermal_properties} are discretised with a tetrahedral mesh shown in Figure~\ref{fig:learning_from_temperatures_mesh}, where half of the square plate is transparent to show the hole in the circular bottom plate. The temperature field is discretised with linear basis functions. 

\begin{figure}[!htbp]
    \centering
    \begin{tikzpicture}
    \begin{axis}[
        name=ax,
        width=0.88\textwidth,
        height=0.26\textwidth,
        xlabel={Time (s)},
        ylabel={Source temperature (K)},
        xmin=-5, xmax=248,
        ymin=270, ymax=900,
        clip=false,
        domain=0:240,
        samples=500,
        xtick={0, 40, 80, 120, 160, 200, 240},
        ytick align=outside,
        xticklabel style={anchor=north west, xshift=2pt},
        grid=major,
        grid style={line width=0.2pt, gray!30},
    ]
    \addplot[black, thick, no markers] {298.15 + (x/240)*600*abs(sin(4*x))};
    \coordinate (t0top) at (axis cs:0, 900);
    \coordinate (t0bot) at (axis cs:0, 270);
    \coordinate (t80top) at (axis cs:80, 900);
    \coordinate (t80bot) at (axis cs:80, 270);
    \coordinate (t160top) at (axis cs:160, 900);
    \coordinate (t160bot) at (axis cs:160, 270);
    \coordinate (t240top) at (axis cs:240, 900);
    \coordinate (t240bot) at (axis cs:240, 270);
    \end{axis}
    \node[anchor=north, inner sep=0] (img0) at ([yshift=-14mm]t0bot)
        {\includegraphics[width=0.22\textwidth]{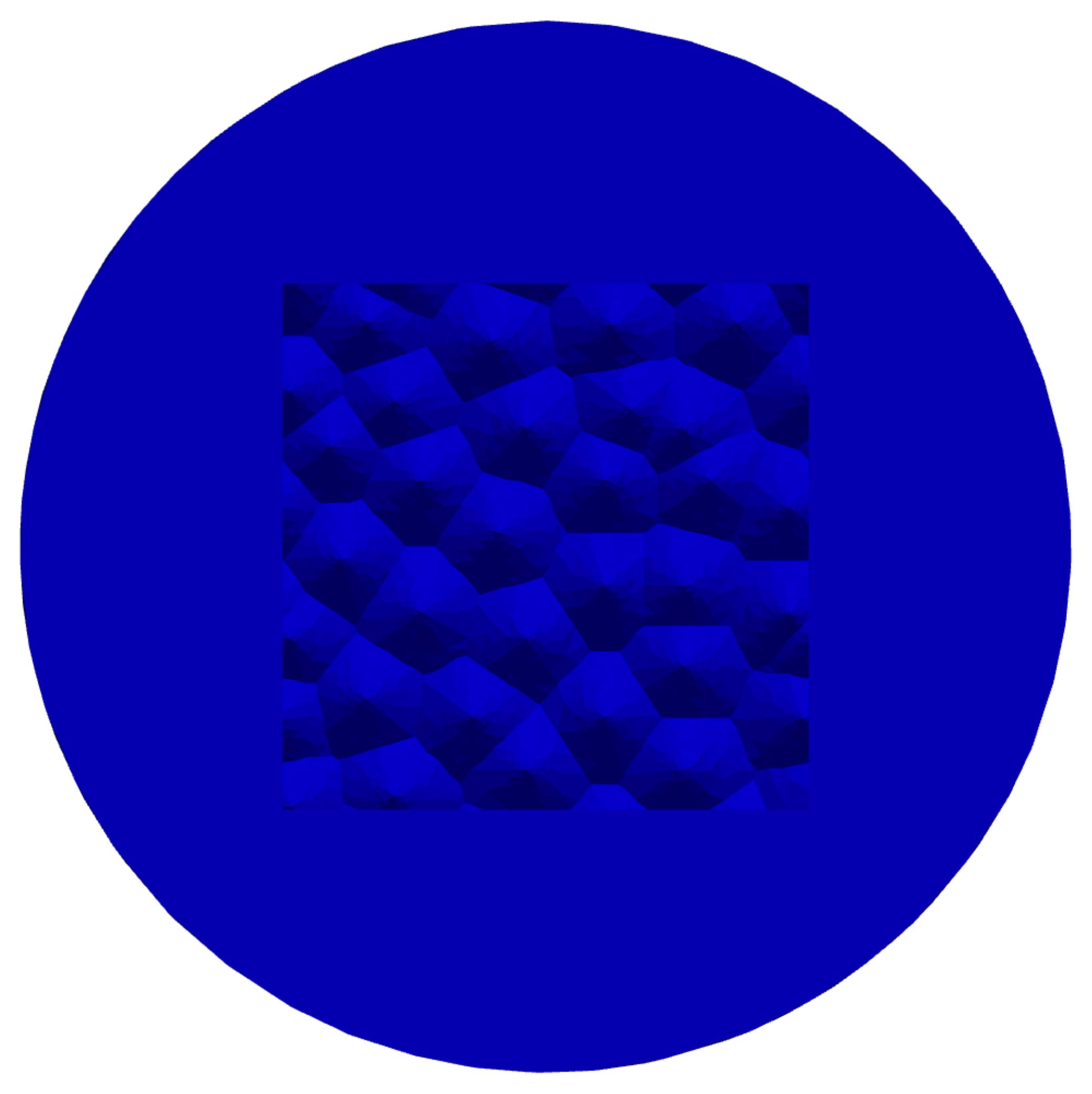}};
    \node[anchor=north, inner sep=0] (img1) at ([yshift=-14mm]t80bot)
        {\includegraphics[width=0.22\textwidth]{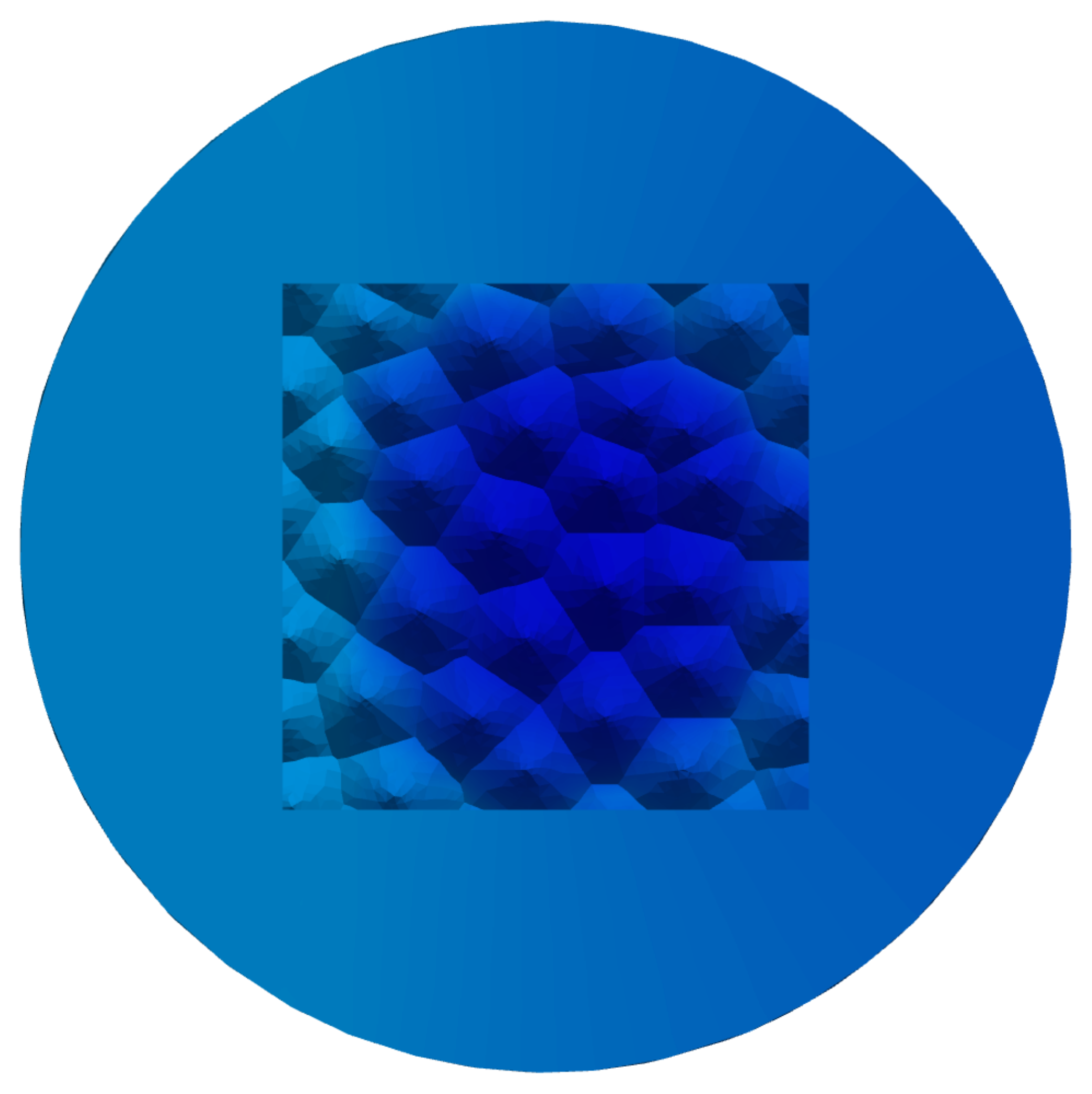}};
    \node[anchor=north, inner sep=0] (img2) at ([yshift=-14mm]t160bot)
        {\includegraphics[width=0.22\textwidth]{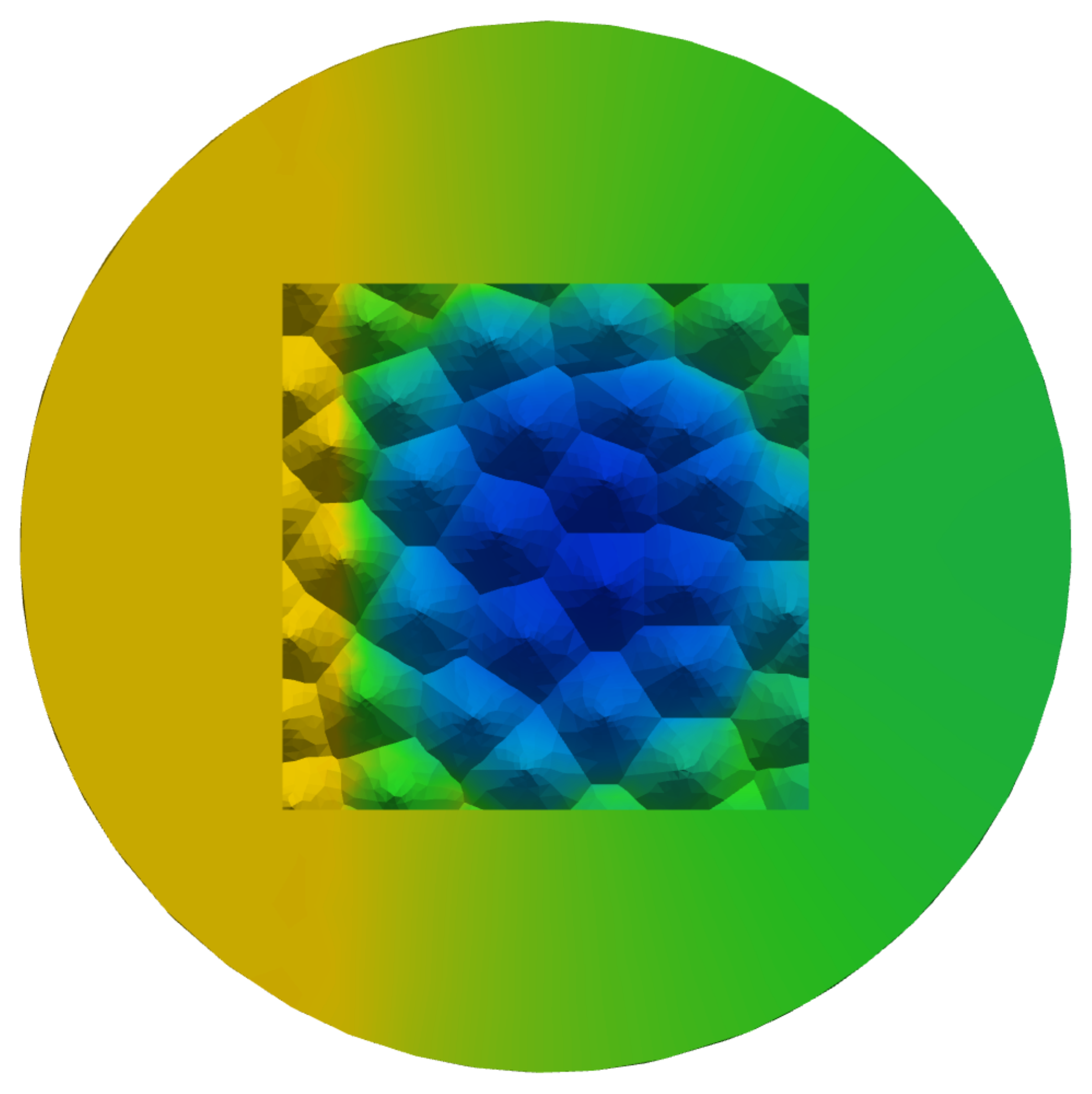}};
    \node[anchor=north, inner sep=0] (img3) at ([yshift=-14mm]t240bot)
        {\includegraphics[width=0.22\textwidth]{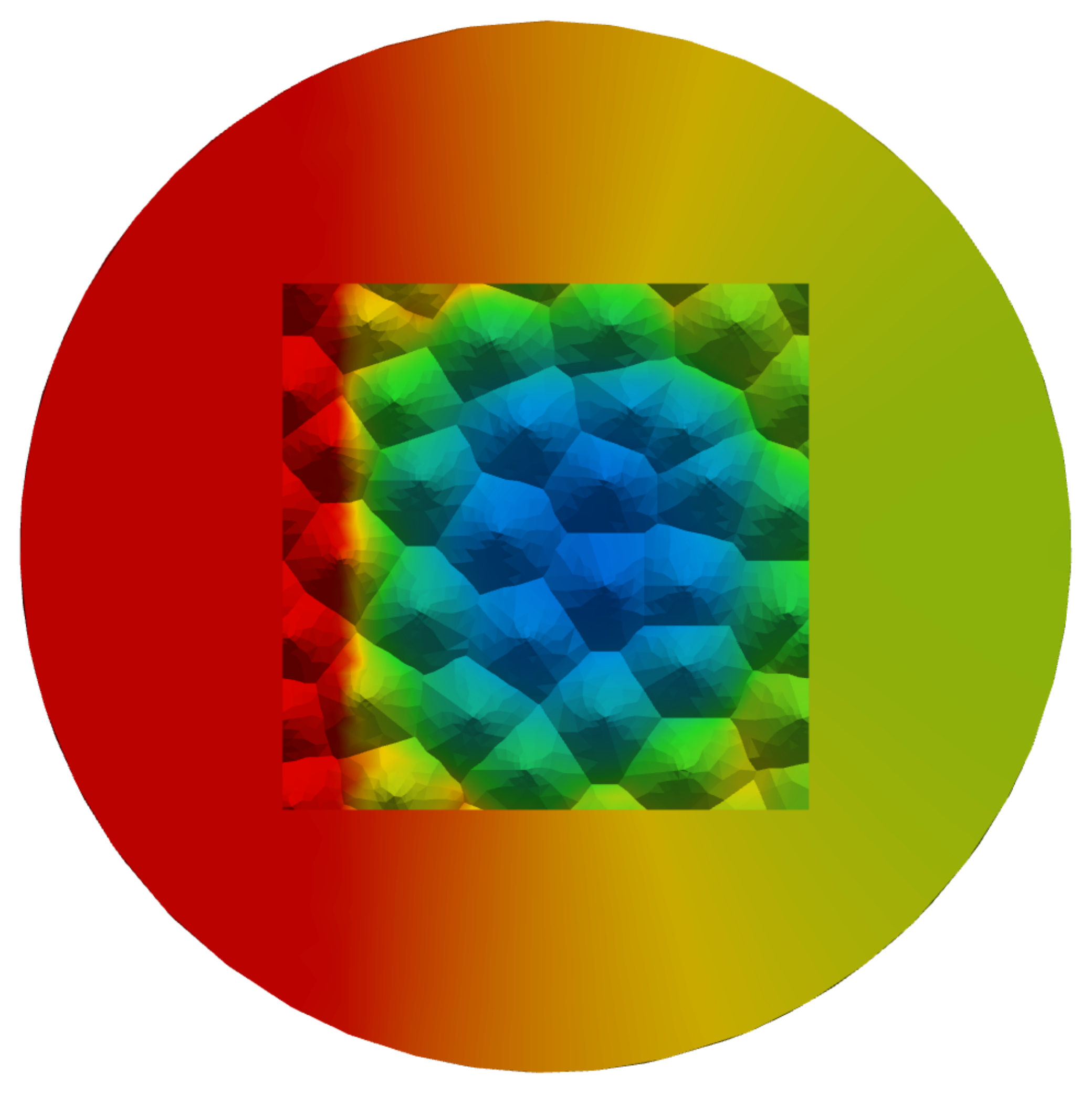}};
    \draw[dashed, black, thin] (t0top) -- (img0.north);
    \draw[dashed, black, thin] (t80top) -- (img1.north);
    \draw[dashed, black, thin] (t160top) -- (img2.north);
    \draw[dashed, black, thin] (t240top) -- (img3.north);
    \node[anchor=north, inner sep=0] at ([yshift=-3mm]$(img0.south)!0.5!(img3.south)$)
        {\includegraphics[width=0.245\textwidth]{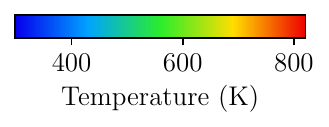}};
    \end{tikzpicture}
    \caption{Fluctuating source temperature boundary condition applied at the left boundary (top) and the resulting temperature field on the simulated domain at four time instants (bottom). The temperature progressively diffuses across the two-body domain.}
    \label{fig:temperature_evolution_snapshots}
\end{figure}

\paragraph{Ground Truth}
The ground truth thermal conductivity law is a nonlinear function of the temperature defined as:

\begin{equation}
k = k_r \left(1 + \beta \frac{T - T_r}{T_r}\right)^{-\delta}
\end{equation}

where \(T\) is the temperature; \(\beta = 1.0\) is a dimensionless constant; \(\delta = 0.62\) is an exponent that characterises the temperature dependence; \(k_r = 2.0\) is a reference thermal conductivity; and \(T_r = 298.0\)~K is the reference temperature. Figure~\ref{fig:temperature_evolution_snapshots} shows the evolution of the temperature field across the domain at four time instants. 

\paragraph{ML Model Architecture}
The ML constitutive model is defined with an MLP with two hidden layers, each containing thirty neurons and using ReLU and Sigmoid as the activation functions. The input of the MLP is the temperature, and the output is the corresponding thermal conductivity, which is then used to solve the PDE.

\subsection{Hyper-parameter summary and sensitivity to training epochs}
\label{sec:appendix_sensitivity}

Table~\ref{tab:hyperparameters} summarises the ML hyper-parameters for the four examples in which an operator is trained (the torsion example of Section~\ref{sec:torque_holed_plate} reuses pretrained operators with no further training). The displacement-controlled example uses a compact MLP with two hidden layers of thirty neurons each, SiLU hidden activations, and the Adam optimiser with a one-cycle learning-rate schedule. The load-controlled plasticity example uses an integral monotone neural network (Section~\ref{sec:appendix_load_controlled}) with a single hidden layer of thirty-two neurons and a one-cycle schedule.

The shear-coupon example (Section~\ref{sec:shear_coupon}, the two shear-coupon columns of Table~\ref{tab:hyperparameters}) is a two-stage, real-data calibration: \emph{Stage~1} learns the hardening law $\sigma_y(p)$ on the rising branch, and \emph{Stage~2} then freezes it and learns the rectified integral network for the damage hazard on the full curve. Both stages minimise the scatter-normalised residual of Eq.~\eqref{eq:shear_loss}, with the residuals grouped into branches that are each averaged internally and combined with fixed weights---elastic, transition, and plateau weighted $2\!:\!1\!:\!1$ in Stage~1, and pre-peak and post-peak weighted $1.5\!:\!2$ in Stage~2---so that branch importance is set explicitly rather than by the number of load steps that fall in each branch. In both stages the gradient is obtained from the firedrake-adjoint, differentiating end-to-end through the finite element solve---including, in Stage~2, the staggered nonlocal-damage update---into the network parameters.

\begin{table}[!htbp]
\centering
\caption{Summary of ML hyper-parameters for the four examples in which an operator is trained; the two-stage shear-coupon calibration is shown as its separate hardening (Stage~1) and damage (Stage~2) stages. In the damage stage the two learning rates apply to the network weights ($0.005$) and to the onset/amplitude scalars ($0.02$).}
\label{tab:hyperparameters}
\footnotesize
\setlength{\tabcolsep}{3pt}
\begin{tabular}{lccccc}
\hline
\textbf{Hyper-parameter} & \textbf{Disp.-controlled} & \textbf{Load-controlled} & \shortstack{\textbf{Shear-coupon}\\\textbf{(hardening)}} & \shortstack{\textbf{Shear-coupon}\\\textbf{(damage)}} & \textbf{Thermal} \\
\hline
Trainable model(s) & 1 MLP & 1 monotone NN & 1 monotone NN & 1 monotone NN & 1 MLP \\
Learned quantity & $E(\kappa)$ & $\sigma_y(p)$ & $\sigma_y(p)$ & $\Phi_{\mathrm{loc}}(p)$ & $k(T)$ \\
Frozen operator(s) & --- & $E(\kappa)$ from Exp.\,1 & $E$ fixed & $E$ fixed;\ $\sigma_y$ & --- \\
Hidden layers & $2 \times 30$ & $1 \times 32$ & $1 \times 32$ & $8$ units & $2 \times 30$ \\
Hidden activation & SiLU & SoftPlus & SoftPlus & rect.-quad. & ReLU \\
Output activation & SoftPlus & SoftPlus (integral) & integral & integral & Sigmoid \\
Optimiser & Adam & Adam & Adam & Adam & Adam \\
Learning rate & 0.03 (max) & 0.05 (max) & 0.03 & 0.005--0.02 & 0.025 \\
Epochs & 200 & 400 & 60 & 60 & 100 \\
Training samples & 6 load steps & 15 load steps & 23 steps & 80 steps & 12 time steps \\
Noise level & 1\% (scalar forces) & 1\% (nodal disp.) & \multicolumn{2}{c}{real data (4 spec.)} & 2\% (nodal temp.) \\
LR scheduler & One-cycle & One-cycle & --- & --- & --- \\
\hline
\end{tabular}
\end{table}

The number of training epochs is the most impactful discrete hyper-parameter, and its adequacy is established directly from the training histories reported for each example. In the displacement-controlled (Figure~\ref{fig:displacement_controlled_plot}), load-controlled (Figure~\ref{fig:plasticity_loss_curves}), and transient heat conduction (Figure~\ref{fig:3d_heat_conduction_plot}) examples, the training and test losses decrease monotonically and then flatten into a plateau over the final epochs. Throughout training the test loss tracks the training loss closely, indicating generalisation rather than overfitting, and the late-epoch plateau shows that the solution has effectively converged well before the final epoch. The reported epoch counts therefore lie within this converged regime, so the learned operators are insensitive to the precise stopping point.

\section*{Acknowledgements}
N.B was supported by the Eric and Wendy Schmidt AI in Science Postdoctoral Fellowship, a Schmidt Futures program. D.A.H. was supported by the Engineering and Physical Sciences Research Council (EPSRC) under grants EP/W029731/1 and EP/W026066/1, and by the Science and Technology Facilities Council (STFC, UKRI) under grant UKRI/ST/B000495/1.

\section*{Data availability}
The experimental data for the real shear-coupon example are from the Second Sandia Fracture Challenge~\cite{boyce2016second} and are included in the code repository below. All other data analysed in this study are synthetic and are generated by the accompanying code (see Code availability).

\section*{Code availability}
The numerical examples were run with the open-source finite element framework Firedrake~\cite{FiredrakeUserManual}. Instructions to run the examples in this paper, together with the Python scripts, meshes, and data required to reproduce all results, are archived on Zenodo (\href{https://doi.org/10.5281/zenodo.21251671}{10.5281/zenodo.21251671}).

\section*{Author contributions}
A.F.; N.B.: Conceptualization; Methodology; Formal analysis; Software; Validation; Writing - original draft; Writing - review \& editing.
D.A.H.: Writing - review \& editing.

\section*{Competing interests}
The authors declare no competing interests.

\bibliography{references}

\end{document}